\documentclass[%
reprint,
superscriptaddress,
showpacs,
amsmath,amssymb,
aps,
prc,
floatfix, ]%
{revtex4-2}

\usepackage{color}
\usepackage{CJK}
\usepackage{graphicx}% Include figure files
\usepackage{dcolumn}% Align table columns on decimal point
\usepackage{bm}% bold math
\usepackage{epstopdf}
\usepackage{multirow}
\usepackage{booktabs}
\usepackage[dvipdfmx,
bookmarks=true,colorlinks,%
citecolor=blue,linkcolor=blue,anchorcolor=blue,filecolor=blue,urlcolor=blue,%
]{hyperref}
\usepackage[toc,titletoc]{appendix}

\begin{document}
\hyphenpenalty=10000
\tolerance=2000
\begin{CJK*}{UTF8}{}
\CJKfamily{gbsn}
		
\title{Angular momentum projection in the deformed relativistic Hartree-Bogoliubov theory in continuum}
		
\author{Xiang-Xiang Sun (孙向向)}
\affiliation{School of Nuclear Science and Technology,
			University of Chinese Academy of Sciences,
			Beijing 100049, China}
\affiliation{CAS Key Laboratory of Theoretical Physics,
			Institute of Theoretical Physics,
			Chinese Academy of Sciences, Beijing 100190, China}

\author{Shan-Gui Zhou (周善贵)}
\email{sgzhou@itp.ac.cn}
\affiliation{CAS Key Laboratory of Theoretical Physics,
			Institute of Theoretical Physics,
			Chinese Academy of Sciences, Beijing 100190, China}
\affiliation{School of Physical Sciences,
			University of Chinese Academy of Sciences,
			Beijing 100049, China}
\affiliation{Center of Theoretical Nuclear Physics, National Laboratory
			of Heavy Ion Accelerator, Lanzhou, 730000, China}
\affiliation{Synergetic Innovation Center for Quantum Effects and Application,
			Hunan Normal University, Changsha, 410081, China}
		
\date{\today}
\begin{abstract}
The angular momentum projection (AMP) method is implemented
in the deformed relativistic Hartree-Bogoliubov theory in continuum (DRHBc)
with the point-coupling density functional.
The wave functions of
angular momentum projected states
are expanded in terms of the Dirac Woods-Saxon (WS) basis,
providing a proper description
of the asymptotic behavior of the wave functions for weakly bound nuclei.
The contribution of continuum induced by the pairing
is considered by treating the pairing correlation
with the Bogoliubov transformation.
We present the formulae and numerical checks for the
DRHBc+AMP approach and use it to study low-lying excited states
of weakly bound deformed nuclei.
Our calculations show that neutron-rich magnesium isotopes
$^{36,38,40}$Mg are all well deformed nuclei.
The low-lying excited states of these three nuclei
are obtained by performing the AMP on the mean-field ground-states.
The ground-state rotational bands of $^{36,38,40}$Mg are reproduced reasonably well
by using this new DRHBc+AMP approach with the density functional PC-F1.
\end{abstract}
\maketitle
\end{CJK*}

\section{Introduction}
With the development of the radioactive-ion-beam facilities,
many exotic nuclear phenomena which differ from the properties
of nuclei close to the $\beta$-stability line have been observed,
including proton or neutron halos
\cite{Tanihata1985_PRL55-2676,Tanihata2013_PPNP68-215},
changes of the nuclear magic numbers
\cite{Ozawa2000_PRL84-5493,Janssens2009_Nature459-1069,
	Wienholtz2013_Nature498-346,Tran2018_NatCommun9-1594},
the island of inversion
\cite{Warburton1990_PRC41-1147},
neutron skin
\cite{Centelles2009_PRL102-122502},
clustering effects
\cite{Ebran2012_Nature487-341,Freer2018_RMP90-035004},
new radioactivities
\cite{Pfuetzner2012_RMP84-567},
nuclear bubble structure
\cite{Mutschler2016_NatPhysP13-152,Yao2013_PLB723-459},
shape coexistence
\cite{Cejnar2010_RMP82-21552212,Heyde2011_RMP83-1467,Li2016_JPG43-024005},
etc.
The study of these exotic structures is at the frontier
of nuclear physics nowadays
\cite{Bender2003_RMP75-121,Cwiok2005_Nature433-705,Meng2006_PPNP57-470,
Heyde2011_RMP83-1467,Meng2015_JPG42-093101,Niksic2011_PPNP66-519,Meng2016_RDFNS,
Zhou2016_PS91-063008,Zhou2017_PoS-INPC2016-373,Freer2018_RMP90-035004,
Otsuka2020_RMP92-015002}.
The description of the structure of exotic nuclei
has been achieved by using many approaches,
e.g., the shell model (SM) approach
\cite{Caurier2005_RMP77-427,Otsuka2020_RMP92-015002},
nuclear density functional theory (NDFT) \cite{Bender2003_RMP75-121,Vretenar2005_PR409-101,Meng2006_PPNP57-470,Meng2016_RDFNS},
antisymmetrized molecular dynamics
\cite{Kimura2016_EPJA52-373},
and few-body models
\cite{Greene2017_RMP89-35006}.
One of the advantages of the NDFT is that it can describe almost all nuclei
in the nuclear chart with global density functionals,
especially for heavy and superheavy nuclei.

The basic implementation of NDFT is achieved by using
self-consistent mean-field (SCMF) methods,
in which the total energy of the system is constructed as a functional of
one-body local nucleon density
\cite{Niksic2011_PPNP66-519}.
The bulk properties of finite nuclei,
including binding energy, radius, deformation, etc.,
have been successfully described by using SCMF methods
\cite{Bender2003_RMP75-121,Meng2006_PPNP57-470,Vretenar2005_PR409-101,
Egido2016_PS91-073003,Meng2016_RDFNS,Nazarewicz2018_NatPhys14-537}.
In general, the wave function obtained from MF calculations in the
intrinsic frame is approximated by a single Slater determinant and
allowed to break symmetries of the Hamiltonian,
such as particle number conservation and
rotational and translation invariances
\cite{Ring1980,Bender2003_RMP75-121,Meng2016_RDFNS,Egido2016_PS91-073003,
Schunck2019_EDF-Nuclei,Robledo2019_JPG46-013001}.
As a consequence, the MF wave function cannot be used to study
correlations corresponding to the spontaneous symmetry breaking,
quantum fluctuation of collective degrees of freedom,
spectroscopic observable in the laboratory frame,
and selection rules of the transitions.
These deficiencies can be complemented
via beyond-mean-field (BMF) calculations based on SCMF methods
\cite{Ring1980,Sheikh2019_arXiv1901.06992}.
The violation of SO(3) symmetry in the intrinsic frame
for deformed nuclei
and U(1) symmetry in the gauge space for superfluid nuclei
can be restored by using the angular momentum projection (AMP)
and particle number projection (PNP), respectively
\cite{Ring1980,Meng2016_RDFNS,Schunck2019_EDF-Nuclei}.
The quantum fluctuation of collective degrees of freedom
is usually treated with the generator coordinate
method (GCM) \cite{Ring1980}.
In principle, these broken symmetries should be restored by using
projection before variation (PBV) calculations \cite{Ring1980},
which are technically very complicated and have been rarely achieved
in NDFT, especially for the case of the AMP,
see, e.g., Ref.~\cite{Sheikh2019_arXiv1901.06992} for a recent review.
Usually the projection after variation (PAV) approach is adopted to
restore the broken symmetries within the framework of NDFT.

The AMP has been successfully implemented in non-relativistic and
relativistic MF models
(see Refs.~\cite{Bender2003_RMP75-121,Niksic2011_PPNP66-519,
Egido2016_PS91-073003,Robledo2019_JPG46-013001,
Sheikh2019_arXiv1901.06992} and references therein) and has
been used to explain or predict many exotic nuclear structures
connected with the nuclear collective excitation,
for instance,
the structure of low-spin and high-spin states \cite{Rodriguez-Guzman2000_PRC62-054308,Bender2003_PRC68-044321,Bender2004_PRC69-064303},
shape coexistence in Kr and Pb isotopes
\cite{Rodriguez-Guzman2004_PRC69-054319,Bender2006_PRC74-024312},
shell evolution in neutron rich Ti and Cr isotopes
\cite{Rodriguez2007_PRL99-062501},
shape transitions
\cite{Niksic2007_PRL99-092502,Rodriguez2008_PLB663-49},
low-lying excitation of hypernuclei
\cite{Cui2015_PRC91-054306,Cui2017_PRC95-024323,
	Mei2018_PRC97-064318,Xia2019_SciChinaPMA62-042011},
excitation of triaxially deformed nuclei
\cite{Bender2008_PRC78-024309,Rodriguez2010_PRC81-064323,
	Yao2009_PRC79-044312,Yao2010_PRC81-044311,Yao2014_PRC89-054306,
	Egido2016_PRL116-052502,Chen2017_PRC95-024307},
and
the structure and fission of superheavy nuclei
\cite{Marevic2020_PRL125-102504,Egido2020_PRL125-192504,Egido2021_PRL126-192501}.
Besides, it is worth mentioning that the BMF calculations
have been performed to study the excitation of odd $N(Z)$ nuclei
\cite{Bally2014_PRL113-162501,Borrajo2016_EPJA52-277,
Borrajo2017_PLB764-328,Borrajo2018_PRC98-044317}.

It should be noted that in
the above-mentioned calculations with the AMP,
the wave functions of the intrinsic and
excited states are almost all expanded in terms of
the harmonic oscillator (HO) wave functions
\cite{Pannert1987_PRL59-2420,Price1987_PRC36-354,Gambhir1990_AP198-132}.
The HO wave functions can be obtained analytically
and have great advantages for numerical treatments.
But the asymptotic behavior of the wave function
in a weakly bound system cannot be described
properly with this basis,
even if the size of the basis space is taken to be very large \cite{Stoitsov1998_PRC58-2086,Zhou2000_CPL17-717,Zhou2003_PRC68-034323,Zhang2013_PRC88-054305}.
Therefore the AMP has been rarely applied to study
loosely bound nuclei, especially for halo nuclei.

Nuclear halos are characterized by the large spatial extension
and formed in loosely bound nuclei when the valence nucleons
close to the threshold of particle emission
occupy low $l$ ($s$- or $p$-wave) orbitals with considerable amplitudes
\cite{Hansen1987_EPL4-409,
Dobaczewski1996_PRC53-2809,
Meng1996_PRL77-3963,
Meng1998_PRL80-460,
Meng1998_NPA635-3,
Jensen2004_RMP76-215,
Riisager2013_PS2013-014001}.
Therefore when studying halo nuclei by employing SCMF approaches,
the single particle wave functions are usually obtained in coordinate ($r$) space by
using the shooting and matching method
\cite{Dobaczewski1984_NPA422-103,Meng1998_NPA635-3},
the finite element solution \cite{Poeschl1997_CPC103-217250},
and the Lagrange-mesh method \cite{Typel2018_FPhys6-73}.
Alternatively, in configuration space, the wave function can be expanded by
a set of proper basis functions,
such as the Woods-Saxon (WS) basis
\cite{Zhou2003_PRC68-034323}
and the transformed HO basis
\cite{Stoitsov1998_PRC58-2086,Stoitsov1998_PRC58-2092}.
Pairing correlations play a vital role in the formation of halos
and are usually treated by using the Bogoliubov transformation \cite{Dobaczewski1984_NPA422-103,Dobaczewski1996_PRC53-2809,Meng1998_NPA635-3}.
For spherical halo nuclei,
by solving the Hartree-Fock-Bogoliubov (HFB) or
relativistic Hartree-Bogoliubov (RHB) equation with spherical potentials,
the ground-state property
\cite{Meng1996_PRL77-3963,Poeschl1997_PRL79-3841,
	Meng1998_NPA635-3,Meng1998_PLB419-1,Long2010_PRC81-031302R}
can be well described.
Deformation-driven halos are common for halo nuclei in medium mass region,
such as those observed in $^{31}$Ne
\cite{Nakamura2009_PRL103-262501,Nakamura2014_PRL112-142501}
and $^{37}$Mg \cite{Kobayashi2014_PRL112-242501}.
Within the framework of SCMF,
the first self-consistent study of deformed halo nuclei has been
achieved by using the deformed relativistic Hartree-Bogoliubov theory
in continuum (DRHBc) \cite{Zhou2010_PRC82-011301R} and after that
many deformed halo nuclei have
been predicted by using MF approaches
\cite{Pei2013_PRC87-051302R,
Chen2014_PRC89-014312,Nakada2018_PRC98-011301R,
Li2012_PRC85-024312,Sun2018_PLB785-530,
Zhang2019_PRC100-034312,Sun2020_NPA1003-122011}.
The establishment of rotational bands of deformed halo nuclei
is helpful to understand the halo structure and configuration
\cite{Nakamura2014_PRL112-142501},
but up to now there are almost no such kind of theoretical investigations within the
framework of the NDFT.

The covariant density functional theory (CDFT)
has become a powerful tool to study the properties of stable and exotic nuclei
over the whole nuclear chart
with universal density functionals
\cite{Reinhard1989_RPP52-439,Ring1996_PPNP37-193,
Vretenar2005_PR409-101,Meng2006_PPNP57-470,
Niksic2011_PPNP66-519,Meng2015_JPG42-093101,
Liang2015_PR570-1,Meng2016_RDFNS}.
For the study of halo nuclei
within the framework of the CDFT,
the relativistic continuum Hartree-Bogoliubov (RCHB)
\cite{Meng1996_PRL77-3963,Meng1998_NPA635-3,Meng1998_PRL80-460,Meng2002_PRC65-041302R}
and relativistic HFB theories
\cite{Long2010_PRC81-031302R,Lu2013_PRC87-034311}
have been developed for spherical halos and the DRHBc theory
based on the Dirac WS basis for deformed halos
\cite{Zhou2010_PRC82-011301R,Li2012_PRC85-024312}.
When studying halos in deformed nuclei,
shape decoupling effects originated from the intrinsic structure of valence levels
have been predicted
by using the DRHBc theory \cite{Zhou2010_PRC82-011301R}.
Deformed halos with shape decoupling effects in C, Ne, and Mg isotopes have been revealed
by using this theory
\cite{Zhou2010_PRC82-011301R,Li2012_PRC85-024312,Sun2018_PLB785-530,
Zhang2019_PRC100-034312,Sun2020_NPA1003-122011}.
Especially, the DRHBc theory can well explain the halo structures in $^{17,19}$B
\cite{Yang2021_PRL126-082501,Sun2021_PRC103-054315}.
In addition, the construction of
the DRHBc nuclear mass table is in progress
\cite{Zhang2020_PRC102-024314,In2021_IJMPE30-2150009,Zhang2021_arXiv2103.08142,
Pan2021_arXiv2104.07337,He2021_arXiv2104.12987}.

The implementation of AMP in the relativistic mean field (RMF)
models with the HO basis has been realized
\cite{Niksic2011_PPNP66-519}.
In Refs.~\cite{Niksic2006_PRC73-034308,Niksic2006_PRC74-064309},
the BMF methods for axially deformed nuclei with spatial reflection symmetry
have been developed.
Three-dimensional (3D) AMP
\cite{Yao2009_PRC79-044312,Yao2010_PRC81-044311}
has been applied to study
low-lying excited states of triaxially deformed nuclei.
Beyond RMF approaches have been also
used to investigate nuclear octupole excitations
\cite{Yao2015_PRC92-041304R}.
Recently, the AMP based on the multidimensionally-constrained (MDC) CDFTs
\cite{Lu2012_PRC85-011301R,Lu2014_PRC89-014323,Zhou2016_PS91-063008,Zhao2017_PRC95-014320}
has been developed
\cite{Wang2021_MDC-RHB+AMP}.
The calculations by using MDC-CDFTs+AMP can describe the
properties of both the ground-state in the MF level
and low-lying excited states in the laboratory frame
for systems with various deformations,
such as $\beta_{20},\beta_{22},\beta_{30},\beta_{32},\beta_{40}$, etc.,
in a microscopic and self-consistent way.
It is desirable to develop the AMP based on the DRHBc theory to study
the properties of the collective motion for weakly bound deformed nuclei.

In the DRHBc theory,
the MF wave function is expanded in terms of the Dirac WS basis,
which can also be used to construct the angular momentum projected states.
In this way, a proper description of the asymptotic behavior of the wave functions in
excited states for a weakly bound nucleus is achieved.
The angular momentum projection after variation has been developed
based on the DRHBc theory,
aiming at a microscopic description of low-lying excitation of the loosely bound nuclei,
especial for deformed halo nuclei.
As a first application of the DRHBc+AMP approach, the rotational excitation of deformed halo nuclei
has been explored and it is found that both the halo structure and shape decoupling
effects can appear in rotational excited states \cite{Sun2021_arXiv2103.10886}.
In this work, we take $^{36,38,40}$Mg as examples and present in detail
how to implement the AMP into the DRHBc theory, careful numerical checks, and
the study of ground-state rotational bands of these three nuclei.

This paper is organized as follows.
The main formulae of the DRHBc+AMP approach are given in Sec.~\ref{Sec:frame}.
We perform numerical checks of this newly developed approach in Sec.~\ref{Num}.
The applications on $^{36,38,40}$Mg are given and discussed in Sec.~\ref{Sec:res}.
We summarize this work in Sec.~\ref{Sec:summ}.

\section{Theoretical Framework}
\label{Sec:frame}
\subsection{The DRHBc theory}
The DRHBc theory with both the meson-exchange
\cite{Zhou2010_PRC82-011301R,Li2012_PRC85-024312,Chen2012_PRC85-067301}
and point-coupling \cite{Zhang2020_PRC102-024314}
effective interactions have been developed.
The AMP is implemented based on the
point-coupling density functionals.
Here we briefly introduce
the main formulae of the DRHBc theory
with the point-coupling density functionals;
more details can be found in
Ref.~\cite{Zhang2020_PRC102-024314}.
We start from the effective Lagrangian
\begin{equation}
\begin{split}
&\mathcal{L} =
\bar{\psi}
\left(
\mathrm{i} \gamma_{\mu} \partial^{\mu} - M
\right)
\psi
-\frac{1}{2} \alpha_{S}(\bar{\psi} \psi)(\bar{\psi} \psi) \\
&
-\frac{1}{2} \alpha_{V}
\left(\bar{\psi} \gamma_{\mu} \psi\right)
\left(\bar{\psi} \gamma^{\mu} \psi\right)
-\frac{1}{2} \alpha_{TS}
(\bar{\psi} \vec{\tau} \psi) \cdot
(\bar{\psi} \vec{\tau} \psi) \\
&
-\frac{1}{2} \alpha_{TV}
\left(\bar{\psi} \vec{\tau} \gamma_{\mu} \psi\right) \cdot
\left(\bar{\psi} \vec{\tau} \gamma^{\mu} \psi\right)
-\frac{1}{2} \delta_{S}
\left(\partial_{\mu} \bar{\psi} \psi\right)
\left(\partial^{\mu} \bar{\psi} \psi\right) \\
&
-\frac{1}{2} \delta_{V}
\left(\partial_{\mu} \bar{\psi} \gamma_{\mu} \psi\right)
\left(\partial^{\mu} \bar{\psi} \gamma^{\mu} \psi\right)
-\frac{1}{2} \delta_{TS}
\left(\partial_{\mu} \bar{\psi} \vec{\tau} \psi\right) \cdot
\left(\partial^{\mu} \bar{\psi} \vec{\tau} \psi\right) \\
&
-\frac{1}{2} \delta_{TV}
\left(\partial_{\mu} \bar{\psi} \vec{\tau} \gamma_{\mu} \psi\right)\cdot
\left(\partial^{\mu} \bar{\psi} \vec{\tau} \gamma^{\mu} \psi\right)
-\frac{1}{3} \beta_{S}(\bar{\psi} \psi)^{3} \\
&
-\frac{1}{4} \gamma_{S}(\bar{\psi} \psi)^{4}
-\frac{1}{4} \gamma_{V}
\left[
\left(\bar{\psi} \gamma_{\mu} \psi\right)
\left(\bar{\psi} \gamma^{\mu} \psi\right)
\right]^{2} \\
&
-e A_{\mu} \bar{\psi}
\frac{\left(1-\tau_{3}\right)}{2}
\gamma^{\mu} \psi
-\frac{1}{4} F_{\mu \nu} F^{\mu \nu},
\end{split}
\end{equation}
where $M$ and $e$ are the mass of nucleon and the unit charge.
$\psi$, $A_\mu$, and $F_{\mu\nu}$ are the Dirac spinor fields of nucleons,
four-vector
potential and field strength tensor for the electromagnetic field.
This Lagrangian contains 11 parameters:
$\alpha_S$, $\alpha_V$, $\alpha_{TS}$, $\alpha_{TV}$,
$\delta_s$, $\delta_V$, $\delta_{TS}$, $\delta_{TV}$,
$\beta_S$, $\gamma_S$, and $\gamma_V$.
In these symbols,
$\alpha$ means the four-fermion coupling terms,
$\delta$ refers to derivative terms,
$\beta$ and $\gamma$ are for the third- and forth-order terms.
The subscripts $S$, $V$, and $T$ mean scalar, vector,
and iso-vector, respectively.

Under the MF and no-sea approximations,
the total energy of the system is constructed as a functional of nucleon densities.
In the DRHBc theory, by using the Bogoliubov transformation,
the MF and pairing correlations are
treated self-consistently
\cite{Dobaczewski1984_NPA422-103,Meng1998_NPA635-3}.
The equation of motion for nucleons is the deformed
RHB equation
\cite{Kucharek1991_ZPA339-23}
and reads
\begin{equation}\label{equ.1}
	\left(
	\begin{array}{cc}
		 h_D - \lambda_\tau &    \Delta             \\
		-\Delta^*           & -{h}^*_D+\lambda_\tau \\
	\end{array}
	\right)
	\left( { U_{k} \atop V_{k} } \right)
	=
	E_{k}
	\left( { U_k \atop V_{k} }  \right),
\end{equation}
where
$\lambda_\tau$ $(\tau={n,p})$ is the Fermi energy.
$(U_k,V_k)^T$ is the quasi particle wave function
with energy $E_k$ and
is expanded in terms of the Dirac WS basis,
\begin{equation}
\begin{split}
	& U_k(\bm rs) = \sum_{n\kappa} u_{k,(n\kappa)}^{(m)}
                    \varphi_{n\kappa m}(\bm r s) ,\\
	& V_k(\bm rs) = \sum_{n\kappa} v_{k,(n\kappa)}^{(m)}
                    \bar\varphi_{n\kappa m}(\bm r s).
\end{split}
\end{equation}
The Dirac WS basis is obtained by solving the Dirac equation
in $r$ space with the spherical WS scalar and vector
potentials \cite{Koepf1991_ZPA339-81,Zhou2003_PRC68-034323}
and the basis function reads
\begin{equation}
\varphi_{n\kappa m}(\bm r s)  = \frac{1}{r}
	\left(
\begin{array}{c}
    iG_{n\kappa}(r)\mathcal{Y}_{jm}^{l}(\Omega s) \\
    -F_{n\kappa}(r)\mathcal{Y}_{jm}^{\tilde l}(\Omega s)
\end{array}
\right),
\end{equation}
where $\mathcal{Y}_{jm}^{l}(\Omega s)$ is the spin spherical harmonics
with the total angular momentum $j$, orbital angular momentum $l$,
and the projection $m$ of the total angular momentum
on the symmetry axis.
$G_{n\kappa}(r)/r$ and $F_{n\kappa}(r)/r$
are radial wave functions for the upper and lower components
of the Dirac spinor with the radial quantum number $n$
and the relativistic quantum number $\kappa = (-)^{j+l+1/2}(j+1/2)$.
$\bar\varphi_{n\kappa m}(\bm r s)$ is the time reversal partner
of $\varphi_{n\kappa m}(\bm r s)$.

The Dirac Hamiltonian reads
\begin{equation}
h_D =
\bm \alpha\cdot \bm p
 + V(\bm r)
 + \beta[M+S(\bm r)],
\end{equation}
where $S(\bm r)$ and $V(\bm r)$ are the scalar and the vector potentials.

The pairing potential is written as
\begin{equation}
\Delta(\bm r_1, \bm r_2)
= V^{pp}(\bm r_1, \bm r_2)
 \kappa(\bm r_1, \bm r_2),
\end{equation}
where $\kappa(\bm{r}_1,\bm{r}_2)$ is the pairing tensor
\cite{Ring1980,Blaizot1985_QTFS}
and a density-dependent zero-range force
\begin{equation}
	V^{pp}(\bm r_1, \bm r_2)=
    \frac{1}{2}V_0(1-\hat P^\sigma)
	\delta(\bm r_1-\bm r_2)
	\left[
    1-\left(\frac{\rho({\bm r_1})}{\rho_{\mathrm{sat}}} \right)
    \right],
	\label{eq:pairing-force}
\end{equation}
is used in the present work.

In the intrinsic frame,
for axially symmetric and spatial reflection symmetric nuclei,
the densities and potentials are expanded in terms of
the Legendre polynomials,
\begin{equation}
	f(\bm r) = \sum_\lambda f_\lambda (r)
    P_\lambda (\cos\theta),\quad \lambda = 0,2,4,\cdots,
	\label{eq:expansion}
\end{equation}
with
\begin{equation}
	f_\lambda (r) = \frac{2\lambda +1}{4\pi}
	\int d\Omega f(\bm r )P_\lambda (\cos\theta).
\end{equation}
The angular averaged density is equal to the spherical component ($\lambda=0$) of
the corresponding density [cf. Eq.~(\ref{eq:expansion})].

After getting the wave functions by solving the RHB equation,
the total energy of the system can be obtained
\begin{equation}
E= E_\mathrm{kin}+ E_{\mathrm{pair}} +
   E_{\mathrm{c.m.}} + E_\mathrm{int}.
\label{Eq:energy}
\end{equation}
For the interaction part, one has
\begin{equation}
\begin{aligned}
E_\mathrm{int} =
& \int d^{3} r
\left\{
\frac{1}{2} \alpha_{S} \rho_{S}^{2}
+
\frac{1}{2} \alpha_{V} \rho_{V}^{2}
+
\frac{1}{2} \alpha_{T V} \rho_{3}^{2}
\right.\\
&+
\frac{2}{3} \beta_{S} \rho_{S}^{3}
+
\frac{3}{4} \gamma_{S} \rho_{S}^{4}
+
\frac{3}{4} \gamma_{V} \rho_{V}^{4}
+
\frac{1}{2} \delta_{S} \rho_{S} \Delta \rho_{S} \\
&\left.+
\frac{1}{2} \delta_{V} \rho_{V} \Delta \rho_{V}
+
\frac{1}{2} \delta_{T V} \rho_{3} \Delta \rho_{3}
+
\frac{1}{2} e A_{0} \rho_{p}\right\},
\end{aligned}
\end{equation}
where the densities read
\begin{equation}
\begin{aligned}
\rho_{S}(\boldsymbol{r})
&=\sum_{k>0} V_{k}^{\dagger}(\boldsymbol{r})
  \gamma_{0} V_{k}(\boldsymbol{r}),
\\
\rho_{V}(\boldsymbol{r})
&=\sum_{k>0} V_{k}^{\dagger}(\boldsymbol{r})
             V_{k}(\boldsymbol{r}) ,\\
\rho_{3}(\boldsymbol{r})
&=\sum_{k>0} V_{k}^{\dagger}(\boldsymbol{r})
    \tau_{3} V_{k}(\boldsymbol{r}).
\end{aligned}
\label{Eq:den}
\end{equation}
The kinetic energy is given by
\begin{equation}
 E_\mathrm{kin} =\mathrm{Tr} [\rho t],
\end{equation}
where $\rho$ and $t$ are the density and kinetic energy matrices.
In the Dirac WS basis, the matrix elements of $\rho$ and $t$ can
be expressed as
\begin{equation}
\begin{aligned}
 &\rho^m_{n\kappa,n'\kappa'} =
 \sum_{k>0} v^{(m)}_{k,n\kappa}v^{(m)}_{k,n'\kappa'}  ,\\
 &t^m_{n\kappa,n'\kappa'} =
 \int d \bm r
 \varphi^\dagger_{n\kappa m}(\bm r s)
 \left( \bm \alpha \cdot \bm p + \beta M \right)
 \varphi_{n'\kappa' m}(\bm r s).
\end{aligned}
\end{equation}
The pairing energy is
\begin{equation}
E_{\mathrm{pair}}=-\frac{1}{2} \operatorname{Tr}[\Delta \kappa].
\end{equation}
The correction energy of center-of-mass spurious motion is considered
after getting the single particle wave functions
\cite{Bender2000_EPJA7-467,Long2004_PRC69-034319,
Zhao2009_CPL26-112102} and reads
\begin{equation}
E_{\mathrm{c.m.}}
=-\frac{1}{2 A M}\left\langle\hat{\bm P}^{2}\right\rangle,
\end{equation}
where $\hat{\bm P}$ is the total momentum for nucleus and the mass number is labeled by $A$.

The root-mean-square (rms) matter radius for the proton ($\tau=-1$)
and neutron ($\tau=1$) are calculated as
\begin{equation}
\begin{aligned}
R_{\tau} &
=\left \langle r^{2} \right \rangle^{1/2}
=\left(
\frac{1}{N_\tau}
\int d^{3} \boldsymbol{r} r^{2} \rho_{V}^{\tau}(\boldsymbol{r})
\right)^{1 / 2} \\
&=\left(
\frac{\sqrt{4 \pi}}{N_\tau}
\int d rr^{4}\rho_{V}^{\tau, \lambda=0}(r)
\right)^{1 / 2},
\end{aligned}
\end{equation}
and the charge radius is
\begin{equation}
R_{\mathrm{ch}} = \sqrt{R_p^2 + 0.64\ \mathrm{fm}^2}.
\end{equation}
The intrinsic quadrupole moment is defined as
\begin{equation}
\begin{aligned}
Q_{\tau} = &
\sqrt{\frac{16 \pi}{5}}
\left\langle
r^{2} Y_{2 0}(\theta, \phi)
\right\rangle \\
=&\frac{8 \pi}{5}
\int \mathrm{d} \bm r
\left[
r^{4} \rho_{V }^{\tau,\lambda=2}(r)
\right],
\end{aligned}
\end{equation}
and then the quadrupole deformation parameters can be written as
\begin{equation}
\beta_{\tau}=
\frac{\sqrt{5 \pi} Q_{\tau}}
{3 N_{\tau}\left\langle R_{\tau}^{2}\right\rangle},
\end{equation}
where $N_{\tau}$ is the number of protons  ($\tau=1$) or neutrons ($\tau=-1$).

The quadrupole deformation constraint calculations can be achieved by replacing
the Dirac Hamiltonian $h_D$ by $\tilde h_D$
\cite{Sun2020_NPA1003-122011,Serot1986_ANP16-1,
Reinhard1989_RPP52-439}
\begin{equation}
	\tilde h_D = h_D +
	  c_1 \left(\langle \hat Q_{2} \rangle -\bar Q_{2}  \right)
	+ c_2 \left(\langle \hat Q_{2} \rangle -\bar Q_{2} \right)^2,
\end{equation}
where $c_1$ and $c_2$ are the Lagrange multiplier and the penalty parameter, respectively.
$\bar Q_{2}$ is the desired expectation value of the quadrupole moment $\hat{Q}_2$.

The canonical basis \cite{Ring1980} can be obtained by diagonalizing
the density matrix in the Dirac WS basis \cite{Li2012_PRC85-024312}
\begin{equation}
\sum_{n'\kappa'} \rho^m_{n\kappa,n'\kappa'} c^i_{n'\kappa'} = v_i^2 c^i_{n\kappa},
\end{equation}
where the eigenvalue $v_i^2$ is the BCS occupation probability
of a single particle level (SPL) and the eigenvector in coordinate
space is constructed as
\begin{equation}
\phi_i(\bm r s) = \sum_{n\kappa} c^i_{n\kappa m}\varphi_{n\kappa m}(\bm r s).
\end{equation}
Here, for axially symmetric and spatial reflection symmetric nuclei,
each SPL can be labeled by $m^\pi$ with the parity $\pi$.

\subsection{Angular momentum projection}
\label{Sec:2.2}
Due to the breaking of spherical symmetry by
the axially deformed MF potential in the intrinsic frame,
the wave function $|\Phi(\beta)\rangle$ with a certain quardupole
deformation parameter $\beta$ is not an eigenvector
of angular momentum operators $\hat{J_z}$ and $\hat{J^2}$.
A low-lying excited state with good angular momentum can be constructed
by performing the AMP on $|\Phi(\beta)\rangle$ given by DRHBc calculations
and reads \cite{Ring1980}
\begin{equation}
   |\Psi^{JM}_\alpha(\beta)\rangle
   = \sum_K f^{JK}_\alpha
    \hat{P}^J_{MK}
    |\Phi(\beta)\rangle,
\label{eq:amp1}
\end{equation}
where $f^{JK}$ is a coefficient and the angular momentum projection operator reads
\begin{equation}
    \hat{P}_{{MK}}^{J}=
    \frac{2 J+1}{8 \pi^{2}} \int d \Omega
    D_{{MK}}^{J *}(\Omega) \hat{R}(\Omega),
\end{equation}
with the Euler angles $\Omega \equiv (\phi,\theta,\varphi)$,
the Wigner function $D_{{MK}}^{J}(\Omega)$,
and the rotational operator
$\hat{R}(\Omega) =
e^{-i\phi    \hat{J_z}}
e^{-i\theta  \hat{J_y}}
e^{-i\varphi \hat{J_z}}$.
The energy $E^{J}$ and $f^{JK}$ of a projected state
can be calculated by solving the Hill-Wheeler (HW) equation
\cite{Ring1980}
\begin{equation}
\begin{aligned}
\sum_{K}f^{JK}_\alpha
&
\left[
\left\langle
\Phi(\beta)|
\hat{H} \hat{P}_{{MK}}^{J}
|\Phi(\beta)
\right\rangle
\right.
\\
&\left.
-E^J_\alpha
\left\langle \Phi(\beta)|
\hat{P}_{{MK}}^{J}
|\Phi(\beta)
\right\rangle\right] =0.
\end{aligned}
\end{equation}

For axially deformed nuclei, the calculation of $E^J$ and $f^{JK}$ can be simplified
because $\hat{J_z}|\Phi(\beta)\rangle = 0$.
The integration over $\phi$ and $\varphi$ can be calculated analytically.
Using the properties of the projection operator and spatial reflection symmetry,
the expectation value of
multipole operator $\hat Q_{\lambda\mu}$ with respect to the projected
state is \cite{Niksic2006_PRC73-034308}
\begin{equation}
\begin{aligned}
&\left\langle\Phi\left(\beta\right)\left|
\hat{Q}_{\lambda \mu}
\hat{P}_{{MK}}^{J}
\right|\Phi\left(\beta\right)\right\rangle
=(2J+1) \delta_{M-\mu} \delta_{K 0}\\
&~~~~\times
\int_{0}^{\pi / 2}
\sin\theta
d_{-\mu 0}^{J *}(\theta)
\left\langle
\Phi\left(\beta\right)
\left |
\hat{Q}_{\lambda \mu}
e^{-i \theta \hat{J}_{y}}
\right|
\Phi\left(\beta\right)
\right\rangle
d \theta,
\end{aligned}
\label{NLK}
\end{equation}
Since $K=0$, $f^{JK}$ can be replaced by $f^J$.
$E^{J}$ and $f^J$ are calculated as
\cite{Hara1995_IJMPE4-637}
\begin{equation}
\begin{aligned}
E^{J} &
= \frac{ \langle\Phi(\beta)
         \left|  \hat{H}\hat{P}^J_{00} \right|
			\Phi(\beta)\rangle}
       {\langle\Phi(\beta)
         \left  |\hat{P}^J_{00}        \right|
			\Phi(\beta)\rangle},
		\\
		f^J &=
		\frac{1}
        {\sqrt{\langle\Phi(\beta)
         \left| \hat{P}^J_{00}         \right|
				\Phi(\beta)\rangle}}.
	\end{aligned}
\label{eq:hw1}
\end{equation}
The normal overlap kernel
\cite{Niksic2006_PRC73-034308}
reads
\begin{equation}
\begin{aligned}
\mathcal{N}^{J}\left(\beta\right)\equiv
&
\left\langle\Phi\left(\beta\right)
\left|\hat{P}_{00}^{J}\right|
\Phi\left(\beta\right)\right\rangle \\
=
&(2 J+1)
\int_{0}^{\pi / 2}
\sin \theta d_{00}^{J *}(\theta) \\
&
\times
\left\langle \Phi\left(\beta\right)
\left|
e^{-i \theta \hat{J}_{y}}
\right|
\Phi\left(\beta\right)
\right\rangle d \theta,
\end{aligned}
\end{equation}
and the Hamiltonian overlap kernel is
\begin{equation}
\begin{aligned}
\mathcal{H}^{J}\left(\beta\right)
\equiv&
\left\langle\Phi\left(\beta\right)
\left|\hat{H} \hat{P}_{00}^{J}\right|
\Phi\left(\beta\right)
\right\rangle \\
=&
(2 J+1)
\int_{0}^{\pi / 2}
\sin \theta d_{00}^{J *}(\theta) \\
&
\times
\left\langle\Phi\left(\beta\right)
\left|\hat{H} e^{-i \theta \hat{J}_{y}}\right|
\Phi\left(\beta\right)\right\rangle
d \theta.
\end{aligned}
\end{equation}

For the calculation of the normal overlap kernel and Hamiltonian overlap kernel,
the generalized Wick's theorem is used
\cite{Valor2000_NPA671-145,Balian1969_INCB164-37,
Onishi1966_NP80-367,Bonche1990_NPA510-466}
and in this work we use the formulae and notation
given in Ref.~\cite{Yao2009_PRC79-044312}.
In practical calculations, the wave functions of single particle states
in the canonical basis with tiny occupation probabilities $v^2$
have negligible contribution to kernels.
Therefore a truncation $\xi$ on the occupation probability is introduced,
which can reduce the numerical computational efforts
\cite{Valor2000_NPA671-145,Yao2009_PRC79-044312} effectively.
We will discuss this truncation on the SPLs in Sec. \ref{Sec3.2}.

The normal overlap defined as
$n(\beta;\theta) \equiv
\langle \Phi(\beta)|\hat{R}(\theta)|\Phi(\beta)\rangle$
with $\hat{R}(\theta) \equiv e^{-i\theta\hat{J}_y}$ is calculated as
\begin{equation}
   n(\beta;\theta) =  \sqrt{\mathrm{det}D \ \mathrm{det} R},
\label{Eq:NOL}
\end{equation}
where $R$ is the rotational matrix and the
matrix elements can be easily obtained
\begin{equation}
\begin{split}
{R}_{mm'}  =
&\langle \phi_m | \hat{R}(\theta)|\phi_{m'}\rangle\\
=
&
\sum_{n\kappa}\sum_{n'\kappa'}
c^m_{n\kappa}c^{m'}_{n'\kappa'}
\delta_{jj'}\delta_{ l l'}d^{j'}_{mm'}(\theta),
\end{split}
\end{equation}
and satisfies
\begin{equation}
    R_{\bar m m'} = -R_{m \bar m'} ^*, \qquad
    R_{\bar m\bar m'} =R_{m m'} ^*.
\end{equation}
We notice that it is simpler to calculate the rotation matrix
elements in the Dirac WS basis than in
the HO basis shown in Refs.~\cite{Niksic2006_PRC73-034308,Yao2009_PRC79-044312}
because the Dirac WS basis functions are eigenvectors of angular momentum operators
and
\begin{equation}
  \langle n\kappa m| \hat{R}(\theta)| n'\kappa'm'
  \rangle
  =
  \delta_{jj'}\delta_{ ll'}d^{j'}_{mm'}(\theta).
\end{equation}
The matrix elements of $D$ are
\cite{Yao2009_PRC79-044312}
\begin{equation}
\begin{split}
& D_{mm'} =
u_m ( R^T)^{-1}_{mm'}u_{m'}+
v_m R^*_{mm'}        v_{m'}, \\
& D_{m  \bar m'} =
u_m ( R^T)^{-1}_{m \bar m'}u_{m'}+
v_m R^*_{m\bar m'}         v_{m'},
\end{split}
\end{equation}
and one finds the following relations
\begin{equation}
    D_{\bar m m'} = -D_{m \bar m'} ^*, \qquad
    D_{\bar m\bar m'} =D_{m m'} ^*.
\end{equation}
The subscript $m$ represents each SPL
in the canonical basis and $\bar m$ is the time reversal state of $m$.

We follow the procedures given in
Refs.~\cite{Niksic2006_PRC73-034308,Yao2009_PRC79-044312}
to calculate the Hamiltonian overlap kernel
\begin{equation}
\begin{split}
 \mathcal{H}^{J}\left(\beta\right)=
 (2 J+1) \int_{0}^{\pi / 2}
 \sin \theta d_{00}^{J *}(\theta)	
  n(\beta;\theta)
 \mathcal{E}(\beta;\theta)d\theta,
\end{split}
\end{equation}
where the mixed energy density has the form of
\begin{equation}
\mathcal{E}(\beta;\theta)
= \int d^3 r
\mathcal{E}\left[\rho(\boldsymbol{r} ; \beta; \theta)
               \kappa(\boldsymbol{r} ; \beta; \theta)
           \right],
\end{equation}
with the mixed density $\rho(\boldsymbol{r}; \beta; \theta)$
and pairing density $\kappa(\boldsymbol{r}; \beta; \theta)$ in $r$
space for each Euler angle $\theta$.
$\mathcal{E}(\beta;\theta) $ has the
similar structure with Eq.~(\ref{Eq:energy}).
The interaction part can be obtained by replacing
the normal densities in Eq.~(\ref{Eq:den}) by the mixed densities.
It should be mentioned that in AMP calculations,
the rotation operation breaks the time reversal symmetry,
therefore the spatial components of the currents
have contribution to the total energy.
In coordinate space,
the mixed densities and currents are
\begin{equation}
\begin{aligned}
\rho_{V}\left(\boldsymbol{r} ; \beta; \theta\right)
& =
\sum_{i, j}
{\bar \phi}_{i}\left(\boldsymbol{r} ; \beta \right)
       \rho_{ji}(\theta)
       \phi_{j}\left(\boldsymbol{r} ; \beta \right),
\\
\rho_{3}\left(\boldsymbol{r} ; \beta; \theta \right)
& =
\sum_{i,j}
{\bar\phi}_{i}\left(\boldsymbol{r} ; \beta \right)
      \tau_{3}
      \rho_{ji}(\theta)
      \phi_{j}\left(\boldsymbol{r} ; \beta \right),
\\
j^{\mu}\left(\boldsymbol{r} ; \beta;\theta\right)
&=
\sum_{i,j}
\bar\phi_{i}\left(\boldsymbol{r} ; \beta\right) \gamma^{\mu}
\rho_{ji} (\theta)
    \phi_{j}\left(\boldsymbol{r} ; \beta\right).
\end{aligned}
\end{equation}
where $\rho_{ji}(\theta)$ is the mixed density matrix
in the canonical basis for each Euler angle and
can be calculated after obtaining $R_{mm'}$ and $D_{mm'}$.
More details can be found in Ref.~\cite{Yao2009_PRC79-044312}.

In the DRHBc theory, the intrinsic densities are axially symmetric
along the $z$-axis and spatial-reflection symmetric.
Therefore the density is expressed as a linear combination of the
Legendre polynomials [cf. Eq.~(\ref{eq:expansion})].
For the mixed densities,
the rotational invariance along the $z$-axis is broken
but kept along the $y$-axis
and the spatial reflection symmetry is also held.
For the currents, the symmetry of the time-component is the same as that of
the mixed densities and the spatial-components are spatial reflection asymmetric.
So in the DRHBc+AMP approach, we expand the mixed densities and currents
in terms of the spherical harmonics
\begin{equation}
f(r,\vartheta,\omega) =
\sum_{l=0}^\infty \sum_{m=-l}^{m=l}
a_{lm}(r)Y_{lm}(\vartheta,\omega),
\label{eq:exp2}
\end{equation}
where
\begin{equation}
a_{lm} (r)=
\int_0^{2\pi} d\omega
\int_0^\pi \sin\vartheta d\vartheta
 Y_{lm}^*(\vartheta,\omega)
f(r,\vartheta,\omega).
\end{equation}
For the mixed scalar density and vector density, we have
\begin{equation}
\rho (\bm r) = \sum_{l}\sum_{m=-l}^{m=l}
\rho_{lm}(r)Y_{lm}(\vartheta,\omega),
l = 0,2,4,\dots,
\label{SH_den}
\end{equation}
and
$\rho_{l-m}(r) = (-1)^m\rho_{lm}(r)$.

For the spatial-components of the mixed currents
\begin{equation}
\vec{j}(\bm r) =
\sum_{l}\sum_{m=-l}^{m=l}
\vec{j}_{lm}(r)Y_{lm}(\vartheta,\omega),
l = 1,3,5,\dots,
\label{SH_cur}
\end{equation}
with
$j_{x(z),l-m}(r) = (-1)^mj_{x(z),lm}(r)$ and $j_{y,l-m}(r) = (-1)^{m+1}j_{y,lm}(r)$.
The details about how to calculate the mixed densities and currents
in coordinate space within the framework of the DRHBc+AMP are given in
Appendix \ref{APP:A1}.

After the calculation of the mixed densities and currents,
the interaction part of $\mathcal{E}(\beta;\theta)$ can be obtained.
The Coulomb part of the mixed energy density is
calculated as
\begin{equation}
\mathcal{E}_{\mathrm{em}}\left(\boldsymbol{r} ;\theta\right)
=\frac{e^{2}}{8 \pi}
 \rho_{p}\left(\boldsymbol{r} ;\theta\right)
 \int d^{3} \boldsymbol{r}^{\prime} \frac{\rho_{p}\left(\boldsymbol{r}^{\prime};\theta\right)}
 {\left|\boldsymbol{r}-\boldsymbol{r}^{\prime}\right|}.
\end{equation}
As what is usually done,
the exchange term of Coulomb energy is neglected.
In Appendix \ref{APP:B1}, we show how to calculate the Coulomb energy in detail.
The pairing part of the mixed energy density is given by
\begin{equation}
\mathcal{E}_{\mathrm{pair}}\left(\boldsymbol{r} ;
\theta\right)
=-\sum_{\tau}
\frac{V_{\tau}(\bm r;\theta)}{4}
\kappa_{\tau}^{*}\left(\boldsymbol{r} ;
\theta\right) \kappa_{\tau}\left(\boldsymbol{r} ;
\theta\right),
\end{equation}
and the mixed densities are used when calculating $V_{\tau}(\bm r;\theta) $.
The correction energy of the center-of-mass spurious motion in the AMP
is taken to be the same as
that in MF calculations.
To consider the correction from the breaking of particle numbers,
following the procedures in Ref.~\cite{Yao2010_PRC81-044311},
a term with the form of
$-\lambda_p[Z(\bm r; \theta)-Z_0] -\lambda_n[N(\bm r; \theta)-N_0]$
is added into the mixed energy density.
$Z_0$ and $N_0$ are the number of protons and neutrons for a given nucleus
and $\lambda_p$ ($\lambda_n$) is the Fermi energy for protons (neutrons) of
the intrinsic state $|\Phi(\beta)\rangle$.
$Z(\bm r; \theta)$ and $N(\bm r; \theta)$
are the mixed vector densities in $r$ space
for protons and neutrons, respectively.
After the calculation of the mixed energy density,
the solution of Eq.~(\ref{eq:hw1}) can be gotten.

In this work, we study even-even nuclei and focus on
excited states with positive parity.
The reduced transition probability from an initial state
$I_i^+$ to a finally state $I_f^+$
is calculated as \cite{Rodriguez-Guzman2002_NPA709-201}
\begin{equation}
\begin{split}
&B\left({E} 2, I_{i}^+\rightarrow I_{f}^+\right)
=
\frac{e^{2}}{2 I_{i}+1}
\left|
 \left\langle I_{f}
 \left\|\widehat{Q}_{2}\right\| I_{i}\right\rangle
\right|^{2},
\end{split}
\end{equation}
where the reduced matrix element of $\hat{Q_2}$ is
\begin{equation}
\begin{split}
&\left\langle I_{f}
\left\|
\widehat{Q}_{2}
\right\| I_{i}
\right\rangle
 =
 \hat{I_i}\hat{I_f}
 \sum_{\mu^{\prime}}
 \left(
\begin{array}{ccc}
	{I_{i}} & {2}   & {I_{f}} \\
    {-\mu^{\prime}} & {\mu^{\prime}} & {0}
\end{array}
\right) \\
&
\times
\int_{0}^{\pi / 2}
d \theta \sin\theta
d_{-\mu^{\prime} 0}^{I_{i} *}(\beta)
\left\langle
\Phi\left(\beta\right)
\left|\widehat{Q}_{2 \mu^{\prime}}
e^{-i \theta \hat{J}_{y}}\right|
\Phi\left(\beta\right)
\right\rangle.
\end{split}
\end{equation}
with $\hat{I_i} =2I_i+1$ and $\widehat{Q}_{2 \mu} = r^2 Y_{2\mu}$.

The spectroscopic quadrupole moment for a state $I^+$ is
\begin{equation}
\begin{split}
Q^{(s)}(I^+)
&=
e \sqrt{\frac{16 \pi}{5}}
\left(\begin{array}{ccc}
I & 2 & I \\
I & 0 & -I
\end{array}
\right)
\left\langle I\left\|\hat{Q}_{2}\right\|
I\right\rangle.
\end{split}
\end{equation}
The dimensionless quardupole deformation parameter $\beta^s$
for a state $I^+$ can be calculated from $Q^{(s)}(I^+)$
\cite{Yao2015_PRC91-024301}
\begin{equation}
\beta^s(I^+) = \sqrt{\frac{5}{16\pi}}
\frac{4\pi}{3ZR^2}
\left(-\frac{2I+3}{I}\right)
Q^{(s)}(I^+),
\end{equation}
with the charge number $Z$ and $R=1.2A^{1/3}$ fm.

\section{Numerical checks}
\label{Num}
In this section, we check the numerical parameters
involved in DRHBc+AMP calculations in detail.
The whole numerical process includes two parts:
The MF (i.e., DRHBc) and AMP calculations.
For MF calculations with the point-coupling
and meson-exchange density functionals,
the numerical details have been presented in
Ref.~\cite{Zhang2020_PRC102-024314} and
Ref. \cite{Li2012_PRC85-024312}, respectively.
In this work, most of the parameters
are taken to be the same as those given in Ref. \cite{Zhang2020_PRC102-024314}
and here we reexamine the energy cut-off in the Fermi sea of the Dirac WS basis
in order to save the computation time.
For AMP calculations, we examine the accuracy of
the integral for the calculations of the normal overlap
and the expansion of the spherical harmonics in Eq.~(\ref{eq:exp2}).
Two truncation parameters are introduced to determine
the number of the SPLs
used in AMP calculations and
the convergence of
energies of projected states and reduced transition probability
with respect to these two cut-off parameters are shown.

\subsection{MF calculations}
In the particle-hole channel, the density function PC-F1
\cite{Burvenich2002_PRC65-044308}
is adopted to compare our results obtained
from DRHBc+AMP calculations
with those in Ref.~\cite{Yao2009_PRC79-044312}.
The box size $R_\mathrm{box}$ used to generate the Dirac WS basis
can be approximated by
$4r_0 A^{1/3}$ with $r_0=1.2$ fm for light nuclei
\cite{Li2012_PRC85-024312} and is taken to be
$20$ fm for other nuclei.
The mesh size $\Delta r$ is equal to 0.1 fm.
The order of the Legendre expansion is up to 6 in Eq.~(\ref{eq:expansion}).
The angular momentum cut-off is taken to be $21/2\hbar$.
By adjusting pairing gaps around $^{38}$Mg (three-point formula),
the pairing strength $V_0$ is taken to be $240$ MeV fm$^{-3}$ and $325$ MeV fm$^{-3}$
for neutrons and protons, respectively, which slightly differ from those values used
in Ref. \cite{Yao2011_PRC83-014308} with a density-independent zero-range  pairing force.
For the pairing window,
the cut-off energy is taken to be 60 MeV in the quasiparticle space
\cite{Li2012_PRC85-024312}.
In the DRHBc theory,
an energy cut-off $E_\mathrm{cut}$ is introduced to determine the
number of basis states in the Fermi sea and the number of basis states in the Dirac sea
is the same as that in the Fermi sea.
For the nuclear mass table calculation with the density functional PC-PK1
\cite{Zhang2020_PRC102-024314},
the energy cut-off for positive energy states $E_\mathrm{cut}=300$ MeV
for the Dirac WS basis,
which can provide an accuracy about $0.001\%$ for
global calculations of the total energy.
If we use this cut-off energy in DRHBc+AMP calculations with PC-F1,
it takes too much CPU-hours due to the very large space size of the single particle basis.
Therefore we recheck the relative accuracy of the bulk properties
with respect to $E_\mathrm{cut}$ for calculations with the density functional PC-F1
and find a relatively small and reasonable value of $E_\mathrm{cut}$
which can ensure the precision, as well as save the computation time,
especially for AMP calculations.

\begin{figure}[h]
	\begin{center}
		\includegraphics[width=0.49\textwidth]{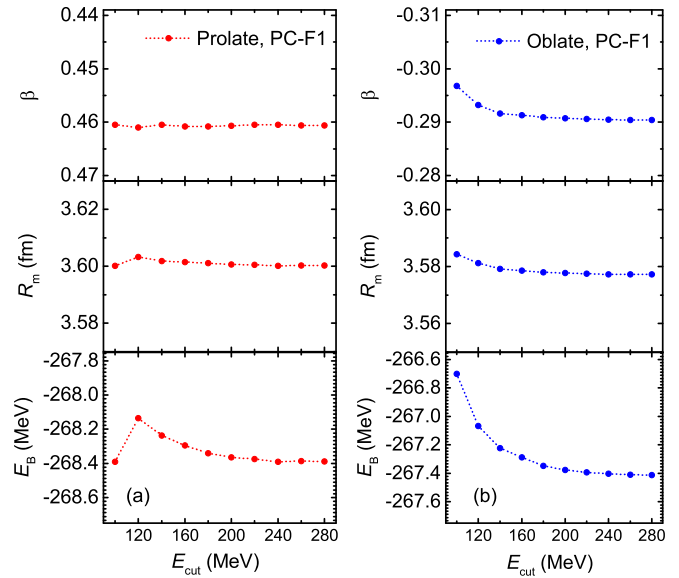}
	\end{center}
	\caption{
	The total energy $E_\mathrm{B}$, rms matter radius $R_m$,
	and quardupole deformation parameter $\beta$ of the ground-state (a) and oblate isomer (b)
	for $^{38}$Mg as a function of $E_\mathrm{cut}$ in DRHBc calculations with PC-F1.
	}
	\label{fig:Mg38_MF_ecut}
\end{figure}

For $^{38}$Mg, two energy minima are found in the potential energy curve
and the ground-state has a prolate shape.
In Fig.~\ref{fig:Mg38_MF_ecut},
we show the calculated bulk properties of $^{38}$Mg in the ground-state
with prolate shape and in the oblate minimum
by using the DRHBc theory with the density functional PC-F1.
From this figure, it is obvious that with the increasing of $E_\mathrm{cut}$,
the total energy, rms matter radius, and deformation parameter all converge well
at  $E_\mathrm{cut}=200$ MeV.
The difference of total energies between $E_\mathrm{cut}=200$ MeV
and $E_\mathrm{cut}=220$ MeV is about 0.01 MeV.
This means that when $E_\mathrm{cut}=200$ MeV
the relative accuracy of binding energy is less than $0.005\%$,
which is enough for the study of ground-state properties.
The relative accuracy of radius and deformation parameter are close to $0.1\%$.
Therefore, in the following calculations, $E_\mathrm{cut} = 200$ MeV is adopted.

\subsection{AMP calculations}
\label{Sec3.2}

\begin{figure}
\begin{center}
\includegraphics[width=0.4\textwidth]{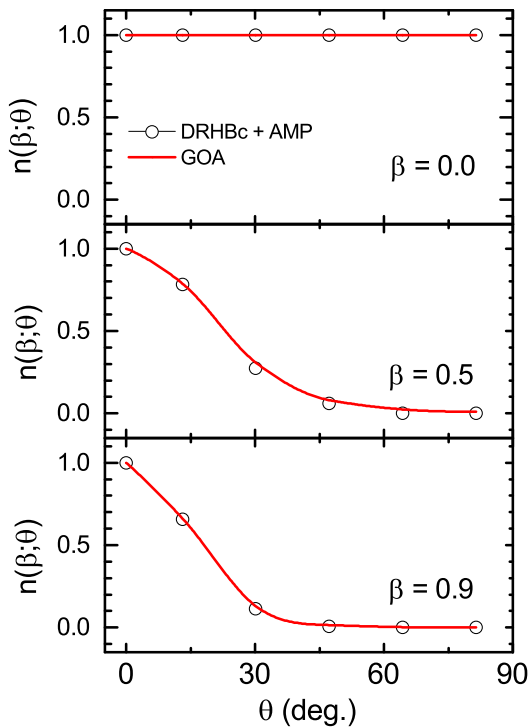}
\end{center}
	\caption{Normal overlap of $^{24}$Mg from the AMP (black circles) and GOA (red lines)
calculations as a function of $\theta$ with PC-F1.
From top to bottom, the deformation parameters of $^{24}$Mg are
constrained to be 0.0, 0.5, and 0.9.
	}
	\label{fig:NOL}
\end{figure}

For axially deformed nuclei, the normal overlap in Eq.~(\ref{Eq:NOL})
can be analytically calculated by using the Gaussian overlap approximation (GOA)
\cite{Bender2003_PRC68-044321,Rodriguez-Guzman2000_PLB474-15}
\begin{equation}
n_{\mathrm{GOA}}(\beta;\theta) =
\exp{\left[-\frac{1}{2}\left\langle \hat{J^2_y}\right\rangle \sin^2\theta\right]},
\end{equation}
with $\left\langle \hat{J^2_y} \right\rangle =
\left\langle\Phi(\beta) \left|\hat{J^2_y} \right|\Phi(\beta)\right\rangle$.
The detailed formulae about the calculation of
$\left\langle \hat{J^2_y}\right \rangle$
can be found in Appendix \ref{APP:A3}.
It has been checked in several works
\cite{Rodriguez-Guzman2000_PLB474-15,Bender2003_PRC68-044321,
Niksic2006_PRC73-034308,Yao2009_PRC79-044312}
that the
GOA is a good approximation for the normal overlap
for both small and large deformation parameters and
can be used to examine the result of the normal overlap in AMP calculations.
In Fig.~\ref{fig:NOL}, we show the $n(\beta;\theta)$
values calculated numerically by using the AMP
and those obtained under the GOA for $^{24}$Mg
with $\beta$ constrained to be 0.0, 0.5, and 0.9.
For the spherical case ($\beta=0$), the calculated values of
$n(\beta;\theta)$ are equal to 1 due to the rotational invariance.
When $\beta=0.5$ and $\beta=0.9$,
it can be seen that the calculated values of $n(\beta;\theta)$
are in good agreement with those obtained under the GOA,
meaning that our calculations for the normal overlap are reliable.

\begin{figure}[h]
		\includegraphics[width=0.4\textwidth]{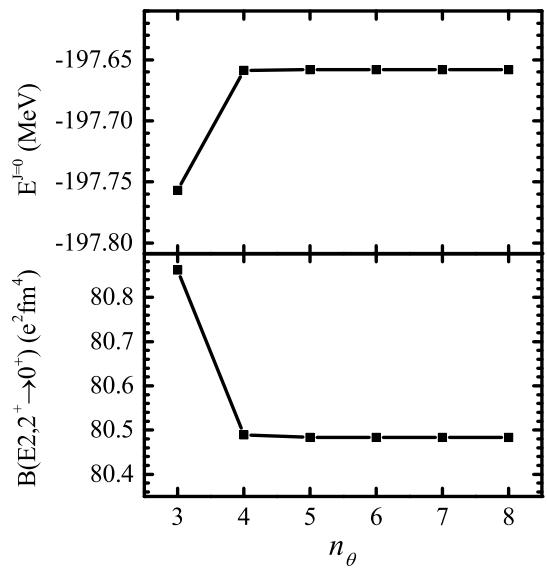}
	\caption{$E^{J=0}$ and
$B(E2,2^+ \rightarrow 0^+)$
obtained from angular momentum projections on the intrinsic state with
$\beta=0.55$ for $^{24}$Mg as a function of $n_\theta$.
	}
	\label{fig:theta}
\end{figure}

For the Hamiltonian and normal overlap kernels,
the one-dimensional integral over $\theta$ is calculated
by using the Gaussian-Legendre quadrature
and the number of the mesh points in the interval $[0,\pi/2]$ is $n_\theta$.
We show the energy of the projected $0^+$ state, $E^{J=0}$, obtained
from the MF states with $\beta=0.55$ for $^{24}$Mg
and the corresponding reduced transition probability
$B(E2,2^+ \rightarrow 0^+)$
as a function of the $n_\theta$ in Fig.~\ref{fig:theta}.
It is clear that these two quantities converge well with
increasing $n_\theta$.
To reach the relative accuracy of $0.0001\%$ for $E^{J=0}$
and $0.001\%$ for $B(E2,2^+\rightarrow 0^+)$,
the number of the mesh point of the Euler angle $\theta$
in the interval $[0,\pi/2]$ should satisfy $n_\theta \ge 5$.
In this work, $n_\theta=6$ is used.

\begin{figure}[h]
\includegraphics[width=.4\textwidth]{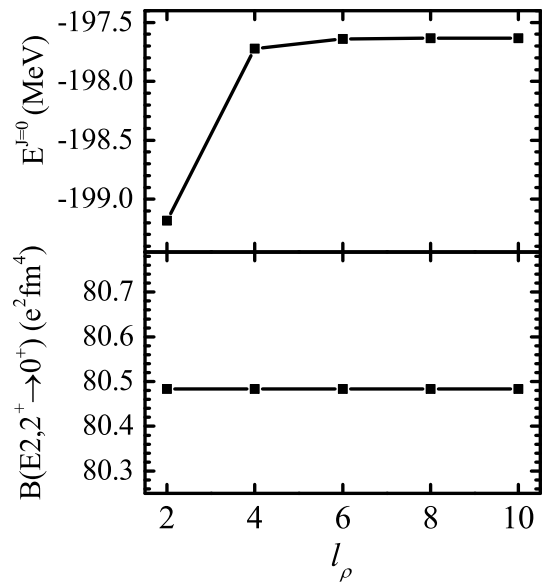}
\caption{
$E^{J=0}$ and $B(E2,2^+ \rightarrow 0^+)$ obtained from angular momentum projections on
the intrinsic state with $\beta=0.55$ for $^{24}$Mg as a function of $l_\rho$.
	}
	\label{fig:lrho}
\end{figure}

\begin{figure}[h]
\includegraphics[width=0.4\textwidth]{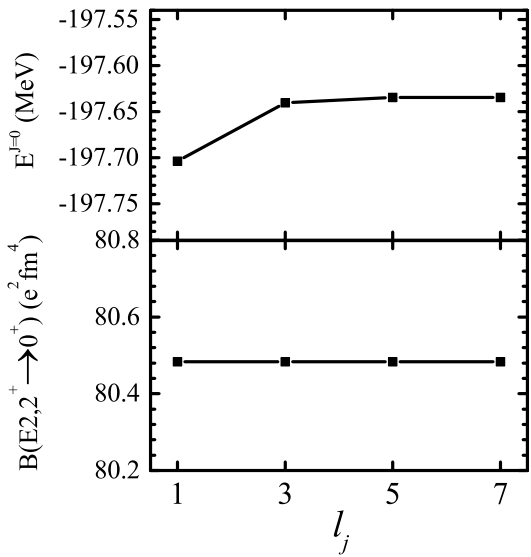}
\caption{$E^{J=0}$ and $B(E2,2^+ \rightarrow 0^+)$ obtained from angular momentum projections on
the intrinsic state with $\beta=0.55$ for $^{24}$Mg as a function of $l_j$.
}
\label{fig:lj}
\end{figure}

The ingredients for the mixed energy density are the mixed densities and currents,
which are expanded in terms of the spherical harmonics [cf. Eq.~(\ref{eq:exp2})].
The convergence with respect to
the maximum expansion orders $l_\rho$ in Eq.~(\ref{SH_den}) and $l_{j}$ in Eq.~(\ref{SH_cur})
for mixed densities and currents should be analyzed.
In Fig.~\ref{fig:lrho}, the calculated values of
$E^{J=0}$ and $B(E2,2^+ \rightarrow 0^+)$
for $^{24}$Mg with $\beta=0.55$ are plotted as a function of $l_\rho$.
We find that to achieve a precision of $0.01\%$
for $E^{J=0}$ and $B(E2,2^+ \rightarrow 0^+)$,
the maximum expansion order $l_\rho$ should fulfill $l_\rho \ge 6$.
With $l_\rho$ fixed to be 6,
the calculated values of $E^{J=0}$ and $B(E2,2^+ \rightarrow 0^+)$
are plotted as a function of $l_j$ in Fig.~\ref{fig:lj}.
One can see that when $l_j=3$ and $l_j=5$,
the calculation of $E^{J=0}$ and $B(E2,2^+ \rightarrow 0^+)$ can reach a relative
accuracy of $0.01\%$ and $0.001\%$, respectively.
The time consuming of the calculations of the currents is much
heavier than that for the mixed densities and
the relative accuracy with $l_j=3$ is good enough for the spectroscopic study.
Therefore for later calculations we choose $l_j=3$ and $l_\rho=6$.

\begin{figure}[h]
\includegraphics[width=0.4\textwidth]{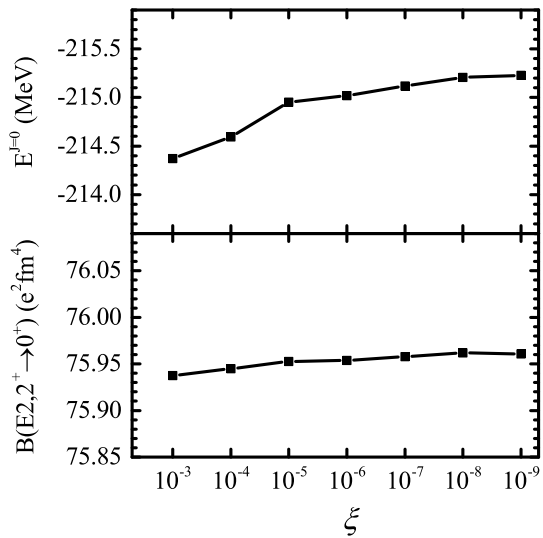}
\caption{$E^{J=0}$ and $B(E2, 2^+ \rightarrow 0^+)$ obtained
	from angular momentum projections on
the intrinsic state with $\beta=0.50$ for $^{26}$Mg as a function of $\xi$.
	}
	\label{fig:v2}
\end{figure}

For the calculation of the rotational matrix,
as mentioned in Sec. \ref{Sec:2.2},
we introduce a truncation $\xi$ on the occupation
probability of SPLs in the canonical basis to
determine the dimension of this matrix.
Therefore, the convergences of $E^{J=0}$ and reduced transition probability
with respect to $\xi$ should be also checked.
In Fig~\ref{fig:v2}, we show the energy of projected $0^+$ and $B(E2,2^+\rightarrow 0^+)$
versus the cut-off parameter $\xi$
and it is clear that the calculation of both quantities can give a very high precision
when $\xi < 10^{-7}$.
With $\xi = 10^{-7}$ and $10^{-8}$, the relative accuracies
for $E^{J=0}$ are about $0.05\%$ and $0.01\%$.
$B(E2,2^+\rightarrow 0^+)$ changes slightly
with decreasing $\xi$ and
the relative accuracy is about $0.001\%$ when $\xi = 10^{-7}$.

\begin{figure}[h]
	\includegraphics[width=0.4\textwidth]{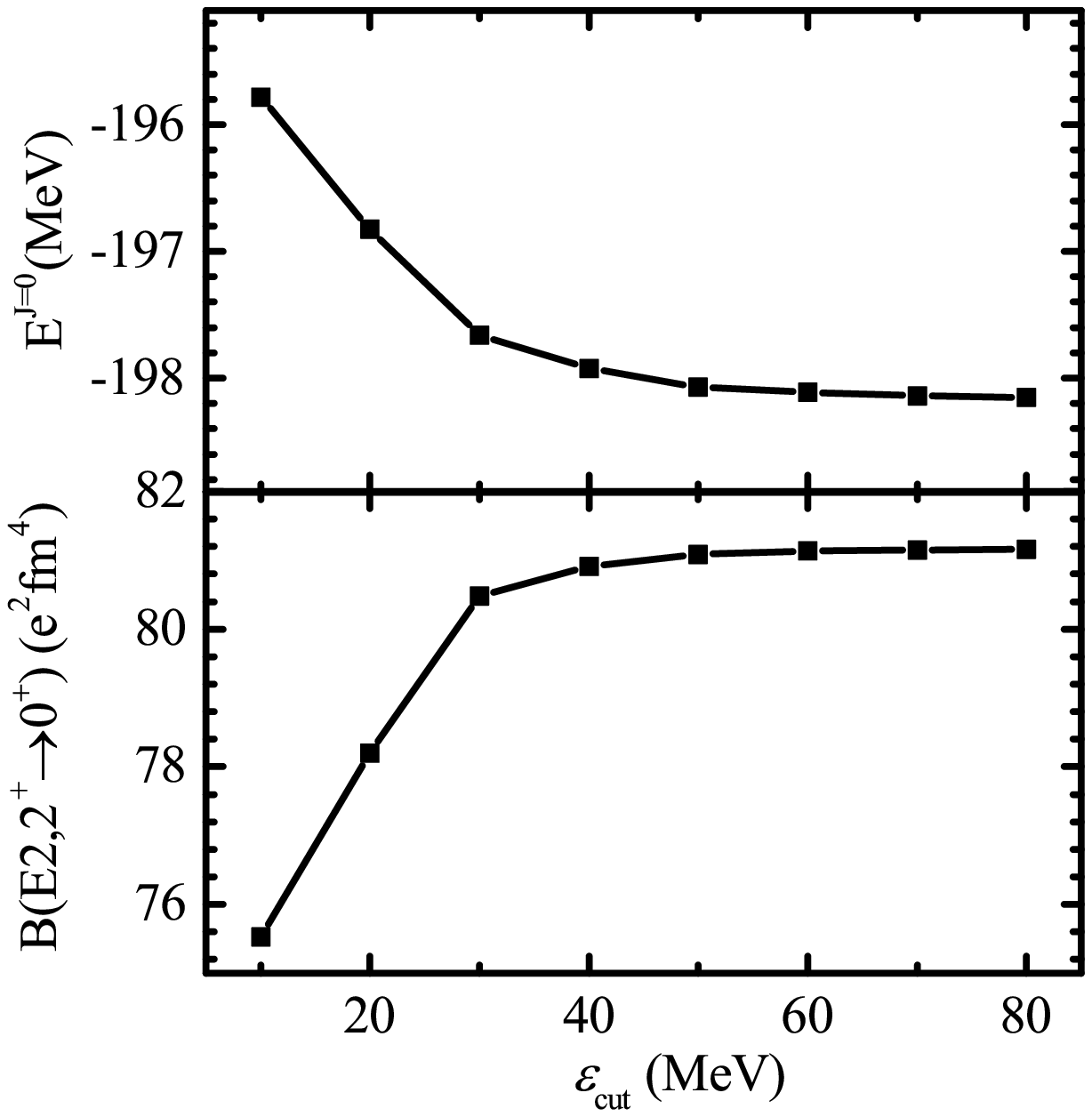}
	\caption{$E^{J=0}$ and $B(E2,2^+ \rightarrow 0^+)$ obtained from
angular momentum projections on
the intrinsic state with $\beta=0.55$ for $^{24}$Mg as a function of  $\epsilon_\mathrm{cut}$.
	}
	\label{fig:ecut}
\end{figure}

But when the calculated pairing energy equals zero
or the pairing strength is taken to be zero,
the SPLs below or above the Fermi level ($\lambda_\tau$)
are fully occupied or empty, respectively.
The truncation on the occupation probability is no longer suitable in this case.
Therefore, we also introduce a cut-off energy on the single particle energy (SPE)
in the canonical basis to determine the total number of SPLs for AMP calculations,
i.e., SPLs with the energy larger than
$\lambda_\tau+\epsilon_\mathrm{cut}$ are neglected
for neutrons ($\tau=1$) and protons ($\tau=-1$).
In Fig.~\ref{fig:ecut}, $E^{J=0}$ and $B(E2, 2^+\rightarrow 0^+)$ are shown as a function
of the cut-off energy $\epsilon_\mathrm{cut}$.
The relative accuracy of
$E^{J=0}$ and $B(E2, 2^+\rightarrow 0^+)$
are about $0.01\%$ when $\epsilon_\mathrm{cut} = 50$ MeV.
In practical calculations, the truncations on SPE and occupation
probability are used simultaneously.

Now we summarize the parameters involved in DRHBc+AMP calculations.
The energy cut-off $E_\mathrm{cut}$ for positive energy states
in the Dirac WS basis is 200 MeV in MF calculations.
The number of mesh points in the Gaussian-Legendre quadruture
for the calculation of the normal overlap kernel and Hamiltonian overlap kernel
is $n_\theta=6$ in the interval $[0,\pi/2]$.
For the mixed density and currents expanded in terms of the
spherical harmonics,
the maximum orders are $l_\rho=6$ and $l_j=3$, respectively.
For determining the number of SPLs,
the truncation of the occupation probability is $\xi=10^{-7}$ and
$\epsilon_\mathrm{cut}=50$ MeV for SPE.

\begin{table}[h]
	\caption{Excitation energies and $B(E2)$ from DRHBc+AMP
(Th. \uppercase\expandafter{\romannumeral1})
and MDC-RHB+AMP (Th. \uppercase\expandafter{\romannumeral2})
calculations for $^{24}$Mg with $\beta_2 =0.55$ and $\beta_2=0.65$.}
	\label{tab:1}
	\centering
	\begin{ruledtabular}
	\begin{tabular}{rrrrr}
		%\toprule[1pt]
		\specialrule{0em}{1pt}{1pt}
		&
		\multicolumn{2}{c}{$\beta_2 = 0.55$} & \multicolumn{2}{c}{$\beta_2 = 0.65$} \\
		\hline
		\specialrule{0em}{2pt}{1pt}
		& Th. \uppercase\expandafter{\romannumeral1}
		& Th. \uppercase\expandafter{\romannumeral2}
		& Th. \uppercase\expandafter{\romannumeral1}
		& Th. \uppercase\expandafter{\romannumeral2} \\
		$E(2^+)$ (MeV) & 1.009 & 1.006 &1.149 &1.128    \\
		$E(4^+)$ (MeV) & 3.553 & 3.556 &3.956 &3.890    \\
		$E(6^+)$ (MeV) & 8.024 & 8.074 &8.627 &8.542    \\
		\hline
				\specialrule{0em}{2pt}{1pt}
		$B(E2, 2^+\rightarrow 0^+)\ (e^2\ \mathrm{fm}^4)$&
		             81.083  & 81.214  &110.606 &110.750 \\
		$B(E2,4^+\rightarrow 2^+)\  (e^2\ \mathrm{fm}^4)$&
		             118.126 & 118.752 &161.459 &162.156 \\
		$B(E2, 6^+\rightarrow 4^+)\  (e^2\ \mathrm{fm}^4)$&
		             135.674 & 136.939 &185.670 &186.912 \\
	\end{tabular}
\end{ruledtabular}
\end{table}

To check this newly developed method further,
the DRHBc+AMP method is applied to stable nuclei and
the calculated results are compared with those from MDC-RHB+AMP calculations
\cite{Wang2021_MDC-RHB+AMP}.
In Table~\ref{tab:1},
we show the excitation energies of $2^+$, $4^+$, and $6^+$ states and $B(E2)$ values
obtained in DRHBc+AMP and MDC-RHB+AMP calculations
for $^{24}$Mg with the quadrupole deformation parameter constrained to be
$0.55$ and $0.65$.
The density functional PC-F1 is used in both methods.
In MDC-RHB+AMP calculations, the number of oscillator
shells is taken to be 14 and the number of mesh points of the Euler angle
$\theta$ in the interval $[0,\pi]$ equals 12.
For both the excitation energies and $B(E2)$ values, the
relative differences between these two methods are around $1\%$.
From this comparison one can conclude that for well bound nuclei, the
low-lying excited spectra and $B(E2)$ values
from DRHBc+AMP calculations are well consistent with
the results from the MDC-RHB+AMP method.

\section{Low-lying excited states of $^{36,38,40}$Mg with DRHBc+AMP}
\label{Sec:res}

Many exotic nuclear structures have been observed
in the magnesium isotopic chain.
The position of proton drip line is $N=8$
and that of neutron drip line is determined to be $N=28$
according to the experimental data so far \cite{Baumann2007_Nature449-1022}
and several theoretical calculations
\cite{Li2012_PRC85-024312,Erler2012_Nature486-509,Chai2020_PRC102-014312,
Tsunoda2020_Nature587-66,Stroberg2021_PRL126-022501,In2021_IJMPE30-2150009}
have predicted that $^{42}$Mg, $^{44}$Mg, and $^{46}$Mg are all possible to be the
last bound nucleus of this chain.
As a proton-rich nucleus, $^{20}$Mg is a Borromean nucleus
and locates just within the proton drip line.
A recent experiment
\cite{Randhawa2019_PRC99-021301R} shows
that this nucleus is well deformed,
indicating the possible quenching of the shell closure at $N=8$.
$\alpha$ cluster structure has been predicted in $^{24}$Mg
\cite{Ebran2017_JPG44-103001,Tohsaki2018_PRC97-011301R} and
some BMF calculations show that the ground-state of $^{24}$Mg
has a triaxial deformation
\cite{Yao2011_PRC83-014308}.
As for $^{32}$Mg, which belongs to the $N=20$ island of inversion,
the ground-state is well deformed
\cite{Motobayashi1995_PLB346-9} and
the valence neutrons are dominated by the intruder $pf$ states
\cite{Warburton1990_PRC41-1147}.
$^{37}$Mg is the heaviest one-neutron halo nucleus observed so far
\cite{Kobayashi2014_PRL112-242501}
and deformation effects play a central role in the halo configuration
\cite{Takechi2014_PRC90-061305R,Watanabe2014_PRC89-044610,Nakada2018_PRC98-011301R}.
$^{39}$Mg is unbound
\cite{Kondev2021_ChinPC45-030001,Huang2021_ChinPC45-030002,Wang2021_ChinPC45-030003}.
As for $^{40}$Mg, the recently established low-lying excited spectrum
\cite{Crawford2019_PRL122-052501} indicates the disappearance of magic number
$N=28$ in this isotopic chain.
$^{42,44}$Mg are predicted to have deformed halo structure
\cite{Zhou2010_PRC82-011301R,Li2012_PRC85-024312,Zhang2019_PRC100-034312}.
The traditional magic numbers $N=8$, $20$, and $28$ may
all disappear in Mg isotopes due to deformation effects.
There are many systematic theoretical investigations on Mg isotopic chain,
including MF calculations
\cite{Li2012_PRC85-024312,Zhang2019_PRC100-034312,Nakada2013_PRC87-014336},
the shell model calculations
\cite{Dong2013_PRC88-024328,Tsunoda2020_Nature587-66},
BMF calculations with Skyrme density functional
\cite{Valor2000_NPA671-145,Bender2003_RMP75-121}
and Gogny force
\cite{Rodriguez-Guzman2002_NPA709-201,Rodriguez2010_PRC81-064323,
Rodriguez2016_EPJA52-190,Shimada2016_PRC93-064314},
and beyond RMF calculations
\cite{Yao2011_PRC83-014308,Wu2015_PRC92-054321}.
In this section, we study the bulk properties
and low-lying excited states of $^{36,38,40}$Mg
by using the newly developed DRHBc+AMP method.

\subsection{Bulk properties}

The calculated bulk properties of $^{36}$Mg, $^{38}$Mg,
and $^{40}$Mg are listed in Table~\ref{tab:2}.
In DRHBc calculations with the density functional PC-F1,
the quadrupole deformation parameters of the ground-states
of $^{36}$Mg, $^{38}$Mg, and $^{40}$Mg are 0.45, 0.49, and 0.48, respectively.
The calculated rms matter radii ($R_m$) of $^{36}$Mg and $^{38}$Mg are 3.49 fm and 3.62 fm,
which are well consistent with the experimental values
\cite{Watanabe2014_PRC89-044610}, $3.49\pm0.01$ fm and $3.60\pm0.04$ fm,
extracted from the measurements of total cross sections \cite{Takechi2014_EWC66-02101}.
The calculated value of $R_m$ for $^{40}$Mg is 3.70 fm.
The two-neutron separation energy
with considering the correction from the AMP of $^{38}$Mg is 3.06 MeV,
which agrees with the experimental value, 2.45(85) MeV
\cite{Wang2017_ChinPhysC41-030003,Wang2021_ChinPC45-030003}
and that of $^{40}$Mg is 2.74 MeV,
which is larger than the experimental values,
1.87(71) MeV in AME2016 \cite{Wang2017_ChinPhysC41-030003} and
0.65(0.71) MeV in AME2020 \cite{Wang2021_ChinPC45-030003}.

\begin{table}[h]
\caption{Ground-state properties
from DRHBc calculations with PC-F1
and energies of the projected $0^+$ state
of $^{36}$Mg, $^{38}$Mg, and $^{40}$Mg.
For each nucleus, we show
the neutron, proton, and total quadrupole deformation parameters
($\beta_n$, $\beta_p$, $\beta_\mathrm{t}$),
neutron, proton, and total rms matter radii ($R_n$, $R_p$, $R_\mathrm{t}$),
the correction energy ($E_\mathrm{c.m.}$) of center-of-mass spurious motion,
the total energy ($E_\mathrm{B}$),
and energy of the projected $0^+$ state ($E^{J=0})$.
}
	\label{tab:2}
	\centering
	\begin{ruledtabular}
	\begin{tabular}{rrrr}
		\specialrule{0em}{2pt}{1pt}
		& $^{36}$Mg & $^{38}$Mg & $^{40}$Mg  \\
		\hline
		\specialrule{0em}{2pt}{1pt}
		$\beta_n$ & 0.4568 & 0.5150& 0.5006  \\
		$\beta_p$ & 0.4331 & 0.4339& 0.4197  \\
		$\beta_\mathrm{t}$ & 0.4489 & 0.4894& 0.4764  \\
		$R_n$ (fm)& 3.6593 & 3.8191& 3.9066  \\
		$R_p$ (fm)& 3.1276 & 3.1568& 3.1815  \\
		$R_t$ (fm)& 3.4911 & 3.6230& 3.7040 \\
		$E_\mathrm{c.m.}$ (MeV)  & $  -9.3127$ & $-  9.1469$ &$  -8.9914$\\
		$E_\mathrm{B}$ (MeV)     & $-265.3905$ & $-267.9706$ &$-270.9131$\\
		$E^{J=0}$ (MeV)          & $-268.0396$ & $-271.1044$ &$-273.8405$\\
		\specialrule{0em}{1pt}{1pt}
	\end{tabular}
	\end{ruledtabular}
\end{table}

\begin{figure}[h]
\begin{center}
\includegraphics[width=0.45\textwidth]{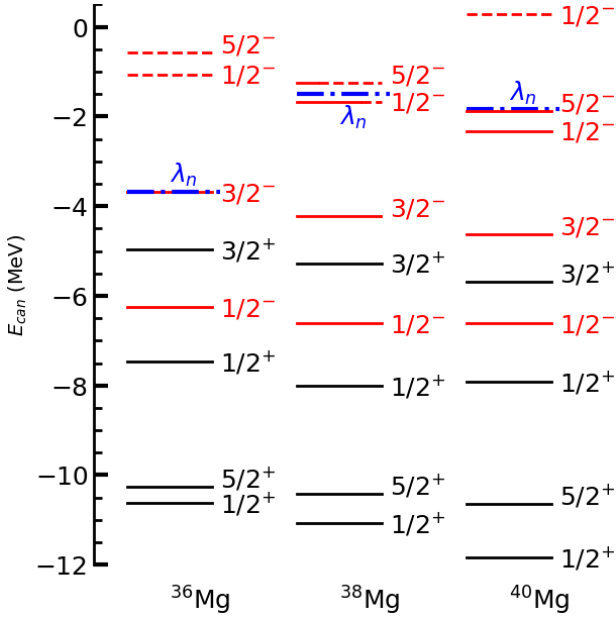}
\end{center}
\caption{SPLs of neutrons around the Fermi energy ($\lambda_n$)
of $^{36}$Mg, $^{38}$Mg, and $^{40}$Mg in the canonical basis.
The length of the solid line is proportional to
the occupation probability $v^2$ of each level labeled by $\Omega^\pi$,
where $\Omega$ and $\pi$ are the projection of total angular momentum on
the symmetry axis in the intrinsic frame and parity.
Red and black lines represent levels with $\pi=-$ and $\pi=+$, respectively.}
\label{fig:SPL}
\end{figure}

In Fig.~\ref{fig:SPL}, we show the SPLs with
$-12\ \mathrm{MeV} < \epsilon_\mathrm{can}<1\ \mathrm{MeV}$
in the canonical basis for $^{36,38,40}$Mg.
It should be noted that near the neutron Fermi energy $\lambda_n$,
the $1/2^-$ and $3/2^-$ levels contain $p$-wave components
and the $5/2^-$ level is totally dominated by $f$-wave components.
Around $\lambda_n$,
SPLs are all fully occupied with $v^2=1$ for $^{36,40}$Mg
and partially occupied for $^{38}$Mg,
meaning the enhancement of pairing in $^{38}$Mg.
$^{37}$Mg is a $p$-wave halo nucleus and the valence neutron is unpaired
\cite{Nakada2018_PRC98-011301R,Kasuya2020_PTEP2021-013D01}.
The configuration of the two valence neutrons for $^{38}$Mg also includes
$p$-wave components and is mainly the mixing of $2p_{1/2}$ and $1f_{7/2}$
with the occupation numbers of 0.72 and 1.18.
For $^{40}$Mg, the fully occupied levels $5/2^-$ and $1/2^-$
near the Fermi energy are close to each other.
It has been shown in Refs.~\cite{Watanabe2014_PRC89-044610,Sun2020_PRC101-014321}
that there is a cross between the $5/2^-$ and $1/2^-$ orbitals
when $\beta \approx 0.5$ around the neutron Fermi energy.
Because of the near degeneracy of ($1/2^-$, $5/2^-$),
it is reasonable to regard $^{40}$Mg as a ``$^{36}$Mg+$4n$" system
instead of ``$^{38}$Mg+$2n$" from the point of view of
the structure of SPLs.
The four valence neutrons are dominated by $p$- and $f$-wave components with
the occupation numbers of 1.2 and 2.8, respectively in DRHBc calculations.
In conclusion,
the configurations of the valence neutrons for $^{38}$Mg and $^{40}$Mg
all have $p$-wave components with considerable occupation,
but they are not halo nuclei
because the valence neutrons are not weakly bound
with calculated two-neutron separation energies
larger than 2 MeV.
In addition,
the study in Ref. \cite{Nakada2018_PRC98-011301R}
shows that $^{40}$Mg is a two-neutron halo nucleus,
which is contrary to the conclusion drawn in our DRHBc calculations.
%Mg38
%1/2^- 0.73: 0.36 2p_1/2 + 0.34 1f_7/2
%5/2^- 0.25:  1f_7/2
%\renewcommand\arraystretch{1.5}

\subsection{Ground-state rotational bands of $^{36,38,40}$Mg}

In this study, low-lying excited spectrum is obtained
by performing the AMP on the deformed ground-state obtained from DRHBc calculations
with the density functional PC-F1,
i.e., for each nucleus, the same MF wave function is used to get the projected states.
For $^{36,38,40}$Mg,
the calculated values of the excitation energy $E({J^+})$,
spectroscopic quarupole moment $Q^{(s)}(J^+)$, and
reduced transition probability $B(E2)$
are summarized in Table~\ref{tab:spectra}.

\begin{table}[h]
\caption{Calculated excitation energy $E({J^+})$,
spectroscopic quarupole moment $Q^{(s)}(J^+)$,
and reduced transition probabilities $B(E2)$
for $^{36,38,40}$Mg with PC-F1.}
	\label{tab:spectra}
	\centering
	\begin{ruledtabular}
	\begin{tabular}{rrrr}
		\specialrule{0em}{1pt}{1pt}
		&$^{36}$Mg & $^{38}$Mg & $^{40}$Mg  \\
		\specialrule{0em}{1pt}{1pt}
		\hline
		\specialrule{0em}{1pt}{1pt}
		$E(2^+)$ (MeV) & 0.46 & 0.66 & 0.53 \\
		$E(4^+)$ (MeV) & 1.61 & 2.15 & 1.82  \\
		$E(6^+)$ (MeV) & 3.65 & 4.44 & 3.94 \\
		\hline
		\specialrule{0em}{1pt}{1pt}
		$Q^{(s)}(2^+)\ (e$ fm$^{2})$ &
		                $-18.21$ & $-18.96$ & $-18.99$ \\
		$Q^{(s)}(4^+)\ (e$ fm$^{2})$ &
		                $-23.22$ & $-24.19$ & $-24.26$ \\
		$Q^{(s)}(6^+)\ (e$ fm$^{2})$ &
		                $-25.64$ & $-26.73$ & $-26.83$ \\
		\hline
		\specialrule{0em}{1pt}{1pt}
		$B(E2,2^+\rightarrow 0^+)\ (e^2\ \mathrm{fm}^4)$ &
		80.91  & 87.66  & 87.89\\
		$B(E2,4^+\rightarrow 2^+)\ (e^2\ \mathrm{fm}^4)$ &
		117.39 & 127.01 & 126.88\\
		$B(E2,6^+\rightarrow 4^+)\ (e^2\ \mathrm{fm}^4)$ &
		133.06 & 143.29 & 142.57\\
	\end{tabular}
	\end{ruledtabular}
\end{table}

\begin{figure}[h]
	\begin{center}
		\includegraphics[width=0.47\textwidth]{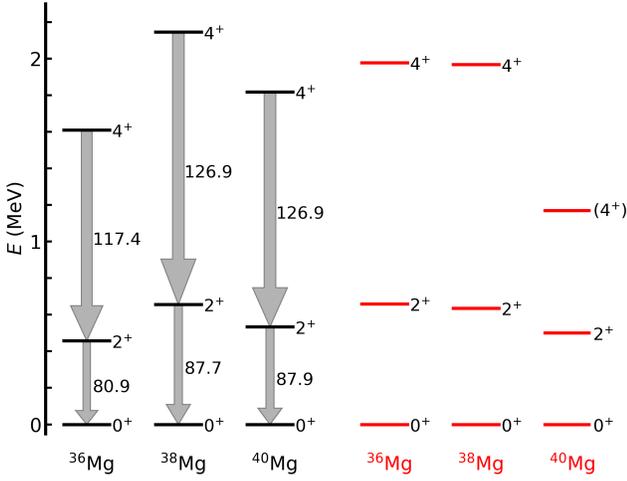}
	\end{center}
\caption{The ground-state rotational bands and
values of $B(E2)$ of $^{36}$Mg, $^{38}$Mg, and $^{40}$Mg.
Black lines and grey arrows represent the results from DRHBc+AMP calculations
and red lines show the experimental data taken from Ref.~\cite{Crawford2019_PRL122-052501}.
Transitions between two states are represented by arrows and the width of each arrow
is proportional to reduced transition probability.}
\label{fig:spectrum}
\end{figure}

In Fig.~\ref{fig:spectrum},
the calculated ground-state bands of $^{36}$Mg, $^{38}$Mg, and $^{40}$Mg are shown
and compared with
the experimental values taken from Ref.~\cite{Crawford2019_PRL122-052501}.
${B(E2\downarrow)}$ values obtained from DRHBc+AMP calculations are also given.
As shown in Refs.~\cite{Rodriguez-Guzman2002_NPA709-201,Yao2011_PRC83-014308},
for well deformed nuclei $^{36,38,40}$Mg
the differences of excitation energies between AMP and AMP+GCM calculations
are relatively small.
Therefore, for these three nuclei,
we can directly compare our results from DRHBc+AMP calculations
with those with GCM.
The overall trend of spectra from
DRHBc+AMP calculations are consistent with
the results obtained from the RMF+1DAMP+GCM calculations
with PC-F1 \cite{Yao2011_PRC83-014308},
in which the HO basis is used, and close to those shown
in Ref.~\cite{Shimada2016_PRC93-064314}.
The excitation energies from BMF calculations
with Gogny force~\cite{Rodriguez-Guzman2002_NPA709-201}
are higher than the results in this work.
The ground-state band of $^{40}$Mg
from our calculations is close to that of
recent Monte Carlo shell model (MCSM) calculations
\cite{Tsunoda2020_Nature587-66}.
The calculated excitation energies of the $2^+$ and $4^+$ states for $^{36}$Mg
are slightly smaller than the experimental values.
The $2^+$ states of $^{38}$Mg and $^{40}$Mg are in line with
the experimental values.
All the BMF calculations mentioned above support that
the shell closure at $N=28$ is quenched and $^{40}$Mg has a prolate shape.
Generally speaking, the DRHBc+AMP calculations reproduce the experimental
low-lying spectra of $^{36,38,40}$Mg reasonably well.

%In addition, in RMF+3DAMP+GCM calculations \cite{Yao2011_PRC83-014308} show
%that the triaxial deformation $\beta_{22}$ also influences the low-lying excitation spectra of
%$^{36,38,40}$ and the energy minimum in projected potential energy surface of $J=0$ has
%$\beta_{22}$ deformation. Although this is beyond the scope of this work,
%it is necessary to investigate whether the halo feature appears or not in $^{40}$Mg when
%including the triaxial deformation.

\begin{figure}[h]
	\begin{center}
		\includegraphics[width=0.49\textwidth]{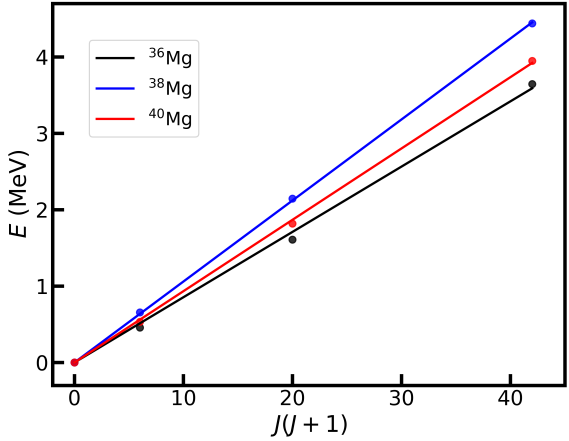}
	\end{center}
\caption{Rotational bands for $^{36,38,40}$Mg.
Excitation energy of collective states $E_{J^+}$
are plotted as a function of $J(J+1)$.
The calculated results are labeled by solid dots,
and the linear fitting of calculated spectrum of each nucleus
is shown by solid line.}
	\label{fig:E_i}
\end{figure}

\begin{figure}[h]
\begin{center}
\includegraphics[width=0.45\textwidth]{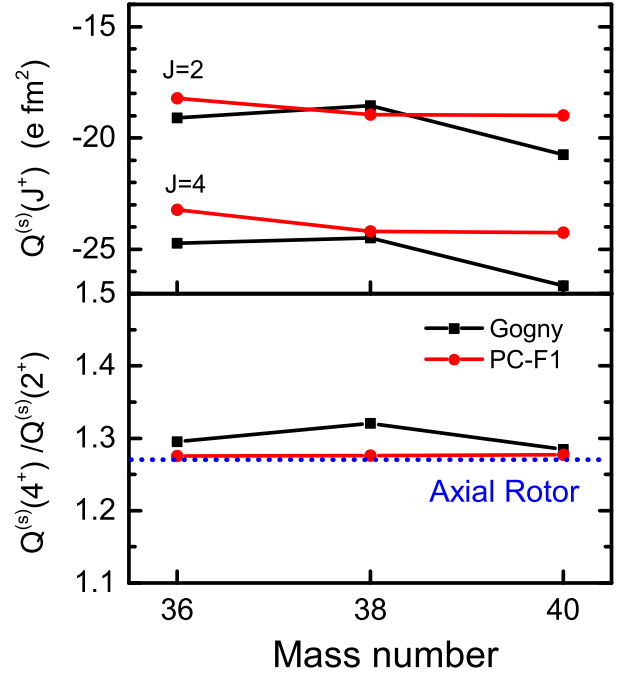}
\end{center}
\caption{Spectroscopic quadrupole moments $Q^{\mathrm(s)}$ of
the $2^+$ and $4^+$ states (the top panel)
and the ratios ${Q^{\mathrm(s)} (4^+)}/{Q^{\mathrm(s)} (2^+)}$ (the bottom panel)
for $^{36,38,40}$Mg. }
\label{fig:Qs}
\end{figure}

In Fig.~\ref{fig:E_i},
we show the calculated excitation energies of the collective states
as a function of $J(J+1)$.
For the calculated spectrum of each nucleus,
we fit the calculated excitation energies to the linear relation
$E({J^+}) =\langle\hat{J}^2\rangle /2\mathcal{J}$
with the moment of inertia $\mathcal{J}$.
It is obvious that the calculated excitation energies
and $J(J+1)$ have a very good linear relation.
This confirms that the ground-state bands are rotational bands
for these three nuclei.
The spectroscopic quadrupole moments
$Q^{(s)}$ in the $2^+$ and $4^+$
states of $^{36,38,40}$Mg
obtained from DRHBc+AMP calculations with PC-F1
are shown in the top panel of Fig.~\ref{fig:Qs}
and compared with the results taken from Ref.~\cite{Rodriguez-Guzman2002_NPA709-201}.
From Fig.~\ref{fig:Qs},
it is found that calculated values of
$Q^{\mathrm(s)}$ with PC-F1 are well consistent with those from
Ref.~\cite{Rodriguez-Guzman2002_NPA709-201},
indicating the prolate shapes of $^{36}$Mg, $^{38}$Mg, and $^{40}$Mg.
Similar conclusion can also be found in Ref. \cite{Yao2011_PRC83-014308}.
The ratios ${Q^{\mathrm(s)} (4^+)}/{Q^{\mathrm(s)} (2^+)}$ obtained
from DRHBc+AMP calculations and Ref.~\cite{Rodriguez-Guzman2002_NPA709-201}
are presented in the bottom panel of Fig.~\ref{fig:Qs}
and compared with the value that corresponds to a rigid axial rotor
without triaxial shapes, 1.27, labeled by the dotted blue line.
One can find that all the calculated ratios are close to that of a rigid rotor.
This indicates that these three nuclei are all good rotors.

In the calculated ground-state bands, the ratio $R_{4/2}=E({4^+})/E({2^+})$
are 3.5, 3.2, and 3.3 and those corresponding to the experimental values
taken from Ref. \cite{Crawford2019_PRL122-052501}
are 3.0, 3.1 and 2.34 for $^{36}$Mg, $^{38}$Mg, and $^{40}$Mg, respectively.
One can conclude that for $^{36}$Mg and $^{38}$Mg, both the experimental and calculated
bands are rotational ones.
For $^{40}$Mg, all above-mentioned theoretical calculations support that
this nucleus is a good rotor.
But this is not the case in Ref. \cite{Crawford2019_PRL122-052501} where
the excitation energy of the second excited state is about 1.2 MeV,
leading to that the ground-state band is no longer a rotational band.
Recently the MCSM calculations
predict that the ground-state band is a rotational band and there is a state
with the excitation energy of 1.2 MeV
belonging to the triaxial rotational band,
showing a nice agreement with the experimental energy levels
\cite{Tsunoda2020_Nature587-66}.
Future detailed studies on both the structure of SPLs and excitation spectra of $^{40}$Mg by
using BMF methods with considering the triaxial deformation are in need, but this
is beyond the scope of the approach in the present work.

In this section, we study the ground-state rotational bands
of $^{36,38,40}$ by using the DRHBc+AMP approach.
Our calculations are consistent with other theoretical results and
describe the low-lying excited spectra of $^{36,38,40}$Mg reasonably well.
The calculated spectra show that these three nuclei are all good rotors.
For $^{40}$Mg, the observed first excited state are well reproduced
but the second excited state can not be understood by the present investigation
if this state belongs to the ground-state rotational band.

\section{Summary and perspective}\label{Sec:summ}

BMF approaches based on SCMF have become general tools for the spectroscopic study.
Previous BMF calculations by using the AMP based on the HO basis
are not suitable for the study of weakly bound nuclei.
The DRHBc theory,
which treats the large spatial extension,
the contribution of continuum induced by pairing correlations,
deformation effects, and the coupling among them self-consistently,
can provide a good description of the ground-state
of axially deformed, weakly bound nuclei
by solving the deformed RHB equation in the Dirac WS basis.
In this work, the AMP method is implemented in this theory
aiming at the study of low-lying excitation of weakly bound nuclei,
especially for deformed halo nuclei.
In the newly developed DRHBc+AMP approach,
the projected wave function, the mixed densities and currents are calculated
in the Dirac WS basis.
We perform carefully numerical checks on convergence
with respect to the parameters involved in the DRHBc+AMP method.

The low-lying excited spectra of $^{36,38,40}$Mg are investigated
by using the DRHBc+AMP method with the density functional PC-F1.
These three nuclei all have pronounced prolate shape
in the ground-states from DRHBc calculations.
The configuration of the valence neutrons for $^{38}$Mg
is the mixing of $p$- and $f$-wave orbitals with occupation amplitudes of $36\%$ and $59\%$.
$^{40}$Mg is not a halo nucleus but the configuration of the four valence neutrons
have $p$-wave components with the occupation amplitude of $30\%$.
The ground-state rotational bands are calculated
by performing the AMP on the deformed ground-state wave function
obtained from DRHBc calculations.
Our results are consistent with other theoretical studies
and reproduce the experimental data reasonably well.
It is found that these three nuclei are all good rotors.
The low-lying excited spectrum of $^{40}$Mg indicates
the breakdown of the shell closure at $N=28$.

In this work,
the density functional PC-F1 is adopted
because it is convenient to compare DRHBc+AMP calculations
with previous BMF calculations in
Refs.~\cite{Yao2009_PRC79-044312,Yao2011_PRC83-014308}.
It is also very interesting to investigate the excitation properties of
weakly bound nuclei with other point coupling density functionals,
such as
PC-PK1 \cite{Zhao2010_PRC82-054319},
PC-X \cite{Taninah2020_PLB800-135065},
and DD-PC1 \cite{Niksic2008_PRC78-034318}
and such studies are ongoing.
It is necessary to develop the PNP and GCM based on
the DRHBc+AMP approach to restore the particle number and take into account of
the quantum fluctuation of collective degrees of freedom in the future.

\acknowledgments
The authors would like to thank Ji-Wei Cui, Kun Wang, Zhen-Hua Zhang,
Peng-Wei Zhao, and the DRHBc Mass Table Collaboration for helpful discussions
and Bing-Nan Lu for sharing the MDC-RHB+AMP codes.
This work has been supported by
the National Key R\&D Program of China (Grant No. 2018YFA0404402),
the National Natural Science Foundation of China (Grants
No. 11525524, No. 12070131001, No. 12047503, No.11975237, and No. 11961141004),
the Key Research Program of Frontier Sciences of Chinese Academy of Sciences (Grant No. QYZDB-SSWSYS013),
and the Strategic Priority Research Program of Chinese Academy of Sciences
(Grant No. XDB34010000 and No. XDPB15).
The results described in this paper are obtained on
the High-performance Computing Cluster of ITP-CAS and
the ScGrid of the Supercomputing Center,
Computer Network Information Center of Chinese Academy of Sciences.

\appendix
\section{Mixed densities and currents in coordinate space }
\label{APP:A1}
In coordinate space, the Dirac spinor of the Dirac WS basis is
\begin{equation}
\varphi_{n\kappa m} (\bm r s p) =
 i^p \frac{R_{n\kappa}(r,p)}{r}
 \mathcal{Y}^{l(p)}_{\kappa m}(\Omega,s),
\end{equation}
where $p=1$ stands for the upper components and $p=2$ for the lower component.
$R_{n\kappa}(r,1) = G_{n\kappa}(r)$ and
$R_{n\kappa}(r,2) = F_{n\kappa}(r)$
are the radial wave functions.
$l(p=1)=j+\frac{1}{2}\mathrm{sgn}(\kappa)$ and
$l(p=2)=j-\frac{1}{2}\mathrm{sgn}(\kappa)$.

In coordinate space,
the mixed vector density for the Euler angle $\theta$
expanded in terms of the spherical harmonics
is written as
\begin{equation}
\begin{split}
\rho_V(\bm r;\beta;\theta) &
= \sum_{kk'} \rho_{k'k}(\theta)
  \phi_k^\dag(\bm r;\beta) \phi_{k'}(\bm r;\beta)
\\ &
= \sum_{\lambda\mu} \rho_{V,\lambda\mu}( r;\beta;\theta)
   Y_{\lambda\mu}(\Omega),
\end{split}
\end{equation}
where
\begin{equation}
\begin{split}
&\rho_{V,\lambda\mu}( r;\beta;\theta) =
 \sum_{n\kappa}\sum_{n'\kappa'}
 \sum_{kk'} \rho_{k'k}(\theta)
 c^k_{n\kappa}c^{k'}_{n'\kappa'}\\
&\times
 \langle \kappa m
 |Y^*_{\lambda\mu}(\Omega)|
 \kappa'm'\rangle
 \frac{1}{r^2}
 \sum_{p=1,2}
 R_{n\kappa}(r,p)
 R_{n'\kappa'}(r,p),
\end{split}
\end{equation}
and
\begin{equation}
\begin{split}
&\langle \kappa m
|Y_{\lambda\mu}|\kappa'm'\rangle \\
& =\sum_{\sigma}
\int d\Omega
\mathcal{Y}_{\kappa'm'}^{l'*} (\Omega,\sigma)
 Y_{\lambda\mu}(\Omega)
\mathcal{Y}_{\kappa m}^{l}    (\Omega,\sigma) \\
&=\sum_{m_l,m'_l}^{m_s,m'_s}
\mathrm{C}^{j'm'}_{l'm'_l \frac{1}{2}m'_s}
\mathrm{C}^{jm}_{lm_l \frac{1}{2}m_s}
\sqrt{\frac{\hat\lambda\hat l}{4\pi\hat l'}}
\mathrm{C}^{l'm'_l}_{\lambda\mu m m_l}
\mathrm{C}^{l'0}_{\lambda 0 l0},
\end{split}
\end{equation}
with $\hat\lambda = 2\lambda+1$, $\hat l = 2l+1$, and $\hat{l'}=2l'+1$.

Similarly, for the mixed scalar density we have
\begin{equation}
\begin{split}
\rho_S(\bm r;\beta;\theta) & =
\sum_{kk'}
\rho_{k'k}(\theta)
\bar\phi_k(\bm r;\beta)
\phi_{k'}(\bm r;\beta)
\\ &
=\sum_{\lambda\mu}
\rho_{S,\lambda\mu}( r;\beta;\theta)
Y_{\lambda\mu}(\Omega),
\end{split}
\end{equation}
where
\begin{equation}
\begin{split}
&\rho_{S,\lambda\mu}( r;\beta;\theta) =
\sum_{n\kappa}\sum_{n'\kappa'}
\sum_{kk'} \rho_{k'k}(\theta)
c^k_{n\kappa}c^{k'}_{n'\kappa'}\\
&\times
 \langle \kappa m
 |Y^*_{\lambda\mu}(\Omega)|
 \kappa'm'\rangle
\sum_{p=1,2}
\frac{i^{2(p-1)}}{r^2}
R_{n\kappa}(r,p)
R_{n'\kappa'}(r,p).
\end{split}
\end{equation}

The spatial components of the current read
\begin{equation}
\begin{aligned}
\vec{j}(\bm r;\beta;\theta)& =
\sum_{ij}
\rho_{k'k}(\theta)
\langle\phi_i
|\bm \alpha |
\phi_j\rangle\\
&=\sum_{\lambda\mu}
\vec{j}_{\lambda\mu}( r;\beta;\theta)
Y_{\lambda\mu}(\Omega),
\end{aligned}
\end{equation}
with
\begin{equation}
\begin{aligned}
&\vec{j}_{\lambda\mu}(r;\beta;\theta) \\
&=
\sum_{ij}
\sum_{n\kappa}^{n'\kappa'}
c_{n\kappa}^{i}c_{n'\kappa'}^{j}
\rho_{k'k}(\theta)
\frac{i}{r^2}
\sum_{m'_l,m'_s}^{m_l,m_s}
\bm{\sigma}_{m_s,m_s'} \\
&
\left\{~
G_{n\kappa}F_{n'\kappa'}
C^{jm}_{lm_l\frac{1}{2}m_s}
C^{j'm'}_{\tilde l'm_l'\frac{1}{2}m_s'}
\langle l m_l
|Y^*_{\lambda\mu}(\Omega)|
\tilde l'm_l'\rangle
\right. \\
&\left.
 - F_{n\kappa}G_{n'\kappa'}
C^{jm}_{\tilde lm_l\frac{1}{2}m_s}
C^{j'm'}_{ l'm_l'\frac{1}{2}m_s'}
\langle\tilde lm_l
|Y^*_{\lambda\mu}(\Omega)|
l'm_l'\rangle
\right\},
\end{aligned}
\end{equation}
and
\begin{equation}
\langle l m_l|Y_{\lambda\mu}|l' m'_l\rangle =
\int d\Omega
Y^*_{l m_l} (\Omega)
Y_{\lambda\mu}(\Omega)
Y_{l' m'_l}(\Omega).
\end{equation}

\section{The Coulomb energy}
\label{APP:B1}
The direct term of the Coulomb energy reads
\begin{equation}
E_{C}^\mathrm{dir}
=\frac{e^{2}}{2}
\int d \bm{r}
\int d \bm{r'}
\frac{
\rho_{p} \left(\bm{r}\right)
\rho_{p} \left(\bm{r'}\right)}
{\left|\bm{r}-\bm{r'}\right|}.
\label{eq:A10}
\end{equation}
The proton density are expanded in terms of the spherical harmonics, Eq.~(\ref{eq:A10})
is rewritten as
\begin{equation}
\begin{split}
E_{C}^\mathrm{dir}
=\frac{e^{2}}{2}
&\int d \bm{r} \int d \bm{r}' \\
&
\sum_{\lambda\mu}^{\lambda'\mu'}
\frac{
\rho_{p,\lambda\mu}(r)
Y_{\lambda\mu}(\Omega)
\rho_{p,\lambda'\mu'}(r')
Y_{\lambda\mu}(\Omega') }
{\left|\bm{r}-\bm{r}'\right|}.
\end{split}
\label{eq:A11}
\end{equation}
For $\frac{1}{\left|\bm{r}-\bm{r}'\right|}$, one has
\begin{equation}
\frac{1}{\left|\bm{r}-\bm{r}'\right|}
=\frac{4\pi}{r_>}
\sum_{l=0}^{\infty}
\frac{1}{\hat l}
\left(\frac{r_<}{r_>}\right)^l
\sum_{m=-l}^l
Y_{lm}^*(\Omega)Y_{lm}(\Omega'),
\end{equation}
where $r_> = r'$ and $r_< = r $ if $r>r'$
and $r_< = r'$ and $r_> = r $ if $r<r'$.
Inserting it into Eq.~(\ref{eq:A11}), we have
\begin{equation}
\begin{aligned}
E_{C}^\mathrm{dir}
& = 2\pi e^{2}
\sum_{\lambda\mu} \int d {r} \int d {r}' \\
&\frac{1}{\hat\lambda}
\rho_{p,\lambda\mu}(r)
\rho_{p,\lambda-\mu}(r')
\left\{
r^2r'
\left(\frac{r}{r'}\right)^\lambda +
r'^2r
\left(\frac{r'}{r}\right)^\lambda \right\},
\end{aligned}
\end{equation}
with the first term in the bracket for the region $r<r'$
and the second one for $r>r'$.

\section{Calculation of $\left\langle\hat{J_y^2} \right\rangle$}
\label{APP:A3}
The expectation of
$ \hat {J_y^2} $ with respect to the BCS-type wave functions is
\begin{equation}
\begin{split}
\left\langle \hat {J_y^2} \right\rangle  & =
\sum_{kk'}
\left\{
(\hat{J}_y)_{kk'}(\hat{J}_y)_{k'k}v_k^2 u_{k'}^2
\right.\\
&
\left.
+2 v_ku_k v_{k'}u_{k'}
\left[
(\hat{J}_y)_{kk'}(\hat{J}_y)_{\bar k\bar k'} -
(\hat{J}_y)_{k\bar k'}(\hat{J}_y)_{\bar kk'}
\right]
\right\},
\end{split}
\end{equation}
where $k$ and $k'$ are used to denote the single particle state in the canonical basis.

For the eigenvector $|jm\rangle$ of angular momentum operator, one has
\begin{equation}
\begin{split}
\langle j m+1 | \hat{J}_y|jm\rangle &= -\frac{i}{2}\sqrt{(j-m)(j+m+1)},\\
\langle j m-1 | \hat{J}_y|jm\rangle &=  \frac{i}{2}\sqrt{(j+m)(j-m+1)}.
\end{split}
\end{equation}

Under the time reversal transformation $\hat T$,
$\hat T\hat{J}\hat T^\dagger = -\hat{J}$.
Therefore we have
\begin{equation}
\begin{split}
\langle k|\hat{J}_y |\bar{k'}\rangle
& =  \langle \bar k|\hat{J}_y |{k'}\rangle ^*, \\
\langle \bar k|\hat{J}_y |\bar{k'}\rangle
& = -\langle \bar k|\hat{J}_y |{k'}\rangle ^*.
\end{split}
\end{equation}
In the single particle basis labeled by $|k\rangle$ or $|l\rangle$ obtained from DRHBc calculations,
the matrix elements of $\hat{J_y}$ read
\begin{equation}
\begin{aligned}
\langle k|\hat{J}_y|l\rangle
= &
\sum_{n\kappa}c^k_{n\kappa}c^l_{n\kappa}\delta_{jj'}\delta_{ll'} \\
&
\left[\delta_{m,m'+1} \left(-\frac{i}{2}\right)\sqrt{(j-m')(j+m'+1)}\right. \\
&
\left.
+\delta_{m,m'-1} \left( \frac{i}{2}\right)\sqrt{(j+m')(j-m'+1)}\right],
\end{aligned}
\end{equation}
and
\begin{equation}
\begin{aligned}
\langle k|\hat{J}_y|\bar l\rangle
 = &
 \sum_{n\kappa}c^k_{n\kappa}c^l_{n\kappa}(-)^{j'-m'+l'}\delta_{jj'}\delta_{ll'} \\
 &
\left[\delta_{m,-m'+1} \left(-\frac{i}{2}\right)\sqrt{(j+m')(j-m'+1)}\right. \\
&
\left.
+\delta_{m,-m'-1} \left( \frac{i}{2}\right)\sqrt{(j-m')(j+m'+1)}\right].
\end{aligned}
\end{equation}

Finally, $\langle \hat{J^2_y}\rangle$ is given by
\begin{equation}
\begin{split}
\langle \hat{J}_y^2\rangle  & =
\sum_{k,k'>0}
\left\{
2(\hat{J}_y)_{kk'}(\hat{J}_y)_{k'k}v_k^2 u_{k'}^2
\right.\\
&
\left.
+2 v_ku_k v_{k'}u_{k'}
\left[
\left(\hat{J}_y\right)_{kk'}^2+
\left(\hat{J}_y\right)_{k\bar k'}^2
\right]
\right\}.
\end{split}
\end{equation}

%\bibliographystyle{apsrev4-2}
%\bibliography{Sunxx_ref}
%\bibliography{../../../jabref/Sunxx_2021.bib}

\begin{thebibliography}{158}%
\makeatletter
\providecommand \@ifxundefined [1]{%
 \@ifx{#1\undefined}
}%
\providecommand \@ifnum [1]{%
 \ifnum #1\expandafter \@firstoftwo
 \else \expandafter \@secondoftwo
 \fi
}%
\providecommand \@ifx [1]{%
 \ifx #1\expandafter \@firstoftwo
 \else \expandafter \@secondoftwo
 \fi
}%
\providecommand \natexlab [1]{#1}%
\providecommand \enquote  [1]{``#1''}%
\providecommand \bibnamefont  [1]{#1}%
\providecommand \bibfnamefont [1]{#1}%
\providecommand \citenamefont [1]{#1}%
\providecommand \href@noop [0]{\@secondoftwo}%
\providecommand \href [0]{\begingroup \@sanitize@url \@href}%
\providecommand \@href[1]{\@@startlink{#1}\@@href}%
\providecommand \@@href[1]{\endgroup#1\@@endlink}%
\providecommand \@sanitize@url [0]{\catcode `\\12\catcode `\$12\catcode
  `\&12\catcode `\#12\catcode `\^12\catcode `\_12\catcode `\%12\relax}%
\providecommand \@@startlink[1]{}%
\providecommand \@@endlink[0]{}%
\providecommand \url  [0]{\begingroup\@sanitize@url \@url }%
\providecommand \@url [1]{\endgroup\@href {#1}{\urlprefix }}%
\providecommand \urlprefix  [0]{URL }%
\providecommand \Eprint [0]{\href }%
\providecommand \doibase [0]{https://doi.org/}%
\providecommand \selectlanguage [0]{\@gobble}%
\providecommand \bibinfo  [0]{\@secondoftwo}%
\providecommand \bibfield  [0]{\@secondoftwo}%
\providecommand \translation [1]{[#1]}%
\providecommand \BibitemOpen [0]{}%
\providecommand \bibitemStop [0]{}%
\providecommand \bibitemNoStop [0]{.\EOS\space}%
\providecommand \EOS [0]{\spacefactor3000\relax}%
\providecommand \BibitemShut  [1]{\csname bibitem#1\endcsname}%
\let\auto@bib@innerbib\@empty
%</preamble>
\bibitem [{\citenamefont {Tanihata}\ \emph {et~al.}(1985)\citenamefont
  {Tanihata}, \citenamefont {Hamagaki}, \citenamefont {Hashimoto},
  \citenamefont {Shida}, \citenamefont {Yoshikawa}, \citenamefont {Sugimoto},
  \citenamefont {Yamakawa}, \citenamefont {Kobayashi},\ and\ \citenamefont
  {Takahashi}}]{Tanihata1985_PRL55-2676}%
  \BibitemOpen
  \bibfield  {author} {\bibinfo {author} {\bibfnamefont {I.}~\bibnamefont
  {Tanihata}}, \bibinfo {author} {\bibfnamefont {H.}~\bibnamefont {Hamagaki}},
  \bibinfo {author} {\bibfnamefont {O.}~\bibnamefont {Hashimoto}}, \bibinfo
  {author} {\bibfnamefont {Y.}~\bibnamefont {Shida}}, \bibinfo {author}
  {\bibfnamefont {N.}~\bibnamefont {Yoshikawa}}, \bibinfo {author}
  {\bibfnamefont {K.}~\bibnamefont {Sugimoto}}, \bibinfo {author}
  {\bibfnamefont {O.}~\bibnamefont {Yamakawa}}, \bibinfo {author}
  {\bibfnamefont {T.}~\bibnamefont {Kobayashi}},\ and\ \bibinfo {author}
  {\bibfnamefont {N.}~\bibnamefont {Takahashi}},\ }\href
  {https://doi.org/10.1103/PhysRevLett.55.2676} {\bibfield  {journal} {\bibinfo
   {journal} {Phys. Rev. Lett.}\ }\textbf {\bibinfo {volume} {55}},\ \bibinfo
  {pages} {2676} (\bibinfo {year} {1985})}\BibitemShut {NoStop}%
\bibitem [{\citenamefont {Tanihata}\ \emph {et~al.}(2013)\citenamefont
  {Tanihata}, \citenamefont {Savajols},\ and\ \citenamefont
  {Kanungo}}]{Tanihata2013_PPNP68-215}%
  \BibitemOpen
  \bibfield  {author} {\bibinfo {author} {\bibfnamefont {I.}~\bibnamefont
  {Tanihata}}, \bibinfo {author} {\bibfnamefont {H.}~\bibnamefont {Savajols}},\
  and\ \bibinfo {author} {\bibfnamefont {R.}~\bibnamefont {Kanungo}},\ }\href
  {https://doi.org/10.1016/j.ppnp.2012.07.001} {\bibfield  {journal} {\bibinfo
  {journal} {Prog. Part. Nucl. Phys.}\ }\textbf {\bibinfo {volume} {68}},\
  \bibinfo {pages} {215 } (\bibinfo {year} {2013})}\BibitemShut {NoStop}%
\bibitem [{\citenamefont {Ozawa}\ \emph {et~al.}(2000)\citenamefont {Ozawa},
  \citenamefont {Kobayashi}, \citenamefont {Suzuki}, \citenamefont {Yoshida},\
  and\ \citenamefont {Tanihata}}]{Ozawa2000_PRL84-5493}%
  \BibitemOpen
  \bibfield  {author} {\bibinfo {author} {\bibfnamefont {A.}~\bibnamefont
  {Ozawa}}, \bibinfo {author} {\bibfnamefont {T.}~\bibnamefont {Kobayashi}},
  \bibinfo {author} {\bibfnamefont {T.}~\bibnamefont {Suzuki}}, \bibinfo
  {author} {\bibfnamefont {K.}~\bibnamefont {Yoshida}},\ and\ \bibinfo {author}
  {\bibfnamefont {I.}~\bibnamefont {Tanihata}},\ }\href
  {https://doi.org/10.1103/PhysRevLett.84.5493} {\bibfield  {journal} {\bibinfo
   {journal} {Phys. Rev. Lett.}\ }\textbf {\bibinfo {volume} {84}},\ \bibinfo
  {pages} {5493} (\bibinfo {year} {2000})}\BibitemShut {NoStop}%
\bibitem [{\citenamefont {Janssens}(2009)}]{Janssens2009_Nature459-1069}%
  \BibitemOpen
  \bibfield  {author} {\bibinfo {author} {\bibfnamefont {R.~V.~F.}\
  \bibnamefont {Janssens}},\ }\href {https://doi.org/10.1038/4591069a}
  {\bibfield  {journal} {\bibinfo  {journal} {Nature}\ }\textbf {\bibinfo
  {volume} {459}},\ \bibinfo {pages} {1069} (\bibinfo {year}
  {2009})}\BibitemShut {NoStop}%
\bibitem [{\citenamefont {Wienholtz}\ \emph {et~al.}(2013)\citenamefont
  {Wienholtz}, \citenamefont {Beck}, \citenamefont {Blaum}, \citenamefont
  {Borgmann}, \citenamefont {Breitenfeldt}, \citenamefont {Cakirli},
  \citenamefont {George}, \citenamefont {Herfurth}, \citenamefont {Holt},
  \citenamefont {Kowalska}, \citenamefont {Kreim}, \citenamefont {Lunney},
  \citenamefont {Manea}, \citenamefont {Men{\'{e}}ndez}, \citenamefont
  {Neidherr}, \citenamefont {Rosenbusch}, \citenamefont {Schweikhard},
  \citenamefont {Schwenk}, \citenamefont {Simonis}, \citenamefont {Stanja},
  \citenamefont {Wolf},\ and\ \citenamefont
  {Zuber}}]{Wienholtz2013_Nature498-346}%
  \BibitemOpen
  \bibfield  {author} {\bibinfo {author} {\bibfnamefont {F.}~\bibnamefont
  {Wienholtz}}, \bibinfo {author} {\bibfnamefont {D.}~\bibnamefont {Beck}},
  \bibinfo {author} {\bibfnamefont {K.}~\bibnamefont {Blaum}}, \bibinfo
  {author} {\bibfnamefont {C.}~\bibnamefont {Borgmann}}, \bibinfo {author}
  {\bibfnamefont {M.}~\bibnamefont {Breitenfeldt}}, \bibinfo {author}
  {\bibfnamefont {R.~B.}\ \bibnamefont {Cakirli}}, \bibinfo {author}
  {\bibfnamefont {S.}~\bibnamefont {George}}, \bibinfo {author} {\bibfnamefont
  {F.}~\bibnamefont {Herfurth}}, \bibinfo {author} {\bibfnamefont {J.~D.}\
  \bibnamefont {Holt}}, \bibinfo {author} {\bibfnamefont {M.}~\bibnamefont
  {Kowalska}}, \bibinfo {author} {\bibfnamefont {S.}~\bibnamefont {Kreim}},
  \bibinfo {author} {\bibfnamefont {D.}~\bibnamefont {Lunney}}, \bibinfo
  {author} {\bibfnamefont {V.}~\bibnamefont {Manea}}, \bibinfo {author}
  {\bibfnamefont {J.}~\bibnamefont {Men{\'{e}}ndez}}, \bibinfo {author}
  {\bibfnamefont {D.}~\bibnamefont {Neidherr}}, \bibinfo {author}
  {\bibfnamefont {M.}~\bibnamefont {Rosenbusch}}, \bibinfo {author}
  {\bibfnamefont {L.}~\bibnamefont {Schweikhard}}, \bibinfo {author}
  {\bibfnamefont {A.}~\bibnamefont {Schwenk}}, \bibinfo {author} {\bibfnamefont
  {J.}~\bibnamefont {Simonis}}, \bibinfo {author} {\bibfnamefont
  {J.}~\bibnamefont {Stanja}}, \bibinfo {author} {\bibfnamefont {R.~N.}\
  \bibnamefont {Wolf}},\ and\ \bibinfo {author} {\bibfnamefont
  {K.}~\bibnamefont {Zuber}},\ }\href
  {https://doi.org/https://doi.org/10.1038/nature12226} {\bibfield  {journal}
  {\bibinfo  {journal} {Nature}\ }\textbf {\bibinfo {volume} {498}},\ \bibinfo
  {pages} {346} (\bibinfo {year} {2013})}\BibitemShut {NoStop}%
\bibitem [{\citenamefont {Tran}\ \emph {et~al.}(2018)\citenamefont {Tran},
  \citenamefont {Ong}, \citenamefont {Hagen}, \citenamefont {Morris},
  \citenamefont {Aoi}, \citenamefont {Suzuki}, \citenamefont {Kanada-En'yo},
  \citenamefont {Geng}, \citenamefont {Terashima}, \citenamefont {Tanihata},
  \citenamefont {Nguyen}, \citenamefont {Ayyad}, \citenamefont {Chan},
  \citenamefont {Fukuda}, \citenamefont {Geissel}, \citenamefont {Harakeh},
  \citenamefont {Hashimoto}, \citenamefont {Hoang}, \citenamefont {Ideguchi},
  \citenamefont {Inoue}, \citenamefont {Jansen}, \citenamefont {Kanungo},
  \citenamefont {Kawabata}, \citenamefont {Khiem}, \citenamefont {Lin},
  \citenamefont {Matsuta}, \citenamefont {Mihara}, \citenamefont {Momota},
  \citenamefont {Nagae}, \citenamefont {Nguyen}, \citenamefont {Nishimura},
  \citenamefont {Otsuka}, \citenamefont {Ozawa}, \citenamefont {Ren},
  \citenamefont {Sakaguchi}, \citenamefont {Scheidenberger}, \citenamefont
  {Tanaka}, \citenamefont {Takechi}, \citenamefont {Wada},\ and\ \citenamefont
  {Yamamoto}}]{Tran2018_NatCommun9-1594}%
  \BibitemOpen
  \bibfield  {author} {\bibinfo {author} {\bibfnamefont {D.~T.}\ \bibnamefont
  {Tran}}, \bibinfo {author} {\bibfnamefont {H.~J.}\ \bibnamefont {Ong}},
  \bibinfo {author} {\bibfnamefont {G.}~\bibnamefont {Hagen}}, \bibinfo
  {author} {\bibfnamefont {T.~D.}\ \bibnamefont {Morris}}, \bibinfo {author}
  {\bibfnamefont {N.}~\bibnamefont {Aoi}}, \bibinfo {author} {\bibfnamefont
  {T.}~\bibnamefont {Suzuki}}, \bibinfo {author} {\bibfnamefont
  {Y.}~\bibnamefont {Kanada-En'yo}}, \bibinfo {author} {\bibfnamefont {L.~S.}\
  \bibnamefont {Geng}}, \bibinfo {author} {\bibfnamefont {S.}~\bibnamefont
  {Terashima}}, \bibinfo {author} {\bibfnamefont {I.}~\bibnamefont {Tanihata}},
  \bibinfo {author} {\bibfnamefont {T.~T.}\ \bibnamefont {Nguyen}}, \bibinfo
  {author} {\bibfnamefont {Y.}~\bibnamefont {Ayyad}}, \bibinfo {author}
  {\bibfnamefont {P.~Y.}\ \bibnamefont {Chan}}, \bibinfo {author}
  {\bibfnamefont {M.}~\bibnamefont {Fukuda}}, \bibinfo {author} {\bibfnamefont
  {H.}~\bibnamefont {Geissel}}, \bibinfo {author} {\bibfnamefont {M.~N.}\
  \bibnamefont {Harakeh}}, \bibinfo {author} {\bibfnamefont {T.}~\bibnamefont
  {Hashimoto}}, \bibinfo {author} {\bibfnamefont {T.~H.}\ \bibnamefont
  {Hoang}}, \bibinfo {author} {\bibfnamefont {E.}~\bibnamefont {Ideguchi}},
  \bibinfo {author} {\bibfnamefont {A.}~\bibnamefont {Inoue}}, \bibinfo
  {author} {\bibfnamefont {G.~R.}\ \bibnamefont {Jansen}}, \bibinfo {author}
  {\bibfnamefont {R.}~\bibnamefont {Kanungo}}, \bibinfo {author} {\bibfnamefont
  {T.}~\bibnamefont {Kawabata}}, \bibinfo {author} {\bibfnamefont {L.~H.}\
  \bibnamefont {Khiem}}, \bibinfo {author} {\bibfnamefont {W.~P.}\ \bibnamefont
  {Lin}}, \bibinfo {author} {\bibfnamefont {K.}~\bibnamefont {Matsuta}},
  \bibinfo {author} {\bibfnamefont {M.}~\bibnamefont {Mihara}}, \bibinfo
  {author} {\bibfnamefont {S.}~\bibnamefont {Momota}}, \bibinfo {author}
  {\bibfnamefont {D.}~\bibnamefont {Nagae}}, \bibinfo {author} {\bibfnamefont
  {N.~D.}\ \bibnamefont {Nguyen}}, \bibinfo {author} {\bibfnamefont
  {D.}~\bibnamefont {Nishimura}}, \bibinfo {author} {\bibfnamefont
  {T.}~\bibnamefont {Otsuka}}, \bibinfo {author} {\bibfnamefont
  {A.}~\bibnamefont {Ozawa}}, \bibinfo {author} {\bibfnamefont {P.~P.}\
  \bibnamefont {Ren}}, \bibinfo {author} {\bibfnamefont {H.}~\bibnamefont
  {Sakaguchi}}, \bibinfo {author} {\bibfnamefont {C.}~\bibnamefont
  {Scheidenberger}}, \bibinfo {author} {\bibfnamefont {J.}~\bibnamefont
  {Tanaka}}, \bibinfo {author} {\bibfnamefont {M.}~\bibnamefont {Takechi}},
  \bibinfo {author} {\bibfnamefont {R.}~\bibnamefont {Wada}},\ and\ \bibinfo
  {author} {\bibfnamefont {T.}~\bibnamefont {Yamamoto}},\ }\href
  {https://doi.org/10.1038/s41467-018-04024-y} {\bibfield  {journal} {\bibinfo
  {journal} {Nat. Commun.}\ }\textbf {\bibinfo {volume} {9}},\ \bibinfo {pages}
  {1594} (\bibinfo {year} {2018})}\BibitemShut {NoStop}%
\bibitem [{\citenamefont {Warburton}\ \emph {et~al.}(1990)\citenamefont
  {Warburton}, \citenamefont {Becker},\ and\ \citenamefont
  {Brown}}]{Warburton1990_PRC41-1147}%
  \BibitemOpen
  \bibfield  {author} {\bibinfo {author} {\bibfnamefont {E.~K.}\ \bibnamefont
  {Warburton}}, \bibinfo {author} {\bibfnamefont {J.~A.}\ \bibnamefont
  {Becker}},\ and\ \bibinfo {author} {\bibfnamefont {B.~A.}\ \bibnamefont
  {Brown}},\ }\href {https://doi.org/10.1103/PhysRevC.41.1147} {\bibfield
  {journal} {\bibinfo  {journal} {Phys. Rev. C}\ }\textbf {\bibinfo {volume}
  {41}},\ \bibinfo {pages} {1147} (\bibinfo {year} {1990})}\BibitemShut
  {NoStop}%
\bibitem [{\citenamefont {Centelles}\ \emph {et~al.}(2009)\citenamefont
  {Centelles}, \citenamefont {Roca-Maza}, \citenamefont {Vi\~nas},\ and\
  \citenamefont {Warda}}]{Centelles2009_PRL102-122502}%
  \BibitemOpen
  \bibfield  {author} {\bibinfo {author} {\bibfnamefont {M.}~\bibnamefont
  {Centelles}}, \bibinfo {author} {\bibfnamefont {X.}~\bibnamefont
  {Roca-Maza}}, \bibinfo {author} {\bibfnamefont {X.}~\bibnamefont {Vi\~nas}},\
  and\ \bibinfo {author} {\bibfnamefont {M.}~\bibnamefont {Warda}},\ }\href
  {https://doi.org/10.1103/PhysRevLett.102.122502} {\bibfield  {journal}
  {\bibinfo  {journal} {Phys. Rev. Lett.}\ }\textbf {\bibinfo {volume} {102}},\
  \bibinfo {pages} {122502} (\bibinfo {year} {2009})}\BibitemShut {NoStop}%
\bibitem [{\citenamefont {Ebran}\ \emph {et~al.}(2012)\citenamefont {Ebran},
  \citenamefont {Khan}, \citenamefont {Niksic},\ and\ \citenamefont
  {Vretenar}}]{Ebran2012_Nature487-341}%
  \BibitemOpen
  \bibfield  {author} {\bibinfo {author} {\bibfnamefont {J.~P.}\ \bibnamefont
  {Ebran}}, \bibinfo {author} {\bibfnamefont {E.}~\bibnamefont {Khan}},
  \bibinfo {author} {\bibfnamefont {T.}~\bibnamefont {Niksic}},\ and\ \bibinfo
  {author} {\bibfnamefont {D.}~\bibnamefont {Vretenar}},\ }\href
  {https://doi.org/10.1038/nature11246} {\bibfield  {journal} {\bibinfo
  {journal} {Nature}\ }\textbf {\bibinfo {volume} {487}},\ \bibinfo {pages}
  {341} (\bibinfo {year} {2012})}\BibitemShut {NoStop}%
%%CITATION = NATUA,487,341;%%
\bibitem [{\citenamefont {Freer}\ \emph {et~al.}(2018)\citenamefont {Freer},
  \citenamefont {Horiuchi}, \citenamefont {Kanada-En'yo}, \citenamefont {Lee},\
  and\ \citenamefont {Mei\ss{}ner}}]{Freer2018_RMP90-035004}%
  \BibitemOpen
  \bibfield  {author} {\bibinfo {author} {\bibfnamefont {M.}~\bibnamefont
  {Freer}}, \bibinfo {author} {\bibfnamefont {H.}~\bibnamefont {Horiuchi}},
  \bibinfo {author} {\bibfnamefont {Y.}~\bibnamefont {Kanada-En'yo}}, \bibinfo
  {author} {\bibfnamefont {D.}~\bibnamefont {Lee}},\ and\ \bibinfo {author}
  {\bibfnamefont {U.-G.}\ \bibnamefont {Mei\ss{}ner}},\ }\href
  {https://doi.org/10.1103/RevModPhys.90.035004} {\bibfield  {journal}
  {\bibinfo  {journal} {Rev. Mod. Phys.}\ }\textbf {\bibinfo {volume} {90}},\
  \bibinfo {pages} {035004} (\bibinfo {year} {2018})}\BibitemShut {NoStop}%
\bibitem [{\citenamefont {Pf\"utzner}\ \emph {et~al.}(2012)\citenamefont
  {Pf\"utzner}, \citenamefont {Karny}, \citenamefont {Grigorenko},\ and\
  \citenamefont {Riisager}}]{Pfuetzner2012_RMP84-567}%
  \BibitemOpen
  \bibfield  {author} {\bibinfo {author} {\bibfnamefont {M.}~\bibnamefont
  {Pf\"utzner}}, \bibinfo {author} {\bibfnamefont {M.}~\bibnamefont {Karny}},
  \bibinfo {author} {\bibfnamefont {L.~V.}\ \bibnamefont {Grigorenko}},\ and\
  \bibinfo {author} {\bibfnamefont {K.}~\bibnamefont {Riisager}},\ }\href
  {https://doi.org/10.1103/RevModPhys.84.567} {\bibfield  {journal} {\bibinfo
  {journal} {Rev. Mod. Phys.}\ }\textbf {\bibinfo {volume} {84}},\ \bibinfo
  {pages} {567} (\bibinfo {year} {2012})}\BibitemShut {NoStop}%
\bibitem [{\citenamefont {Mutschler}\ \emph {et~al.}(2016)\citenamefont
  {Mutschler}, \citenamefont {Lemasson}, \citenamefont {Sorlin}, \citenamefont
  {Bazin}, \citenamefont {Borcea}, \citenamefont {Borcea}, \citenamefont
  {Dombr{\'{a}}di}, \citenamefont {Ebran}, \citenamefont {Gade}, \citenamefont
  {Iwasaki}, \citenamefont {Khan}, \citenamefont {Lepailleur}, \citenamefont
  {Recchia}, \citenamefont {Roger}, \citenamefont {Rotaru}, \citenamefont
  {Sohler}, \citenamefont {Stanoiu}, \citenamefont {Stroberg}, \citenamefont
  {Tostevin}, \citenamefont {Vandebrouck}, \citenamefont {Weisshaar},\ and\
  \citenamefont {Wimmer}}]{Mutschler2016_NatPhysP13-152}%
  \BibitemOpen
  \bibfield  {author} {\bibinfo {author} {\bibfnamefont {A.}~\bibnamefont
  {Mutschler}}, \bibinfo {author} {\bibfnamefont {A.}~\bibnamefont {Lemasson}},
  \bibinfo {author} {\bibfnamefont {O.}~\bibnamefont {Sorlin}}, \bibinfo
  {author} {\bibfnamefont {D.}~\bibnamefont {Bazin}}, \bibinfo {author}
  {\bibfnamefont {C.}~\bibnamefont {Borcea}}, \bibinfo {author} {\bibfnamefont
  {R.}~\bibnamefont {Borcea}}, \bibinfo {author} {\bibfnamefont
  {Z.}~\bibnamefont {Dombr{\'{a}}di}}, \bibinfo {author} {\bibfnamefont
  {J.-P.}\ \bibnamefont {Ebran}}, \bibinfo {author} {\bibfnamefont
  {A.}~\bibnamefont {Gade}}, \bibinfo {author} {\bibfnamefont {H.}~\bibnamefont
  {Iwasaki}}, \bibinfo {author} {\bibfnamefont {E.}~\bibnamefont {Khan}},
  \bibinfo {author} {\bibfnamefont {A.}~\bibnamefont {Lepailleur}}, \bibinfo
  {author} {\bibfnamefont {F.}~\bibnamefont {Recchia}}, \bibinfo {author}
  {\bibfnamefont {T.}~\bibnamefont {Roger}}, \bibinfo {author} {\bibfnamefont
  {F.}~\bibnamefont {Rotaru}}, \bibinfo {author} {\bibfnamefont
  {D.}~\bibnamefont {Sohler}}, \bibinfo {author} {\bibfnamefont
  {M.}~\bibnamefont {Stanoiu}}, \bibinfo {author} {\bibfnamefont {S.~R.}\
  \bibnamefont {Stroberg}}, \bibinfo {author} {\bibfnamefont {J.~A.}\
  \bibnamefont {Tostevin}}, \bibinfo {author} {\bibfnamefont {M.}~\bibnamefont
  {Vandebrouck}}, \bibinfo {author} {\bibfnamefont {D.}~\bibnamefont
  {Weisshaar}},\ and\ \bibinfo {author} {\bibfnamefont {K.}~\bibnamefont
  {Wimmer}},\ }\href {https://doi.org/10.1038/nphys3916} {\bibfield  {journal}
  {\bibinfo  {journal} {Nat. Phys.}\ }\textbf {\bibinfo {volume} {13}},\
  \bibinfo {pages} {152} (\bibinfo {year} {2016})}\BibitemShut {NoStop}%
\bibitem [{\citenamefont {Yao}\ \emph {et~al.}(2013)\citenamefont {Yao},
  \citenamefont {Mei},\ and\ \citenamefont {Li}}]{Yao2013_PLB723-459}%
  \BibitemOpen
  \bibfield  {author} {\bibinfo {author} {\bibfnamefont {J.}~\bibnamefont
  {Yao}}, \bibinfo {author} {\bibfnamefont {H.}~\bibnamefont {Mei}},\ and\
  \bibinfo {author} {\bibfnamefont {Z.}~\bibnamefont {Li}},\ }\href
  {https://doi.org/10.1016/j.physletb.2013.05.049} {\bibfield  {journal}
  {\bibinfo  {journal} {Phys. Lett. B}\ }\textbf {\bibinfo {volume} {723}},\
  \bibinfo {pages} {459} (\bibinfo {year} {2013})}\BibitemShut {NoStop}%
\bibitem [{\citenamefont {Cejnar}\ \emph {et~al.}(2010)\citenamefont {Cejnar},
  \citenamefont {Jolie},\ and\ \citenamefont
  {Casten}}]{Cejnar2010_RMP82-21552212}%
  \BibitemOpen
  \bibfield  {author} {\bibinfo {author} {\bibfnamefont {P.}~\bibnamefont
  {Cejnar}}, \bibinfo {author} {\bibfnamefont {J.}~\bibnamefont {Jolie}},\ and\
  \bibinfo {author} {\bibfnamefont {R.~F.}\ \bibnamefont {Casten}},\ }\href
  {https://doi.org/10.1103/RevModPhys.82.2155} {\bibfield  {journal} {\bibinfo
  {journal} {Rev. Mod. Phys.}\ }\textbf {\bibinfo {volume} {82}},\ \bibinfo
  {pages} {2155} (\bibinfo {year} {2010})}\BibitemShut {NoStop}%
\bibitem [{\citenamefont {Heyde}\ and\ \citenamefont
  {Wood}(2011)}]{Heyde2011_RMP83-1467}%
  \BibitemOpen
  \bibfield  {author} {\bibinfo {author} {\bibfnamefont {K.}~\bibnamefont
  {Heyde}}\ and\ \bibinfo {author} {\bibfnamefont {J.~L.}\ \bibnamefont
  {Wood}},\ }\href {https://doi.org/10.1103/RevModPhys.83.1467} {\bibfield
  {journal} {\bibinfo  {journal} {Rev. Mod. Phys.}\ }\textbf {\bibinfo {volume}
  {83}},\ \bibinfo {pages} {1467} (\bibinfo {year} {2011})}\BibitemShut
  {NoStop}%
\bibitem [{\citenamefont {Li}\ \emph {et~al.}(2016)\citenamefont {Li},
  \citenamefont {Nik\v{s}i\'{c}},\ and\ \citenamefont
  {Vretenar}}]{Li2016_JPG43-024005}%
  \BibitemOpen
  \bibfield  {author} {\bibinfo {author} {\bibfnamefont {Z.~P.}\ \bibnamefont
  {Li}}, \bibinfo {author} {\bibfnamefont {T.}~\bibnamefont {Nik\v{s}i\'{c}}},\
  and\ \bibinfo {author} {\bibfnamefont {D.}~\bibnamefont {Vretenar}},\ }\href
  {http://stacks.iop.org/0954-3899/43/i=2/a=024005} {\bibfield  {journal}
  {\bibinfo  {journal} {J. Phys. G: Nucl. Part. Phys.}\ }\textbf {\bibinfo
  {volume} {43}},\ \bibinfo {pages} {024005} (\bibinfo {year}
  {2016})}\BibitemShut {NoStop}%
\bibitem [{\citenamefont {Bender}\ \emph
  {et~al.}(2003{\natexlab{a}})\citenamefont {Bender}, \citenamefont {Heenen},\
  and\ \citenamefont {Reinhard}}]{Bender2003_RMP75-121}%
  \BibitemOpen
  \bibfield  {author} {\bibinfo {author} {\bibfnamefont {M.}~\bibnamefont
  {Bender}}, \bibinfo {author} {\bibfnamefont {P.-H.}\ \bibnamefont {Heenen}},\
  and\ \bibinfo {author} {\bibfnamefont {P.-G.}\ \bibnamefont {Reinhard}},\
  }\href {https://doi.org/10.1103/RevModPhys.75.121} {\bibfield  {journal}
  {\bibinfo  {journal} {Rev. Mod. Phys.}\ }\textbf {\bibinfo {volume} {75}},\
  \bibinfo {pages} {121} (\bibinfo {year} {2003}{\natexlab{a}})}\BibitemShut
  {NoStop}%
\bibitem [{\citenamefont {{\'{C}}wiok}\ \emph {et~al.}(2005)\citenamefont
  {{\'{C}}wiok}, \citenamefont {Heenen},\ and\ \citenamefont
  {Nazarewicz}}]{Cwiok2005_Nature433-705}%
  \BibitemOpen
  \bibfield  {author} {\bibinfo {author} {\bibfnamefont {S.}~\bibnamefont
  {{\'{C}}wiok}}, \bibinfo {author} {\bibfnamefont {P.-H.}\ \bibnamefont
  {Heenen}},\ and\ \bibinfo {author} {\bibfnamefont {W.}~\bibnamefont
  {Nazarewicz}},\ }\href {https://doi.org/10.1038/nature03336} {\bibfield
  {journal} {\bibinfo  {journal} {Nature}\ }\textbf {\bibinfo {volume} {433}},\
  \bibinfo {pages} {705} (\bibinfo {year} {2005})}\BibitemShut {NoStop}%
\bibitem [{\citenamefont {Meng}\ \emph {et~al.}(2006)\citenamefont {Meng},
  \citenamefont {Toki}, \citenamefont {Zhou}, \citenamefont {Zhang},
  \citenamefont {Long},\ and\ \citenamefont {Geng}}]{Meng2006_PPNP57-470}%
  \BibitemOpen
  \bibfield  {author} {\bibinfo {author} {\bibfnamefont {J.}~\bibnamefont
  {Meng}}, \bibinfo {author} {\bibfnamefont {H.}~\bibnamefont {Toki}}, \bibinfo
  {author} {\bibfnamefont {S.}~\bibnamefont {Zhou}}, \bibinfo {author}
  {\bibfnamefont {S.}~\bibnamefont {Zhang}}, \bibinfo {author} {\bibfnamefont
  {W.}~\bibnamefont {Long}},\ and\ \bibinfo {author} {\bibfnamefont
  {L.}~\bibnamefont {Geng}},\ }\href
  {https://doi.org/10.1016/j.ppnp.2005.06.001} {\bibfield  {journal} {\bibinfo
  {journal} {Prog. Part. Nucl. Phys.}\ }\textbf {\bibinfo {volume} {57}},\
  \bibinfo {pages} {470} (\bibinfo {year} {2006})}\BibitemShut {NoStop}%
\bibitem [{\citenamefont {Meng}\ and\ \citenamefont
  {Zhou}(2015)}]{Meng2015_JPG42-093101}%
  \BibitemOpen
  \bibfield  {author} {\bibinfo {author} {\bibfnamefont {J.}~\bibnamefont
  {Meng}}\ and\ \bibinfo {author} {\bibfnamefont {S.-G.}\ \bibnamefont
  {Zhou}},\ }\href {https://doi.org/10.1088/0954-3899/42/9/093101} {\bibfield
  {journal} {\bibinfo  {journal} {J. Phys. G: Nucl. Part. Phys.}\ }\textbf
  {\bibinfo {volume} {42}},\ \bibinfo {pages} {093101} (\bibinfo {year}
  {2015})}\BibitemShut {NoStop}%
\bibitem [{\citenamefont {Nik{\v{s}}i{\'{c}}}\ \emph
  {et~al.}(2011)\citenamefont {Nik{\v{s}}i{\'{c}}}, \citenamefont {Vretenar},\
  and\ \citenamefont {Ring}}]{Niksic2011_PPNP66-519}%
  \BibitemOpen
  \bibfield  {author} {\bibinfo {author} {\bibfnamefont {T.}~\bibnamefont
  {Nik{\v{s}}i{\'{c}}}}, \bibinfo {author} {\bibfnamefont {D.}~\bibnamefont
  {Vretenar}},\ and\ \bibinfo {author} {\bibfnamefont {P.}~\bibnamefont
  {Ring}},\ }\href {https://doi.org/10.1016/j.ppnp.2011.01.055} {\bibfield
  {journal} {\bibinfo  {journal} {Prog. Part. Nucl. Phys.}\ }\textbf {\bibinfo
  {volume} {66}},\ \bibinfo {pages} {519} (\bibinfo {year} {2011})}\BibitemShut
  {NoStop}%
\bibitem [{\citenamefont {Meng}(2016)}]{Meng2016_RDFNS}%
  \BibitemOpen
  \bibinfo {editor} {\bibfnamefont {J.}~\bibnamefont {Meng}},\ ed.,\ \href
  {https://doi.org/10.1142/9872} {\emph {\bibinfo {title} {Relativistic Density
  Functional for Nuclear Structure}}}\ (\bibinfo  {publisher} {World
  Scientific},\ \bibinfo {year} {2016})\BibitemShut {NoStop}%
\bibitem [{\citenamefont {Zhou}(2016)}]{Zhou2016_PS91-063008}%
  \BibitemOpen
  \bibfield  {author} {\bibinfo {author} {\bibfnamefont {S.-G.}\ \bibnamefont
  {Zhou}},\ }\href {https://doi.org/10.1088/0031-8949/91/6/063008} {\bibfield
  {journal} {\bibinfo  {journal} {Phys. Scr.}\ }\textbf {\bibinfo {volume}
  {91}},\ \bibinfo {pages} {063008} (\bibinfo {year} {2016})}\BibitemShut
  {NoStop}%
\bibitem [{\citenamefont {Zhou}(2017)}]{Zhou2017_PoS-INPC2016-373}%
  \BibitemOpen
  \bibfield  {author} {\bibinfo {author} {\bibfnamefont {S.-G.}\ \bibnamefont
  {Zhou}},\ }\href {https://doi.org/10.22323/1.281.0373} {\bibfield  {journal}
  {\bibinfo  {journal} {PoS}\ }\textbf {\bibinfo {volume} {INPC2016}},\
  \bibinfo {pages} {373} (\bibinfo {year} {2017})}\BibitemShut {NoStop}%
\bibitem [{\citenamefont {Otsuka}\ \emph {et~al.}(2020)\citenamefont {Otsuka},
  \citenamefont {Gade}, \citenamefont {Sorlin}, \citenamefont {Suzuki},\ and\
  \citenamefont {Utsuno}}]{Otsuka2020_RMP92-015002}%
  \BibitemOpen
  \bibfield  {author} {\bibinfo {author} {\bibfnamefont {T.}~\bibnamefont
  {Otsuka}}, \bibinfo {author} {\bibfnamefont {A.}~\bibnamefont {Gade}},
  \bibinfo {author} {\bibfnamefont {O.}~\bibnamefont {Sorlin}}, \bibinfo
  {author} {\bibfnamefont {T.}~\bibnamefont {Suzuki}},\ and\ \bibinfo {author}
  {\bibfnamefont {Y.}~\bibnamefont {Utsuno}},\ }\href
  {https://doi.org/10.1103/RevModPhys.92.015002} {\bibfield  {journal}
  {\bibinfo  {journal} {Rev. Mod. Phys.}\ }\textbf {\bibinfo {volume} {92}},\
  \bibinfo {pages} {015002} (\bibinfo {year} {2020})}\BibitemShut {NoStop}%
\bibitem [{\citenamefont {Caurier}\ \emph {et~al.}(2005)\citenamefont
  {Caurier}, \citenamefont {Mart\'{\i}nez-Pinedo}, \citenamefont {Nowack},
  \citenamefont {Poves},\ and\ \citenamefont {Zuker}}]{Caurier2005_RMP77-427}%
  \BibitemOpen
  \bibfield  {author} {\bibinfo {author} {\bibfnamefont {E.}~\bibnamefont
  {Caurier}}, \bibinfo {author} {\bibfnamefont {G.}~\bibnamefont
  {Mart\'{\i}nez-Pinedo}}, \bibinfo {author} {\bibfnamefont {F.}~\bibnamefont
  {Nowack}}, \bibinfo {author} {\bibfnamefont {A.}~\bibnamefont {Poves}},\ and\
  \bibinfo {author} {\bibfnamefont {A.~P.}\ \bibnamefont {Zuker}},\ }\href
  {https://doi.org/10.1103/RevModPhys.77.427} {\bibfield  {journal} {\bibinfo
  {journal} {Rev. Mod. Phys.}\ }\textbf {\bibinfo {volume} {77}},\ \bibinfo
  {pages} {427} (\bibinfo {year} {2005})}\BibitemShut {NoStop}%
\bibitem [{\citenamefont {Vretenar}\ \emph {et~al.}(2005)\citenamefont
  {Vretenar}, \citenamefont {Afanasjev}, \citenamefont {Lalazissis},\ and\
  \citenamefont {Ring}}]{Vretenar2005_PR409-101}%
  \BibitemOpen
  \bibfield  {author} {\bibinfo {author} {\bibfnamefont {D.}~\bibnamefont
  {Vretenar}}, \bibinfo {author} {\bibfnamefont {A.~V.}\ \bibnamefont
  {Afanasjev}}, \bibinfo {author} {\bibfnamefont {G.~A.}\ \bibnamefont
  {Lalazissis}},\ and\ \bibinfo {author} {\bibfnamefont {P.}~\bibnamefont
  {Ring}},\ }\href {https://doi.org/10.1016/j.physrep.2004.10.001} {\bibfield
  {journal} {\bibinfo  {journal} {Phys. Rep.}\ }\textbf {\bibinfo {volume}
  {409}},\ \bibinfo {pages} {101} (\bibinfo {year} {2005})}\BibitemShut
  {NoStop}%
\bibitem [{\citenamefont {Kimura}\ \emph {et~al.}(2016)\citenamefont {Kimura},
  \citenamefont {Suhara},\ and\ \citenamefont
  {Kanada-En'yo}}]{Kimura2016_EPJA52-373}%
  \BibitemOpen
  \bibfield  {author} {\bibinfo {author} {\bibfnamefont {M.}~\bibnamefont
  {Kimura}}, \bibinfo {author} {\bibfnamefont {T.}~\bibnamefont {Suhara}},\
  and\ \bibinfo {author} {\bibfnamefont {Y.}~\bibnamefont {Kanada-En'yo}},\
  }\href {https://doi.org/10.1140/epja/i2016-16373-9} {\bibfield  {journal}
  {\bibinfo  {journal} {Eur. Phys. J. A}\ }\textbf {\bibinfo {volume} {52}},\
  \bibinfo {pages} {373} (\bibinfo {year} {2016})}\BibitemShut {NoStop}%
\bibitem [{\citenamefont {Greene}\ \emph {et~al.}(2017)\citenamefont {Greene},
  \citenamefont {Giannakeas},\ and\ \citenamefont
  {P\'erez-R\'{\i}os}}]{Greene2017_RMP89-35006}%
  \BibitemOpen
  \bibfield  {author} {\bibinfo {author} {\bibfnamefont {C.~H.}\ \bibnamefont
  {Greene}}, \bibinfo {author} {\bibfnamefont {P.}~\bibnamefont {Giannakeas}},\
  and\ \bibinfo {author} {\bibfnamefont {J.}~\bibnamefont
  {P\'erez-R\'{\i}os}},\ }\href {https://doi.org/10.1103/RevModPhys.89.035006}
  {\bibfield  {journal} {\bibinfo  {journal} {Rev. Mod. Phys.}\ }\textbf
  {\bibinfo {volume} {89}},\ \bibinfo {pages} {035006} (\bibinfo {year}
  {2017})}\BibitemShut {NoStop}%
\bibitem [{\citenamefont {Egido}(2016)}]{Egido2016_PS91-073003}%
  \BibitemOpen
  \bibfield  {author} {\bibinfo {author} {\bibfnamefont {J.~L.}\ \bibnamefont
  {Egido}},\ }\href {https://doi.org/10.1088/0031-8949/91/7/073003} {\bibfield
  {journal} {\bibinfo  {journal} {Phys. Scr.}\ }\textbf {\bibinfo {volume}
  {91}},\ \bibinfo {pages} {073003} (\bibinfo {year} {2016})}\BibitemShut
  {NoStop}%
\bibitem [{\citenamefont {Nazarewicz}(2018)}]{Nazarewicz2018_NatPhys14-537}%
  \BibitemOpen
  \bibfield  {author} {\bibinfo {author} {\bibfnamefont {W.}~\bibnamefont
  {Nazarewicz}},\ }\href {https://doi.org/10.1038/s41567-018-0163-3} {\bibfield
   {journal} {\bibinfo  {journal} {Nat. Phys.}\ }\textbf {\bibinfo {volume}
  {14}},\ \bibinfo {pages} {537} (\bibinfo {year} {2018})}\BibitemShut
  {NoStop}%
\bibitem [{\citenamefont {Ring}\ and\ \citenamefont {Schuck}(1980)}]{Ring1980}%
  \BibitemOpen
  \bibfield  {author} {\bibinfo {author} {\bibfnamefont {P.}~\bibnamefont
  {Ring}}\ and\ \bibinfo {author} {\bibfnamefont {P.}~\bibnamefont {Schuck}},\
  }\href {https://www.springer.com/us/book/9783540212065} {\emph {\bibinfo
  {title} {The Nuclear Many-Body Problem}}}\ (\bibinfo  {publisher}
  {Springer-Verlag Berlin Heidelberg},\ \bibinfo {year} {1980})\BibitemShut
  {NoStop}%
\bibitem [{\citenamefont {Schunck}(2019)}]{Schunck2019_EDF-Nuclei}%
  \BibitemOpen
  \bibinfo {editor} {\bibfnamefont {N.}~\bibnamefont {Schunck}},\ ed.,\ \href
  {https://doi.org/10.1088/2053-2563/aae0ed} {\emph {\bibinfo {title} {Energy
  Density Functional Methods for Atomic Nuclei}}},\ {IOP} Expanding Physics\
  (\bibinfo  {publisher} {{IOP} Publishing},\ \bibinfo {year}
  {2019})\BibitemShut {NoStop}%
\bibitem [{\citenamefont {Robledo}\ \emph {et~al.}(2019)\citenamefont
  {Robledo}, \citenamefont {Rodr{\'i}guez},\ and\ \citenamefont
  {Rodr{\'i}guez-Guzm{\'a}n}}]{Robledo2019_JPG46-013001}%
  \BibitemOpen
  \bibfield  {author} {\bibinfo {author} {\bibfnamefont {L.~M.}\ \bibnamefont
  {Robledo}}, \bibinfo {author} {\bibfnamefont {T.~R.}\ \bibnamefont
  {Rodr{\'i}guez}},\ and\ \bibinfo {author} {\bibfnamefont {R.~R.}\
  \bibnamefont {Rodr{\'i}guez-Guzm{\'a}n}},\ }\href
  {https://doi.org/10.1088/1361-6471/aadebd} {\bibfield  {journal} {\bibinfo
  {journal} {J. Phys. G: Nucl. Part. Phys.}\ }\textbf {\bibinfo {volume}
  {46}},\ \bibinfo {pages} {013001} (\bibinfo {year} {2019})}\BibitemShut
  {NoStop}%
\bibitem [{\citenamefont {Sheikh}\ \emph {et~al.}(2019)\citenamefont {Sheikh},
  \citenamefont {Dobaczewski}, \citenamefont {Ring}, \citenamefont {Robledo},\
  and\ \citenamefont {Yannouleas}}]{Sheikh2019_arXiv1901.06992}%
  \BibitemOpen
  \bibfield  {author} {\bibinfo {author} {\bibfnamefont {J.~A.}\ \bibnamefont
  {Sheikh}}, \bibinfo {author} {\bibfnamefont {J.}~\bibnamefont {Dobaczewski}},
  \bibinfo {author} {\bibfnamefont {P.}~\bibnamefont {Ring}}, \bibinfo {author}
  {\bibfnamefont {L.~M.}\ \bibnamefont {Robledo}},\ and\ \bibinfo {author}
  {\bibfnamefont {C.}~\bibnamefont {Yannouleas}},\ }\href
  {https://arxiv.org/abs/1901.06992v1} {\bibinfo {title} {Symmetry restoration
  in mean-field approaches}},\ \bibinfo {howpublished} {arXiv:1901.06992
  [nucl-th]} (\bibinfo {year} {2019}),\ \Eprint
  {https://arxiv.org/abs/1901.06992v1} {1901.06992v1} \BibitemShut {NoStop}%
\bibitem [{\citenamefont {Rodr\'{i}guez-Guzm\'an}\ \emph
  {et~al.}(2000{\natexlab{a}})\citenamefont {Rodr\'{i}guez-Guzm\'an},
  \citenamefont {Egido},\ and\ \citenamefont
  {Robledo}}]{Rodriguez-Guzman2000_PRC62-054308}%
  \BibitemOpen
  \bibfield  {author} {\bibinfo {author} {\bibfnamefont {R.~R.}\ \bibnamefont
  {Rodr\'{i}guez-Guzm\'an}}, \bibinfo {author} {\bibfnamefont {J.~L.}\
  \bibnamefont {Egido}},\ and\ \bibinfo {author} {\bibfnamefont {L.~M.}\
  \bibnamefont {Robledo}},\ }\href {https://doi.org/10.1103/PhysRevC.62.054308}
  {\bibfield  {journal} {\bibinfo  {journal} {Phys. Rev. C}\ }\textbf {\bibinfo
  {volume} {62}},\ \bibinfo {pages} {054308} (\bibinfo {year}
  {2000}{\natexlab{a}})}\BibitemShut {NoStop}%
\bibitem [{\citenamefont {Bender}\ \emph
  {et~al.}(2003{\natexlab{b}})\citenamefont {Bender}, \citenamefont {Flocard},\
  and\ \citenamefont {Heenen}}]{Bender2003_PRC68-044321}%
  \BibitemOpen
  \bibfield  {author} {\bibinfo {author} {\bibfnamefont {M.}~\bibnamefont
  {Bender}}, \bibinfo {author} {\bibfnamefont {H.}~\bibnamefont {Flocard}},\
  and\ \bibinfo {author} {\bibfnamefont {P.~H.}\ \bibnamefont {Heenen}},\
  }\href {https://doi.org/10.1103/PhysRevC.68.044321} {\bibfield  {journal}
  {\bibinfo  {journal} {Phys. Rev. C}\ }\textbf {\bibinfo {volume} {68}},\
  \bibinfo {pages} {044321} (\bibinfo {year} {2003}{\natexlab{b}})}\BibitemShut
  {NoStop}%
\bibitem [{\citenamefont {Bender}\ \emph {et~al.}(2004)\citenamefont {Bender},
  \citenamefont {Bonche}, \citenamefont {Duguet},\ and\ \citenamefont
  {Heenen}}]{Bender2004_PRC69-064303}%
  \BibitemOpen
  \bibfield  {author} {\bibinfo {author} {\bibfnamefont {M.}~\bibnamefont
  {Bender}}, \bibinfo {author} {\bibfnamefont {P.}~\bibnamefont {Bonche}},
  \bibinfo {author} {\bibfnamefont {T.}~\bibnamefont {Duguet}},\ and\ \bibinfo
  {author} {\bibfnamefont {P.-H.}\ \bibnamefont {Heenen}},\ }\href
  {https://doi.org/10.1103/PhysRevC.69.064303} {\bibfield  {journal} {\bibinfo
  {journal} {Phys. Rev. C}\ }\textbf {\bibinfo {volume} {69}},\ \bibinfo
  {pages} {064303} (\bibinfo {year} {2004})}\BibitemShut {NoStop}%
\bibitem [{\citenamefont {Rodr\'{i}guez-Guzm\'an}\ \emph
  {et~al.}(2004)\citenamefont {Rodr\'{i}guez-Guzm\'an}, \citenamefont {Egido},\
  and\ \citenamefont {Robledo}}]{Rodriguez-Guzman2004_PRC69-054319}%
  \BibitemOpen
  \bibfield  {author} {\bibinfo {author} {\bibfnamefont {R.~R.}\ \bibnamefont
  {Rodr\'{i}guez-Guzm\'an}}, \bibinfo {author} {\bibfnamefont {J.~L.}\
  \bibnamefont {Egido}},\ and\ \bibinfo {author} {\bibfnamefont {L.~M.}\
  \bibnamefont {Robledo}},\ }\href {https://doi.org/10.1103/PhysRevC.69.054319}
  {\bibfield  {journal} {\bibinfo  {journal} {Phys. Rev. C}\ }\textbf {\bibinfo
  {volume} {69}},\ \bibinfo {pages} {054319} (\bibinfo {year}
  {2004})}\BibitemShut {NoStop}%
\bibitem [{\citenamefont {Bender}\ \emph {et~al.}(2006)\citenamefont {Bender},
  \citenamefont {Bonche},\ and\ \citenamefont
  {Heenen}}]{Bender2006_PRC74-024312}%
  \BibitemOpen
  \bibfield  {author} {\bibinfo {author} {\bibfnamefont {M.}~\bibnamefont
  {Bender}}, \bibinfo {author} {\bibfnamefont {P.}~\bibnamefont {Bonche}},\
  and\ \bibinfo {author} {\bibfnamefont {P.-H.}\ \bibnamefont {Heenen}},\
  }\href {https://doi.org/10.1103/PhysRevC.74.024312} {\bibfield  {journal}
  {\bibinfo  {journal} {Phys. Rev. C}\ }\textbf {\bibinfo {volume} {74}},\
  \bibinfo {pages} {024312} (\bibinfo {year} {2006})}\BibitemShut {NoStop}%
\bibitem [{\citenamefont {Rodr\'{i}guez}\ and\ \citenamefont
  {Egido}(2007)}]{Rodriguez2007_PRL99-062501}%
  \BibitemOpen
  \bibfield  {author} {\bibinfo {author} {\bibfnamefont {T.~R.}\ \bibnamefont
  {Rodr\'{i}guez}}\ and\ \bibinfo {author} {\bibfnamefont {J.~L.}\ \bibnamefont
  {Egido}},\ }\href {https://doi.org/10.1103/PhysRevLett.99.062501} {\bibfield
  {journal} {\bibinfo  {journal} {Phys. Rev. Lett.}\ }\textbf {\bibinfo
  {volume} {99}},\ \bibinfo {pages} {062501} (\bibinfo {year}
  {2007})}\BibitemShut {NoStop}%
\bibitem [{\citenamefont {Nik{\v{s}}i{\'{c}}}\ \emph
  {et~al.}(2007)\citenamefont {Nik{\v{s}}i{\'{c}}}, \citenamefont {Vretenar},
  \citenamefont {Lalazissis},\ and\ \citenamefont
  {Ring}}]{Niksic2007_PRL99-092502}%
  \BibitemOpen
  \bibfield  {author} {\bibinfo {author} {\bibfnamefont {T.}~\bibnamefont
  {Nik{\v{s}}i{\'{c}}}}, \bibinfo {author} {\bibfnamefont {D.}~\bibnamefont
  {Vretenar}}, \bibinfo {author} {\bibfnamefont {G.~A.}\ \bibnamefont
  {Lalazissis}},\ and\ \bibinfo {author} {\bibfnamefont {P.}~\bibnamefont
  {Ring}},\ }\href {https://doi.org/10.1103/PhysRevLett.99.092502} {\bibfield
  {journal} {\bibinfo  {journal} {Phys. Rev. Lett.}\ }\textbf {\bibinfo
  {volume} {99}},\ \bibinfo {pages} {092502} (\bibinfo {year}
  {2007})}\BibitemShut {NoStop}%
\bibitem [{\citenamefont {Rodr\'{i}guez}\ and\ \citenamefont
  {Egido}(2008)}]{Rodriguez2008_PLB663-49}%
  \BibitemOpen
  \bibfield  {author} {\bibinfo {author} {\bibfnamefont {T.~R.}\ \bibnamefont
  {Rodr\'{i}guez}}\ and\ \bibinfo {author} {\bibfnamefont {J.~L.}\ \bibnamefont
  {Egido}},\ }\href {https://doi.org/10.1016/j.physletb.2008.03.061} {\bibfield
   {journal} {\bibinfo  {journal} {Phys. Lett. B}\ }\textbf {\bibinfo {volume}
  {663}},\ \bibinfo {pages} {49} (\bibinfo {year} {2008})}\BibitemShut
  {NoStop}%
\bibitem [{\citenamefont {Cui}\ \emph {et~al.}(2015)\citenamefont {Cui},
  \citenamefont {Zhou},\ and\ \citenamefont {Schulze}}]{Cui2015_PRC91-054306}%
  \BibitemOpen
  \bibfield  {author} {\bibinfo {author} {\bibfnamefont {J.-W.}\ \bibnamefont
  {Cui}}, \bibinfo {author} {\bibfnamefont {X.-R.}\ \bibnamefont {Zhou}},\ and\
  \bibinfo {author} {\bibfnamefont {H.-J.}\ \bibnamefont {Schulze}},\ }\href
  {https://doi.org/10.1103/PhysRevC.91.054306} {\bibfield  {journal} {\bibinfo
  {journal} {Phys. Rev. C}\ }\textbf {\bibinfo {volume} {91}},\ \bibinfo
  {pages} {054306} (\bibinfo {year} {2015})}\BibitemShut {NoStop}%
\bibitem [{\citenamefont {Cui}\ \emph {et~al.}(2017)\citenamefont {Cui},
  \citenamefont {Zhou}, \citenamefont {Guo},\ and\ \citenamefont
  {Schulze}}]{Cui2017_PRC95-024323}%
  \BibitemOpen
  \bibfield  {author} {\bibinfo {author} {\bibfnamefont {J.-W.}\ \bibnamefont
  {Cui}}, \bibinfo {author} {\bibfnamefont {X.-R.}\ \bibnamefont {Zhou}},
  \bibinfo {author} {\bibfnamefont {L.-X.}\ \bibnamefont {Guo}},\ and\ \bibinfo
  {author} {\bibfnamefont {H.-J.}\ \bibnamefont {Schulze}},\ }\href
  {https://doi.org/10.1103/PhysRevC.95.024323} {\bibfield  {journal} {\bibinfo
  {journal} {Phys. Rev. C}\ }\textbf {\bibinfo {volume} {95}},\ \bibinfo
  {pages} {024323} (\bibinfo {year} {2017})}\BibitemShut {NoStop}%
\bibitem [{\citenamefont {Mei}\ \emph {et~al.}(2018)\citenamefont {Mei},
  \citenamefont {Hagino}, \citenamefont {Yao},\ and\ \citenamefont
  {Motoba}}]{Mei2018_PRC97-064318}%
  \BibitemOpen
  \bibfield  {author} {\bibinfo {author} {\bibfnamefont {H.}~\bibnamefont
  {Mei}}, \bibinfo {author} {\bibfnamefont {K.}~\bibnamefont {Hagino}},
  \bibinfo {author} {\bibfnamefont {J.~M.}\ \bibnamefont {Yao}},\ and\ \bibinfo
  {author} {\bibfnamefont {T.}~\bibnamefont {Motoba}},\ }\href
  {https://doi.org/10.1103/PhysRevC.97.064318} {\bibfield  {journal} {\bibinfo
  {journal} {Phys. Rev. C}\ }\textbf {\bibinfo {volume} {97}},\ \bibinfo
  {pages} {064318} (\bibinfo {year} {2018})}\BibitemShut {NoStop}%
\bibitem [{\citenamefont {Xia}\ \emph {et~al.}(2018)\citenamefont {Xia},
  \citenamefont {Wu}, \citenamefont {Mei},\ and\ \citenamefont
  {Yao}}]{Xia2019_SciChinaPMA62-042011}%
  \BibitemOpen
  \bibfield  {author} {\bibinfo {author} {\bibfnamefont {H.-J.}\ \bibnamefont
  {Xia}}, \bibinfo {author} {\bibfnamefont {X.-Y.}\ \bibnamefont {Wu}},
  \bibinfo {author} {\bibfnamefont {H.}~\bibnamefont {Mei}},\ and\ \bibinfo
  {author} {\bibfnamefont {J.-M.}\ \bibnamefont {Yao}},\ }\href
  {https://doi.org/10.1007/s11433-018-9308-0} {\bibfield  {journal} {\bibinfo
  {journal} {Sci. China G}\ }\textbf {\bibinfo {volume} {62}},\ \bibinfo
  {pages} {42011} (\bibinfo {year} {2018})}\BibitemShut {NoStop}%
\bibitem [{\citenamefont {Bender}\ and\ \citenamefont
  {Heenen}(2008)}]{Bender2008_PRC78-024309}%
  \BibitemOpen
  \bibfield  {author} {\bibinfo {author} {\bibfnamefont {M.}~\bibnamefont
  {Bender}}\ and\ \bibinfo {author} {\bibfnamefont {P.-H.}\ \bibnamefont
  {Heenen}},\ }\href {https://doi.org/10.1103/PhysRevC.78.024309} {\bibfield
  {journal} {\bibinfo  {journal} {Phys. Rev. C}\ }\textbf {\bibinfo {volume}
  {78}},\ \bibinfo {pages} {024309} (\bibinfo {year} {2008})}\BibitemShut
  {NoStop}%
\bibitem [{\citenamefont {Rodr\'{\i}guez}\ and\ \citenamefont
  {Egido}(2010)}]{Rodriguez2010_PRC81-064323}%
  \BibitemOpen
  \bibfield  {author} {\bibinfo {author} {\bibfnamefont {T.~R.}\ \bibnamefont
  {Rodr\'{\i}guez}}\ and\ \bibinfo {author} {\bibfnamefont {J.~L.}\
  \bibnamefont {Egido}},\ }\href {https://doi.org/10.1103/PhysRevC.81.064323}
  {\bibfield  {journal} {\bibinfo  {journal} {Phys. Rev. C}\ }\textbf {\bibinfo
  {volume} {81}},\ \bibinfo {pages} {064323} (\bibinfo {year}
  {2010})}\BibitemShut {NoStop}%
\bibitem [{\citenamefont {Yao}\ \emph {et~al.}(2009)\citenamefont {Yao},
  \citenamefont {Meng}, \citenamefont {Ring},\ and\ \citenamefont
  {Arteaga}}]{Yao2009_PRC79-044312}%
  \BibitemOpen
  \bibfield  {author} {\bibinfo {author} {\bibfnamefont {J.~M.}\ \bibnamefont
  {Yao}}, \bibinfo {author} {\bibfnamefont {J.}~\bibnamefont {Meng}}, \bibinfo
  {author} {\bibfnamefont {P.}~\bibnamefont {Ring}},\ and\ \bibinfo {author}
  {\bibfnamefont {D.~P.}\ \bibnamefont {Arteaga}},\ }\href
  {https://doi.org/10.1103/PhysRevC.79.044312} {\bibfield  {journal} {\bibinfo
  {journal} {Phys. Rev. C}\ }\textbf {\bibinfo {volume} {79}},\ \bibinfo
  {pages} {044312} (\bibinfo {year} {2009})}\BibitemShut {NoStop}%
\bibitem [{\citenamefont {Yao}\ \emph {et~al.}(2010)\citenamefont {Yao},
  \citenamefont {Meng}, \citenamefont {Ring},\ and\ \citenamefont
  {Vretenar}}]{Yao2010_PRC81-044311}%
  \BibitemOpen
  \bibfield  {author} {\bibinfo {author} {\bibfnamefont {J.~M.}\ \bibnamefont
  {Yao}}, \bibinfo {author} {\bibfnamefont {J.}~\bibnamefont {Meng}}, \bibinfo
  {author} {\bibfnamefont {P.}~\bibnamefont {Ring}},\ and\ \bibinfo {author}
  {\bibfnamefont {D.}~\bibnamefont {Vretenar}},\ }\href
  {https://doi.org/10.1103/PhysRevC.81.044311} {\bibfield  {journal} {\bibinfo
  {journal} {Phys. Rev. C}\ }\textbf {\bibinfo {volume} {81}},\ \bibinfo
  {pages} {044311} (\bibinfo {year} {2010})}\BibitemShut {NoStop}%
\bibitem [{\citenamefont {Yao}\ \emph {et~al.}(2014)\citenamefont {Yao},
  \citenamefont {Hagino}, \citenamefont {Li}, \citenamefont {Meng},\ and\
  \citenamefont {Ring}}]{Yao2014_PRC89-054306}%
  \BibitemOpen
  \bibfield  {author} {\bibinfo {author} {\bibfnamefont {J.~M.}\ \bibnamefont
  {Yao}}, \bibinfo {author} {\bibfnamefont {K.}~\bibnamefont {Hagino}},
  \bibinfo {author} {\bibfnamefont {Z.~P.}\ \bibnamefont {Li}}, \bibinfo
  {author} {\bibfnamefont {J.}~\bibnamefont {Meng}},\ and\ \bibinfo {author}
  {\bibfnamefont {P.}~\bibnamefont {Ring}},\ }\href
  {https://doi.org/10.1103/PhysRevC.89.054306} {\bibfield  {journal} {\bibinfo
  {journal} {Phys. Rev. C}\ }\textbf {\bibinfo {volume} {89}},\ \bibinfo
  {pages} {054306} (\bibinfo {year} {2014})}\BibitemShut {NoStop}%
\bibitem [{\citenamefont {Egido}\ \emph {et~al.}(2016)\citenamefont {Egido},
  \citenamefont {Borrajo},\ and\ \citenamefont
  {Rodr\'{\i}guez}}]{Egido2016_PRL116-052502}%
  \BibitemOpen
  \bibfield  {author} {\bibinfo {author} {\bibfnamefont {J.~L.}\ \bibnamefont
  {Egido}}, \bibinfo {author} {\bibfnamefont {M.}~\bibnamefont {Borrajo}},\
  and\ \bibinfo {author} {\bibfnamefont {T.~R.}\ \bibnamefont
  {Rodr\'{\i}guez}},\ }\href {https://doi.org/10.1103/PhysRevLett.116.052502}
  {\bibfield  {journal} {\bibinfo  {journal} {Phys. Rev. Lett.}\ }\textbf
  {\bibinfo {volume} {116}},\ \bibinfo {pages} {052502} (\bibinfo {year}
  {2016})}\BibitemShut {NoStop}%
\bibitem [{\citenamefont {Chen}\ and\ \citenamefont
  {Egido}(2017)}]{Chen2017_PRC95-024307}%
  \BibitemOpen
  \bibfield  {author} {\bibinfo {author} {\bibfnamefont {F.-Q.}\ \bibnamefont
  {Chen}}\ and\ \bibinfo {author} {\bibfnamefont {J.~L.}\ \bibnamefont
  {Egido}},\ }\href {https://doi.org/10.1103/PhysRevC.95.024307} {\bibfield
  {journal} {\bibinfo  {journal} {Phys. Rev. C}\ }\textbf {\bibinfo {volume}
  {95}},\ \bibinfo {pages} {024307} (\bibinfo {year} {2017})}\BibitemShut
  {NoStop}%
\bibitem [{\citenamefont {Marevi\'{c}}\ and\ \citenamefont
  {Schunck}(2020)}]{Marevic2020_PRL125-102504}%
  \BibitemOpen
  \bibfield  {author} {\bibinfo {author} {\bibfnamefont {P.}~\bibnamefont
  {Marevi\'{c}}}\ and\ \bibinfo {author} {\bibfnamefont {N.}~\bibnamefont
  {Schunck}},\ }\href {https://doi.org/10.1103/PhysRevLett.125.102504}
  {\bibfield  {journal} {\bibinfo  {journal} {Phys. Rev. Lett.}\ }\textbf
  {\bibinfo {volume} {125}},\ \bibinfo {pages} {102504} (\bibinfo {year}
  {2020})}\BibitemShut {NoStop}%
\bibitem [{\citenamefont {Egido}\ and\ \citenamefont
  {Jungclaus}(2020)}]{Egido2020_PRL125-192504}%
  \BibitemOpen
  \bibfield  {author} {\bibinfo {author} {\bibfnamefont {J.~L.}\ \bibnamefont
  {Egido}}\ and\ \bibinfo {author} {\bibfnamefont {A.}~\bibnamefont
  {Jungclaus}},\ }\href {https://doi.org/10.1103/PhysRevLett.125.192504}
  {\bibfield  {journal} {\bibinfo  {journal} {Phys. Rev. Lett.}\ }\textbf
  {\bibinfo {volume} {125}},\ \bibinfo {pages} {192504} (\bibinfo {year}
  {2020})}\BibitemShut {NoStop}%
\bibitem [{\citenamefont {Egido}\ and\ \citenamefont
  {Jungclaus}(2021)}]{Egido2021_PRL126-192501}%
  \BibitemOpen
  \bibfield  {author} {\bibinfo {author} {\bibfnamefont {J.~L.}\ \bibnamefont
  {Egido}}\ and\ \bibinfo {author} {\bibfnamefont {A.}~\bibnamefont
  {Jungclaus}},\ }\href {https://doi.org/10.1103/PhysRevLett.126.192501}
  {\bibfield  {journal} {\bibinfo  {journal} {Phys. Rev. Lett.}\ }\textbf
  {\bibinfo {volume} {126}},\ \bibinfo {pages} {192501} (\bibinfo {year}
  {2021})}\BibitemShut {NoStop}%
\bibitem [{\citenamefont {Bally}\ \emph {et~al.}(2014)\citenamefont {Bally},
  \citenamefont {Avez}, \citenamefont {Bender},\ and\ \citenamefont
  {Heenen}}]{Bally2014_PRL113-162501}%
  \BibitemOpen
  \bibfield  {author} {\bibinfo {author} {\bibfnamefont {B.}~\bibnamefont
  {Bally}}, \bibinfo {author} {\bibfnamefont {B.}~\bibnamefont {Avez}},
  \bibinfo {author} {\bibfnamefont {M.}~\bibnamefont {Bender}},\ and\ \bibinfo
  {author} {\bibfnamefont {P.-H.}\ \bibnamefont {Heenen}},\ }\href
  {https://doi.org/10.1103/PhysRevLett.113.162501} {\bibfield  {journal}
  {\bibinfo  {journal} {Phys. Rev. Lett.}\ }\textbf {\bibinfo {volume} {113}},\
  \bibinfo {pages} {162501} (\bibinfo {year} {2014})}\BibitemShut {NoStop}%
\bibitem [{\citenamefont {Borrajo}\ and\ \citenamefont
  {Egido}(2016)}]{Borrajo2016_EPJA52-277}%
  \BibitemOpen
  \bibfield  {author} {\bibinfo {author} {\bibfnamefont {M.}~\bibnamefont
  {Borrajo}}\ and\ \bibinfo {author} {\bibfnamefont {J.~L.}\ \bibnamefont
  {Egido}},\ }\href {https://doi.org/10.1140/epja/i2016-16277-8} {\bibfield
  {journal} {\bibinfo  {journal} {Euro. Phys. J. A}\ }\textbf {\bibinfo
  {volume} {52}},\ \bibinfo {pages} {277} (\bibinfo {year} {2016})}\BibitemShut
  {NoStop}%
\bibitem [{\citenamefont {Borrajo}\ and\ \citenamefont
  {Egido}(2017)}]{Borrajo2017_PLB764-328}%
  \BibitemOpen
  \bibfield  {author} {\bibinfo {author} {\bibfnamefont {M.}~\bibnamefont
  {Borrajo}}\ and\ \bibinfo {author} {\bibfnamefont {J.~L.}\ \bibnamefont
  {Egido}},\ }\href {https://doi.org/10.1016/j.physletb.2016.11.037} {\bibfield
   {journal} {\bibinfo  {journal} {Phys. Lett. B}\ }\textbf {\bibinfo {volume}
  {764}},\ \bibinfo {pages} {328 } (\bibinfo {year} {2017})}\BibitemShut
  {NoStop}%
\bibitem [{\citenamefont {Borrajo}\ and\ \citenamefont
  {Egido}(2018)}]{Borrajo2018_PRC98-044317}%
  \BibitemOpen
  \bibfield  {author} {\bibinfo {author} {\bibfnamefont {M.}~\bibnamefont
  {Borrajo}}\ and\ \bibinfo {author} {\bibfnamefont {J.~L.}\ \bibnamefont
  {Egido}},\ }\href {https://doi.org/10.1103/PhysRevC.98.044317} {\bibfield
  {journal} {\bibinfo  {journal} {Phys. Rev. C}\ }\textbf {\bibinfo {volume}
  {98}},\ \bibinfo {pages} {044317} (\bibinfo {year} {2018})}\BibitemShut
  {NoStop}%
\bibitem [{\citenamefont {Pannert}\ \emph {et~al.}(1987)\citenamefont
  {Pannert}, \citenamefont {Ring},\ and\ \citenamefont
  {Boguta}}]{Pannert1987_PRL59-2420}%
  \BibitemOpen
  \bibfield  {author} {\bibinfo {author} {\bibfnamefont {W.}~\bibnamefont
  {Pannert}}, \bibinfo {author} {\bibfnamefont {P.}~\bibnamefont {Ring}},\ and\
  \bibinfo {author} {\bibfnamefont {J.}~\bibnamefont {Boguta}},\ }\href
  {https://doi.org/10.1103/PhysRevLett.59.2420} {\bibfield  {journal} {\bibinfo
   {journal} {Phys. Rev. Lett.}\ }\textbf {\bibinfo {volume} {59}},\ \bibinfo
  {pages} {2420} (\bibinfo {year} {1987})}\BibitemShut {NoStop}%
\bibitem [{\citenamefont {Price}\ and\ \citenamefont
  {Walker}(1987)}]{Price1987_PRC36-354}%
  \BibitemOpen
  \bibfield  {author} {\bibinfo {author} {\bibfnamefont {C.~E.}\ \bibnamefont
  {Price}}\ and\ \bibinfo {author} {\bibfnamefont {G.~E.}\ \bibnamefont
  {Walker}},\ }\href {https://doi.org/10.1103/PhysRevC.36.354} {\bibfield
  {journal} {\bibinfo  {journal} {Phys. Rev. C}\ }\textbf {\bibinfo {volume}
  {36}},\ \bibinfo {pages} {354} (\bibinfo {year} {1987})}\BibitemShut
  {NoStop}%
\bibitem [{\citenamefont {Gambhir}\ \emph {et~al.}(1990)\citenamefont
  {Gambhir}, \citenamefont {Ring},\ and\ \citenamefont
  {Thimet}}]{Gambhir1990_AP198-132}%
  \BibitemOpen
  \bibfield  {author} {\bibinfo {author} {\bibfnamefont {Y.}~\bibnamefont
  {Gambhir}}, \bibinfo {author} {\bibfnamefont {P.}~\bibnamefont {Ring}},\ and\
  \bibinfo {author} {\bibfnamefont {A.}~\bibnamefont {Thimet}},\ }\href
  {https://doi.org/10.1016/0003-4916(90)90330-Q} {\bibfield  {journal}
  {\bibinfo  {journal} {Ann. Phys.}\ }\textbf {\bibinfo {volume} {198}},\
  \bibinfo {pages} {132} (\bibinfo {year} {1990})}\BibitemShut {NoStop}%
\bibitem [{\citenamefont {Stoitsov}\ \emph
  {et~al.}(1998{\natexlab{a}})\citenamefont {Stoitsov}, \citenamefont {Ring},
  \citenamefont {Vretenar},\ and\ \citenamefont
  {Lalazissis}}]{Stoitsov1998_PRC58-2086}%
  \BibitemOpen
  \bibfield  {author} {\bibinfo {author} {\bibfnamefont {M.}~\bibnamefont
  {Stoitsov}}, \bibinfo {author} {\bibfnamefont {P.}~\bibnamefont {Ring}},
  \bibinfo {author} {\bibfnamefont {D.}~\bibnamefont {Vretenar}},\ and\
  \bibinfo {author} {\bibfnamefont {G.~A.}\ \bibnamefont {Lalazissis}},\ }\href
  {https://doi.org/10.1103/PhysRevC.58.2086} {\bibfield  {journal} {\bibinfo
  {journal} {Phys. Rev. C}\ }\textbf {\bibinfo {volume} {58}},\ \bibinfo
  {pages} {2086} (\bibinfo {year} {1998}{\natexlab{a}})}\BibitemShut {NoStop}%
\bibitem [{\citenamefont {Zhou}\ \emph {et~al.}(2000)\citenamefont {Zhou},
  \citenamefont {Meng}, \citenamefont {Yamaji},\ and\ \citenamefont
  {Yang}}]{Zhou2000_CPL17-717}%
  \BibitemOpen
  \bibfield  {author} {\bibinfo {author} {\bibfnamefont {S.-G.}\ \bibnamefont
  {Zhou}}, \bibinfo {author} {\bibfnamefont {J.}~\bibnamefont {Meng}}, \bibinfo
  {author} {\bibfnamefont {S.}~\bibnamefont {Yamaji}},\ and\ \bibinfo {author}
  {\bibfnamefont {S.-C.}\ \bibnamefont {Yang}},\ }\href
  {https://doi.org/10.1088/0256-307X/17/10/006} {\bibfield  {journal} {\bibinfo
   {journal} {Chin. Phys. Lett.}\ }\textbf {\bibinfo {volume} {17}},\ \bibinfo
  {pages} {717} (\bibinfo {year} {2000})}\BibitemShut {NoStop}%
\bibitem [{\citenamefont {Zhou}\ \emph {et~al.}(2003)\citenamefont {Zhou},
  \citenamefont {Meng},\ and\ \citenamefont {Ring}}]{Zhou2003_PRC68-034323}%
  \BibitemOpen
  \bibfield  {author} {\bibinfo {author} {\bibfnamefont {S.-G.}\ \bibnamefont
  {Zhou}}, \bibinfo {author} {\bibfnamefont {J.}~\bibnamefont {Meng}},\ and\
  \bibinfo {author} {\bibfnamefont {P.}~\bibnamefont {Ring}},\ }\href
  {https://doi.org/10.1103/PhysRevC.68.034323} {\bibfield  {journal} {\bibinfo
  {journal} {Phys. Rev. C}\ }\textbf {\bibinfo {volume} {68}},\ \bibinfo
  {pages} {034323} (\bibinfo {year} {2003})}\BibitemShut {NoStop}%
\bibitem [{\citenamefont {Zhang}\ \emph {et~al.}(2013)\citenamefont {Zhang},
  \citenamefont {Pei},\ and\ \citenamefont {Xu}}]{Zhang2013_PRC88-054305}%
  \BibitemOpen
  \bibfield  {author} {\bibinfo {author} {\bibfnamefont {Y.~N.}\ \bibnamefont
  {Zhang}}, \bibinfo {author} {\bibfnamefont {J.~C.}\ \bibnamefont {Pei}},\
  and\ \bibinfo {author} {\bibfnamefont {F.~R.}\ \bibnamefont {Xu}},\ }\href
  {https://doi.org/10.1103/PhysRevC.88.054305} {\bibfield  {journal} {\bibinfo
  {journal} {Phys. Rev. C}\ }\textbf {\bibinfo {volume} {88}},\ \bibinfo
  {pages} {054305} (\bibinfo {year} {2013})}\BibitemShut {NoStop}%
\bibitem [{\citenamefont {Hansen}\ and\ \citenamefont
  {Jonson}(1987)}]{Hansen1987_EPL4-409}%
  \BibitemOpen
  \bibfield  {author} {\bibinfo {author} {\bibfnamefont {P.~G.}\ \bibnamefont
  {Hansen}}\ and\ \bibinfo {author} {\bibfnamefont {B.}~\bibnamefont
  {Jonson}},\ }\href {https://doi.org/10.1209/0295-5075/4/4/005} {\bibfield
  {journal} {\bibinfo  {journal} {Europhys. Lett.}\ }\textbf {\bibinfo {volume}
  {4}},\ \bibinfo {pages} {409} (\bibinfo {year} {1987})}\BibitemShut {NoStop}%
\bibitem [{\citenamefont {Dobaczewski}\ \emph {et~al.}(1996)\citenamefont
  {Dobaczewski}, \citenamefont {Nazarewicz}, \citenamefont {Werner},
  \citenamefont {Berger}, \citenamefont {Chinn},\ and\ \citenamefont
  {Decharg\'e}}]{Dobaczewski1996_PRC53-2809}%
  \BibitemOpen
  \bibfield  {author} {\bibinfo {author} {\bibfnamefont {J.}~\bibnamefont
  {Dobaczewski}}, \bibinfo {author} {\bibfnamefont {W.}~\bibnamefont
  {Nazarewicz}}, \bibinfo {author} {\bibfnamefont {T.~R.}\ \bibnamefont
  {Werner}}, \bibinfo {author} {\bibfnamefont {J.~F.}\ \bibnamefont {Berger}},
  \bibinfo {author} {\bibfnamefont {C.~R.}\ \bibnamefont {Chinn}},\ and\
  \bibinfo {author} {\bibfnamefont {J.}~\bibnamefont {Decharg\'e}},\ }\href
  {https://doi.org/10.1103/PhysRevC.53.2809} {\bibfield  {journal} {\bibinfo
  {journal} {Phys. Rev. C}\ }\textbf {\bibinfo {volume} {53}},\ \bibinfo
  {pages} {2809} (\bibinfo {year} {1996})}\BibitemShut {NoStop}%
\bibitem [{\citenamefont {Meng}\ and\ \citenamefont
  {Ring}(1996)}]{Meng1996_PRL77-3963}%
  \BibitemOpen
  \bibfield  {author} {\bibinfo {author} {\bibfnamefont {J.}~\bibnamefont
  {Meng}}\ and\ \bibinfo {author} {\bibfnamefont {P.}~\bibnamefont {Ring}},\
  }\href {https://doi.org/10.1103/PhysRevLett.77.3963} {\bibfield  {journal}
  {\bibinfo  {journal} {Phys. Rev. Lett.}\ }\textbf {\bibinfo {volume} {77}},\
  \bibinfo {pages} {3963} (\bibinfo {year} {1996})}\BibitemShut {NoStop}%
\bibitem [{\citenamefont {Meng}\ and\ \citenamefont
  {Ring}(1998)}]{Meng1998_PRL80-460}%
  \BibitemOpen
  \bibfield  {author} {\bibinfo {author} {\bibfnamefont {J.}~\bibnamefont
  {Meng}}\ and\ \bibinfo {author} {\bibfnamefont {P.}~\bibnamefont {Ring}},\
  }\href {https://doi.org/10.1103/PhysRevLett.80.460} {\bibfield  {journal}
  {\bibinfo  {journal} {Phys. Rev. Lett.}\ }\textbf {\bibinfo {volume} {80}},\
  \bibinfo {pages} {460} (\bibinfo {year} {1998})}\BibitemShut {NoStop}%
\bibitem [{\citenamefont {Meng}(1998)}]{Meng1998_NPA635-3}%
  \BibitemOpen
  \bibfield  {author} {\bibinfo {author} {\bibfnamefont {J.}~\bibnamefont
  {Meng}},\ }\href {https://doi.org/10.1016/S0375-9474(98)00178-X} {\bibfield
  {journal} {\bibinfo  {journal} {Nucl. Phys. A}\ }\textbf {\bibinfo {volume}
  {635}},\ \bibinfo {pages} {3} (\bibinfo {year} {1998})}\BibitemShut {NoStop}%
\bibitem [{\citenamefont {Jensen}\ \emph {et~al.}(2004)\citenamefont {Jensen},
  \citenamefont {Riisager}, \citenamefont {Fedorov},\ and\ \citenamefont
  {Garrido}}]{Jensen2004_RMP76-215}%
  \BibitemOpen
  \bibfield  {author} {\bibinfo {author} {\bibfnamefont {A.~S.}\ \bibnamefont
  {Jensen}}, \bibinfo {author} {\bibfnamefont {K.}~\bibnamefont {Riisager}},
  \bibinfo {author} {\bibfnamefont {D.~V.}\ \bibnamefont {Fedorov}},\ and\
  \bibinfo {author} {\bibfnamefont {E.}~\bibnamefont {Garrido}},\ }\href
  {https://doi.org/10.1103/RevModPhys.76.215} {\bibfield  {journal} {\bibinfo
  {journal} {Rev. Mod. Phys.}\ }\textbf {\bibinfo {volume} {76}},\ \bibinfo
  {pages} {215} (\bibinfo {year} {2004})}\BibitemShut {NoStop}%
\bibitem [{\citenamefont {Riisager}(2013)}]{Riisager2013_PS2013-014001}%
  \BibitemOpen
  \bibfield  {author} {\bibinfo {author} {\bibfnamefont {K.}~\bibnamefont
  {Riisager}},\ }\href {https://doi.org/10.1088/0031-8949/2013/T152/014001}
  {\bibfield  {journal} {\bibinfo  {journal} {Phys. Scr.}\ }\textbf {\bibinfo
  {volume} {2013}},\ \bibinfo {pages} {014001} (\bibinfo {year}
  {2013})}\BibitemShut {NoStop}%
\bibitem [{\citenamefont {Dobaczewski}\ \emph {et~al.}(1984)\citenamefont
  {Dobaczewski}, \citenamefont {Flocard},\ and\ \citenamefont
  {Treiner}}]{Dobaczewski1984_NPA422-103}%
  \BibitemOpen
  \bibfield  {author} {\bibinfo {author} {\bibfnamefont {J.}~\bibnamefont
  {Dobaczewski}}, \bibinfo {author} {\bibfnamefont {H.}~\bibnamefont
  {Flocard}},\ and\ \bibinfo {author} {\bibfnamefont {J.}~\bibnamefont
  {Treiner}},\ }\href {https://doi.org/10.1016/0375-9474(84)90433-0} {\bibfield
   {journal} {\bibinfo  {journal} {Nucl. Phys. A}\ }\textbf {\bibinfo {volume}
  {422}},\ \bibinfo {pages} {103} (\bibinfo {year} {1984})}\BibitemShut
  {NoStop}%
\bibitem [{\citenamefont {P\"oschl}\ \emph
  {et~al.}(1997{\natexlab{a}})\citenamefont {P\"oschl}, \citenamefont
  {Vretenar},\ and\ \citenamefont {Ring}}]{Poeschl1997_CPC103-217250}%
  \BibitemOpen
  \bibfield  {author} {\bibinfo {author} {\bibfnamefont {W.}~\bibnamefont
  {P\"oschl}}, \bibinfo {author} {\bibfnamefont {D.}~\bibnamefont {Vretenar}},\
  and\ \bibinfo {author} {\bibfnamefont {P.}~\bibnamefont {Ring}},\ }\href
  {https://doi.org/10.1016/S0010-4655(97)00042-8} {\bibfield  {journal}
  {\bibinfo  {journal} {Comput. Phys. Commun.}\ }\textbf {\bibinfo {volume}
  {103}},\ \bibinfo {pages} {217} (\bibinfo {year}
  {1997}{\natexlab{a}})}\BibitemShut {NoStop}%
\bibitem [{\citenamefont {Typel}(2018)}]{Typel2018_FPhys6-73}%
  \BibitemOpen
  \bibfield  {author} {\bibinfo {author} {\bibfnamefont {S.}~\bibnamefont
  {Typel}},\ }\href {https://doi.org/10.3389/fphy.2018.00073} {\bibfield
  {journal} {\bibinfo  {journal} {Front. Phys.}\ }\textbf {\bibinfo {volume}
  {6}},\ \bibinfo {pages} {73} (\bibinfo {year} {2018})}\BibitemShut {NoStop}%
\bibitem [{\citenamefont {Stoitsov}\ \emph
  {et~al.}(1998{\natexlab{b}})\citenamefont {Stoitsov}, \citenamefont
  {Nazarewicz},\ and\ \citenamefont {Pittel}}]{Stoitsov1998_PRC58-2092}%
  \BibitemOpen
  \bibfield  {author} {\bibinfo {author} {\bibfnamefont {M.~V.}\ \bibnamefont
  {Stoitsov}}, \bibinfo {author} {\bibfnamefont {W.}~\bibnamefont
  {Nazarewicz}},\ and\ \bibinfo {author} {\bibfnamefont {S.}~\bibnamefont
  {Pittel}},\ }\href {https://doi.org/10.1103/PhysRevC.58.2092} {\bibfield
  {journal} {\bibinfo  {journal} {Phys. Rev. C}\ }\textbf {\bibinfo {volume}
  {58}},\ \bibinfo {pages} {2092} (\bibinfo {year}
  {1998}{\natexlab{b}})}\BibitemShut {NoStop}%
\bibitem [{\citenamefont {P\"oschl}\ \emph
  {et~al.}(1997{\natexlab{b}})\citenamefont {P\"oschl}, \citenamefont
  {Vretenar}, \citenamefont {Lalazissis},\ and\ \citenamefont
  {Ring}}]{Poeschl1997_PRL79-3841}%
  \BibitemOpen
  \bibfield  {author} {\bibinfo {author} {\bibfnamefont {W.}~\bibnamefont
  {P\"oschl}}, \bibinfo {author} {\bibfnamefont {D.}~\bibnamefont {Vretenar}},
  \bibinfo {author} {\bibfnamefont {G.~A.}\ \bibnamefont {Lalazissis}},\ and\
  \bibinfo {author} {\bibfnamefont {P.}~\bibnamefont {Ring}},\ }\href
  {https://doi.org/10.1103/PhysRevLett.79.3841} {\bibfield  {journal} {\bibinfo
   {journal} {Phys. Rev. Lett.}\ }\textbf {\bibinfo {volume} {79}},\ \bibinfo
  {pages} {3841} (\bibinfo {year} {1997}{\natexlab{b}})}\BibitemShut {NoStop}%
\bibitem [{\citenamefont {Meng}\ \emph {et~al.}(1998)\citenamefont {Meng},
  \citenamefont {Tanihata},\ and\ \citenamefont {Yamaji}}]{Meng1998_PLB419-1}%
  \BibitemOpen
  \bibfield  {author} {\bibinfo {author} {\bibfnamefont {J.}~\bibnamefont
  {Meng}}, \bibinfo {author} {\bibfnamefont {I.}~\bibnamefont {Tanihata}},\
  and\ \bibinfo {author} {\bibfnamefont {S.}~\bibnamefont {Yamaji}},\ }\href
  {https://doi.org/10.1016/s0370-2693(97)01386-5} {\bibfield  {journal}
  {\bibinfo  {journal} {Phys. Lett. B}\ }\textbf {\bibinfo {volume} {419}},\
  \bibinfo {pages} {1} (\bibinfo {year} {1998})}\BibitemShut {NoStop}%
\bibitem [{\citenamefont {Long}\ \emph {et~al.}(2010)\citenamefont {Long},
  \citenamefont {Ring}, \citenamefont {Meng}, \citenamefont {Van~Giai},\ and\
  \citenamefont {Bertulani}}]{Long2010_PRC81-031302R}%
  \BibitemOpen
  \bibfield  {author} {\bibinfo {author} {\bibfnamefont {W.~H.}\ \bibnamefont
  {Long}}, \bibinfo {author} {\bibfnamefont {P.}~\bibnamefont {Ring}}, \bibinfo
  {author} {\bibfnamefont {J.}~\bibnamefont {Meng}}, \bibinfo {author}
  {\bibfnamefont {N.}~\bibnamefont {Van~Giai}},\ and\ \bibinfo {author}
  {\bibfnamefont {C.~A.}\ \bibnamefont {Bertulani}},\ }\href
  {https://doi.org/10.1103/PhysRevC.81.031302} {\bibfield  {journal} {\bibinfo
  {journal} {Phys. Rev. C}\ }\textbf {\bibinfo {volume} {81}},\ \bibinfo
  {pages} {031302(R)} (\bibinfo {year} {2010})}\BibitemShut {NoStop}%
\bibitem [{\citenamefont {Nakamura}\ \emph {et~al.}(2009)\citenamefont
  {Nakamura}, \citenamefont {Kobayashi}, \citenamefont {Kondo}, \citenamefont
  {Satou}, \citenamefont {Aoi}, \citenamefont {Baba}, \citenamefont {Deguchi},
  \citenamefont {Fukuda}, \citenamefont {Gibelin}, \citenamefont {Inabe},
  \citenamefont {Ishihara}, \citenamefont {Kameda}, \citenamefont {Kawada},
  \citenamefont {Kubo}, \citenamefont {Kusaka}, \citenamefont {Mengoni},
  \citenamefont {Motobayashi}, \citenamefont {Ohnishi}, \citenamefont {Ohtake},
  \citenamefont {Orr}, \citenamefont {Otsu}, \citenamefont {Otsuka},
  \citenamefont {Saito}, \citenamefont {Sakurai}, \citenamefont {Shimoura},
  \citenamefont {Sumikama}, \citenamefont {Takeda}, \citenamefont {Takeshita},
  \citenamefont {Takechi}, \citenamefont {Takeuchi}, \citenamefont {Tanaka},
  \citenamefont {Tanaka}, \citenamefont {Tanaka}, \citenamefont {Togano},
  \citenamefont {Utsuno}, \citenamefont {Yoneda}, \citenamefont {Yoshida},\
  and\ \citenamefont {Yoshida}}]{Nakamura2009_PRL103-262501}%
  \BibitemOpen
  \bibfield  {author} {\bibinfo {author} {\bibfnamefont {T.}~\bibnamefont
  {Nakamura}}, \bibinfo {author} {\bibfnamefont {N.}~\bibnamefont {Kobayashi}},
  \bibinfo {author} {\bibfnamefont {Y.}~\bibnamefont {Kondo}}, \bibinfo
  {author} {\bibfnamefont {Y.}~\bibnamefont {Satou}}, \bibinfo {author}
  {\bibfnamefont {N.}~\bibnamefont {Aoi}}, \bibinfo {author} {\bibfnamefont
  {H.}~\bibnamefont {Baba}}, \bibinfo {author} {\bibfnamefont {S.}~\bibnamefont
  {Deguchi}}, \bibinfo {author} {\bibfnamefont {N.}~\bibnamefont {Fukuda}},
  \bibinfo {author} {\bibfnamefont {J.}~\bibnamefont {Gibelin}}, \bibinfo
  {author} {\bibfnamefont {N.}~\bibnamefont {Inabe}}, \bibinfo {author}
  {\bibfnamefont {M.}~\bibnamefont {Ishihara}}, \bibinfo {author}
  {\bibfnamefont {D.}~\bibnamefont {Kameda}}, \bibinfo {author} {\bibfnamefont
  {Y.}~\bibnamefont {Kawada}}, \bibinfo {author} {\bibfnamefont
  {T.}~\bibnamefont {Kubo}}, \bibinfo {author} {\bibfnamefont {K.}~\bibnamefont
  {Kusaka}}, \bibinfo {author} {\bibfnamefont {A.}~\bibnamefont {Mengoni}},
  \bibinfo {author} {\bibfnamefont {T.}~\bibnamefont {Motobayashi}}, \bibinfo
  {author} {\bibfnamefont {T.}~\bibnamefont {Ohnishi}}, \bibinfo {author}
  {\bibfnamefont {M.}~\bibnamefont {Ohtake}}, \bibinfo {author} {\bibfnamefont
  {N.~A.}\ \bibnamefont {Orr}}, \bibinfo {author} {\bibfnamefont
  {H.}~\bibnamefont {Otsu}}, \bibinfo {author} {\bibfnamefont {T.}~\bibnamefont
  {Otsuka}}, \bibinfo {author} {\bibfnamefont {A.}~\bibnamefont {Saito}},
  \bibinfo {author} {\bibfnamefont {H.}~\bibnamefont {Sakurai}}, \bibinfo
  {author} {\bibfnamefont {S.}~\bibnamefont {Shimoura}}, \bibinfo {author}
  {\bibfnamefont {T.}~\bibnamefont {Sumikama}}, \bibinfo {author}
  {\bibfnamefont {H.}~\bibnamefont {Takeda}}, \bibinfo {author} {\bibfnamefont
  {E.}~\bibnamefont {Takeshita}}, \bibinfo {author} {\bibfnamefont
  {M.}~\bibnamefont {Takechi}}, \bibinfo {author} {\bibfnamefont
  {S.}~\bibnamefont {Takeuchi}}, \bibinfo {author} {\bibfnamefont
  {K.}~\bibnamefont {Tanaka}}, \bibinfo {author} {\bibfnamefont {K.~N.}\
  \bibnamefont {Tanaka}}, \bibinfo {author} {\bibfnamefont {N.}~\bibnamefont
  {Tanaka}}, \bibinfo {author} {\bibfnamefont {Y.}~\bibnamefont {Togano}},
  \bibinfo {author} {\bibfnamefont {Y.}~\bibnamefont {Utsuno}}, \bibinfo
  {author} {\bibfnamefont {K.}~\bibnamefont {Yoneda}}, \bibinfo {author}
  {\bibfnamefont {A.}~\bibnamefont {Yoshida}},\ and\ \bibinfo {author}
  {\bibfnamefont {K.}~\bibnamefont {Yoshida}},\ }\href
  {https://doi.org/10.1103/PhysRevLett.103.262501} {\bibfield  {journal}
  {\bibinfo  {journal} {Phys. Rev. Lett.}\ }\textbf {\bibinfo {volume} {103}},\
  \bibinfo {pages} {262501} (\bibinfo {year} {2009})}\BibitemShut {NoStop}%
\bibitem [{\citenamefont {Nakamura}\ \emph {et~al.}(2014)\citenamefont
  {Nakamura}, \citenamefont {Kobayashi}, \citenamefont {Kondo}, \citenamefont
  {Satou}, \citenamefont {Tostevin}, \citenamefont {Utsuno}, \citenamefont
  {Aoi}, \citenamefont {Baba}, \citenamefont {Fukuda}, \citenamefont {Gibelin},
  \citenamefont {Inabe}, \citenamefont {Ishihara}, \citenamefont {Kameda},
  \citenamefont {Kubo}, \citenamefont {Motobayashi}, \citenamefont {Ohnishi},
  \citenamefont {Orr}, \citenamefont {Otsu}, \citenamefont {Otsuka},
  \citenamefont {Sakurai}, \citenamefont {Sumikama}, \citenamefont {Takeda},
  \citenamefont {Takeshita}, \citenamefont {Takechi}, \citenamefont {Takeuchi},
  \citenamefont {Togano},\ and\ \citenamefont
  {Yoneda}}]{Nakamura2014_PRL112-142501}%
  \BibitemOpen
  \bibfield  {author} {\bibinfo {author} {\bibfnamefont {T.}~\bibnamefont
  {Nakamura}}, \bibinfo {author} {\bibfnamefont {N.}~\bibnamefont {Kobayashi}},
  \bibinfo {author} {\bibfnamefont {Y.}~\bibnamefont {Kondo}}, \bibinfo
  {author} {\bibfnamefont {Y.}~\bibnamefont {Satou}}, \bibinfo {author}
  {\bibfnamefont {J.~A.}\ \bibnamefont {Tostevin}}, \bibinfo {author}
  {\bibfnamefont {Y.}~\bibnamefont {Utsuno}}, \bibinfo {author} {\bibfnamefont
  {N.}~\bibnamefont {Aoi}}, \bibinfo {author} {\bibfnamefont {H.}~\bibnamefont
  {Baba}}, \bibinfo {author} {\bibfnamefont {N.}~\bibnamefont {Fukuda}},
  \bibinfo {author} {\bibfnamefont {J.}~\bibnamefont {Gibelin}}, \bibinfo
  {author} {\bibfnamefont {N.}~\bibnamefont {Inabe}}, \bibinfo {author}
  {\bibfnamefont {M.}~\bibnamefont {Ishihara}}, \bibinfo {author}
  {\bibfnamefont {D.}~\bibnamefont {Kameda}}, \bibinfo {author} {\bibfnamefont
  {T.}~\bibnamefont {Kubo}}, \bibinfo {author} {\bibfnamefont {T.}~\bibnamefont
  {Motobayashi}}, \bibinfo {author} {\bibfnamefont {T.}~\bibnamefont
  {Ohnishi}}, \bibinfo {author} {\bibfnamefont {N.~A.}\ \bibnamefont {Orr}},
  \bibinfo {author} {\bibfnamefont {H.}~\bibnamefont {Otsu}}, \bibinfo {author}
  {\bibfnamefont {T.}~\bibnamefont {Otsuka}}, \bibinfo {author} {\bibfnamefont
  {H.}~\bibnamefont {Sakurai}}, \bibinfo {author} {\bibfnamefont
  {T.}~\bibnamefont {Sumikama}}, \bibinfo {author} {\bibfnamefont
  {H.}~\bibnamefont {Takeda}}, \bibinfo {author} {\bibfnamefont
  {E.}~\bibnamefont {Takeshita}}, \bibinfo {author} {\bibfnamefont
  {M.}~\bibnamefont {Takechi}}, \bibinfo {author} {\bibfnamefont
  {S.}~\bibnamefont {Takeuchi}}, \bibinfo {author} {\bibfnamefont
  {Y.}~\bibnamefont {Togano}},\ and\ \bibinfo {author} {\bibfnamefont
  {K.}~\bibnamefont {Yoneda}},\ }\href
  {https://doi.org/10.1103/PhysRevLett.112.142501} {\bibfield  {journal}
  {\bibinfo  {journal} {Phys. Rev. Lett.}\ }\textbf {\bibinfo {volume} {112}},\
  \bibinfo {pages} {142501} (\bibinfo {year} {2014})}\BibitemShut {NoStop}%
\bibitem [{\citenamefont {Kobayashi}\ \emph {et~al.}(2014)\citenamefont
  {Kobayashi}, \citenamefont {Nakamura}, \citenamefont {Kondo}, \citenamefont
  {Tostevin}, \citenamefont {Utsuno}, \citenamefont {Aoi}, \citenamefont
  {Baba}, \citenamefont {Barthelemy}, \citenamefont {Famiano}, \citenamefont
  {Fukuda}, \citenamefont {Inabe}, \citenamefont {Ishihara}, \citenamefont
  {Kanungo}, \citenamefont {Kim}, \citenamefont {Kubo}, \citenamefont {Lee},
  \citenamefont {Lee}, \citenamefont {Matsushita}, \citenamefont {Motobayashi},
  \citenamefont {Ohnishi}, \citenamefont {Orr}, \citenamefont {Otsu},
  \citenamefont {Otsuka}, \citenamefont {Sako}, \citenamefont {Sakurai},
  \citenamefont {Satou}, \citenamefont {Sumikama}, \citenamefont {Takeda},
  \citenamefont {Takeuchi}, \citenamefont {Tanaka}, \citenamefont {Togano},\
  and\ \citenamefont {Yoneda}}]{Kobayashi2014_PRL112-242501}%
  \BibitemOpen
  \bibfield  {author} {\bibinfo {author} {\bibfnamefont {N.}~\bibnamefont
  {Kobayashi}}, \bibinfo {author} {\bibfnamefont {T.}~\bibnamefont {Nakamura}},
  \bibinfo {author} {\bibfnamefont {Y.}~\bibnamefont {Kondo}}, \bibinfo
  {author} {\bibfnamefont {J.~A.}\ \bibnamefont {Tostevin}}, \bibinfo {author}
  {\bibfnamefont {Y.}~\bibnamefont {Utsuno}}, \bibinfo {author} {\bibfnamefont
  {N.}~\bibnamefont {Aoi}}, \bibinfo {author} {\bibfnamefont {H.}~\bibnamefont
  {Baba}}, \bibinfo {author} {\bibfnamefont {R.}~\bibnamefont {Barthelemy}},
  \bibinfo {author} {\bibfnamefont {M.~A.}\ \bibnamefont {Famiano}}, \bibinfo
  {author} {\bibfnamefont {N.}~\bibnamefont {Fukuda}}, \bibinfo {author}
  {\bibfnamefont {N.}~\bibnamefont {Inabe}}, \bibinfo {author} {\bibfnamefont
  {M.}~\bibnamefont {Ishihara}}, \bibinfo {author} {\bibfnamefont
  {R.}~\bibnamefont {Kanungo}}, \bibinfo {author} {\bibfnamefont
  {S.}~\bibnamefont {Kim}}, \bibinfo {author} {\bibfnamefont {T.}~\bibnamefont
  {Kubo}}, \bibinfo {author} {\bibfnamefont {G.~S.}\ \bibnamefont {Lee}},
  \bibinfo {author} {\bibfnamefont {H.~S.}\ \bibnamefont {Lee}}, \bibinfo
  {author} {\bibfnamefont {M.}~\bibnamefont {Matsushita}}, \bibinfo {author}
  {\bibfnamefont {T.}~\bibnamefont {Motobayashi}}, \bibinfo {author}
  {\bibfnamefont {T.}~\bibnamefont {Ohnishi}}, \bibinfo {author} {\bibfnamefont
  {N.~A.}\ \bibnamefont {Orr}}, \bibinfo {author} {\bibfnamefont
  {H.}~\bibnamefont {Otsu}}, \bibinfo {author} {\bibfnamefont {T.}~\bibnamefont
  {Otsuka}}, \bibinfo {author} {\bibfnamefont {T.}~\bibnamefont {Sako}},
  \bibinfo {author} {\bibfnamefont {H.}~\bibnamefont {Sakurai}}, \bibinfo
  {author} {\bibfnamefont {Y.}~\bibnamefont {Satou}}, \bibinfo {author}
  {\bibfnamefont {T.}~\bibnamefont {Sumikama}}, \bibinfo {author}
  {\bibfnamefont {H.}~\bibnamefont {Takeda}}, \bibinfo {author} {\bibfnamefont
  {S.}~\bibnamefont {Takeuchi}}, \bibinfo {author} {\bibfnamefont
  {R.}~\bibnamefont {Tanaka}}, \bibinfo {author} {\bibfnamefont
  {Y.}~\bibnamefont {Togano}},\ and\ \bibinfo {author} {\bibfnamefont
  {K.}~\bibnamefont {Yoneda}},\ }\href
  {https://doi.org/10.1103/PhysRevLett.112.242501} {\bibfield  {journal}
  {\bibinfo  {journal} {Phys. Rev. Lett.}\ }\textbf {\bibinfo {volume} {112}},\
  \bibinfo {pages} {242501} (\bibinfo {year} {2014})}\BibitemShut {NoStop}%
\bibitem [{\citenamefont {Zhou}\ \emph {et~al.}(2010)\citenamefont {Zhou},
  \citenamefont {Meng}, \citenamefont {Ring},\ and\ \citenamefont
  {Zhao}}]{Zhou2010_PRC82-011301R}%
  \BibitemOpen
  \bibfield  {author} {\bibinfo {author} {\bibfnamefont {S.-G.}\ \bibnamefont
  {Zhou}}, \bibinfo {author} {\bibfnamefont {J.}~\bibnamefont {Meng}}, \bibinfo
  {author} {\bibfnamefont {P.}~\bibnamefont {Ring}},\ and\ \bibinfo {author}
  {\bibfnamefont {E.-G.}\ \bibnamefont {Zhao}},\ }\href
  {https://doi.org/10.1103/physrevc.82.011301} {\bibfield  {journal} {\bibinfo
  {journal} {Phys. Rev. C}\ }\textbf {\bibinfo {volume} {82}},\ \bibinfo
  {pages} {011301(R)} (\bibinfo {year} {2010})}\BibitemShut {NoStop}%
\bibitem [{\citenamefont {Pei}\ \emph {et~al.}(2013)\citenamefont {Pei},
  \citenamefont {Zhang},\ and\ \citenamefont {Xu}}]{Pei2013_PRC87-051302R}%
  \BibitemOpen
  \bibfield  {author} {\bibinfo {author} {\bibfnamefont {J.~C.}\ \bibnamefont
  {Pei}}, \bibinfo {author} {\bibfnamefont {Y.~N.}\ \bibnamefont {Zhang}},\
  and\ \bibinfo {author} {\bibfnamefont {F.~R.}\ \bibnamefont {Xu}},\ }\href
  {https://doi.org/10.1103/PhysRevC.87.051302} {\bibfield  {journal} {\bibinfo
  {journal} {Phys. Rev. C}\ }\textbf {\bibinfo {volume} {87}},\ \bibinfo
  {pages} {051302(R)} (\bibinfo {year} {2013})}\BibitemShut {NoStop}%
\bibitem [{\citenamefont {Chen}\ \emph {et~al.}(2014)\citenamefont {Chen},
  \citenamefont {Ring},\ and\ \citenamefont {Meng}}]{Chen2014_PRC89-014312}%
  \BibitemOpen
  \bibfield  {author} {\bibinfo {author} {\bibfnamefont {Y.}~\bibnamefont
  {Chen}}, \bibinfo {author} {\bibfnamefont {P.}~\bibnamefont {Ring}},\ and\
  \bibinfo {author} {\bibfnamefont {J.}~\bibnamefont {Meng}},\ }\href
  {https://doi.org/10.1103/PhysRevC.89.014312} {\bibfield  {journal} {\bibinfo
  {journal} {Phys. Rev. C}\ }\textbf {\bibinfo {volume} {89}},\ \bibinfo
  {pages} {014312} (\bibinfo {year} {2014})}\BibitemShut {NoStop}%
\bibitem [{\citenamefont {Nakada}\ and\ \citenamefont
  {Takayama}(2018)}]{Nakada2018_PRC98-011301R}%
  \BibitemOpen
  \bibfield  {author} {\bibinfo {author} {\bibfnamefont {H.}~\bibnamefont
  {Nakada}}\ and\ \bibinfo {author} {\bibfnamefont {K.}~\bibnamefont
  {Takayama}},\ }\href {https://doi.org/10.1103/PhysRevC.98.011301} {\bibfield
  {journal} {\bibinfo  {journal} {Phys. Rev. C}\ }\textbf {\bibinfo {volume}
  {98}},\ \bibinfo {pages} {011301(R)} (\bibinfo {year} {2018})}\BibitemShut
  {NoStop}%
\bibitem [{\citenamefont {Li}\ \emph {et~al.}(2012)\citenamefont {Li},
  \citenamefont {Meng}, \citenamefont {Ring}, \citenamefont {Zhao},\ and\
  \citenamefont {Zhou}}]{Li2012_PRC85-024312}%
  \BibitemOpen
  \bibfield  {author} {\bibinfo {author} {\bibfnamefont {L.-L.}\ \bibnamefont
  {Li}}, \bibinfo {author} {\bibfnamefont {J.}~\bibnamefont {Meng}}, \bibinfo
  {author} {\bibfnamefont {P.}~\bibnamefont {Ring}}, \bibinfo {author}
  {\bibfnamefont {E.-G.}\ \bibnamefont {Zhao}},\ and\ \bibinfo {author}
  {\bibfnamefont {S.-G.}\ \bibnamefont {Zhou}},\ }\href
  {https://doi.org/10.1103/PhysRevC.85.024312} {\bibfield  {journal} {\bibinfo
  {journal} {Phys. Rev. C}\ }\textbf {\bibinfo {volume} {85}},\ \bibinfo
  {pages} {024312} (\bibinfo {year} {2012})}\BibitemShut {NoStop}%
\bibitem [{\citenamefont {Sun}\ \emph {et~al.}(2018)\citenamefont {Sun},
  \citenamefont {Zhao},\ and\ \citenamefont {Zhou}}]{Sun2018_PLB785-530}%
  \BibitemOpen
  \bibfield  {author} {\bibinfo {author} {\bibfnamefont {X.-X.}\ \bibnamefont
  {Sun}}, \bibinfo {author} {\bibfnamefont {J.}~\bibnamefont {Zhao}},\ and\
  \bibinfo {author} {\bibfnamefont {S.-G.}\ \bibnamefont {Zhou}},\ }\href
  {https://doi.org/10.1016/j.physletb.2018.08.071} {\bibfield  {journal}
  {\bibinfo  {journal} {Phys. Lett. B}\ }\textbf {\bibinfo {volume} {785}},\
  \bibinfo {pages} {530} (\bibinfo {year} {2018})}\BibitemShut {NoStop}%
\bibitem [{\citenamefont {Zhang}\ \emph {et~al.}(2019)\citenamefont {Zhang},
  \citenamefont {Wang},\ and\ \citenamefont {Zhang}}]{Zhang2019_PRC100-034312}%
  \BibitemOpen
  \bibfield  {author} {\bibinfo {author} {\bibfnamefont {K.~Y.}\ \bibnamefont
  {Zhang}}, \bibinfo {author} {\bibfnamefont {D.~Y.}\ \bibnamefont {Wang}},\
  and\ \bibinfo {author} {\bibfnamefont {S.~Q.}\ \bibnamefont {Zhang}},\ }\href
  {https://doi.org/10.1103/PhysRevC.100.034312} {\bibfield  {journal} {\bibinfo
   {journal} {Phys. Rev. C}\ }\textbf {\bibinfo {volume} {100}},\ \bibinfo
  {pages} {034312} (\bibinfo {year} {2019})}\BibitemShut {NoStop}%
\bibitem [{\citenamefont {Sun}\ \emph {et~al.}(2020{\natexlab{a}})\citenamefont
  {Sun}, \citenamefont {Zhao},\ and\ \citenamefont
  {Zhou}}]{Sun2020_NPA1003-122011}%
  \BibitemOpen
  \bibfield  {author} {\bibinfo {author} {\bibfnamefont {X.-X.}\ \bibnamefont
  {Sun}}, \bibinfo {author} {\bibfnamefont {J.}~\bibnamefont {Zhao}},\ and\
  \bibinfo {author} {\bibfnamefont {S.-G.}\ \bibnamefont {Zhou}},\ }\href
  {https://doi.org/10.1016/j.nuclphysa.2020.122011} {\bibfield  {journal}
  {\bibinfo  {journal} {Nucl. Phys. A}\ }\textbf {\bibinfo {volume} {1003}},\
  \bibinfo {pages} {122011} (\bibinfo {year} {2020}{\natexlab{a}})}\BibitemShut
  {NoStop}%
\bibitem [{\citenamefont {Reinhard}(1989)}]{Reinhard1989_RPP52-439}%
  \BibitemOpen
  \bibfield  {author} {\bibinfo {author} {\bibfnamefont {P.~G.}\ \bibnamefont
  {Reinhard}},\ }\href {https://doi.org/10.1088/0034-4885/52/4/002} {\bibfield
  {journal} {\bibinfo  {journal} {Rep. Prog. Phys.}\ }\textbf {\bibinfo
  {volume} {52}},\ \bibinfo {pages} {439} (\bibinfo {year} {1989})}\BibitemShut
  {NoStop}%
\bibitem [{\citenamefont {Ring}(1996)}]{Ring1996_PPNP37-193}%
  \BibitemOpen
  \bibfield  {author} {\bibinfo {author} {\bibfnamefont {P.}~\bibnamefont
  {Ring}},\ }\href {https://doi.org/10.1016/0146-6410(96)00054-3} {\bibfield
  {journal} {\bibinfo  {journal} {Prog. Part. Nucl. Phys.}\ }\textbf {\bibinfo
  {volume} {37}},\ \bibinfo {pages} {193} (\bibinfo {year} {1996})}\BibitemShut
  {NoStop}%
\bibitem [{\citenamefont {Liang}\ \emph {et~al.}(2015)\citenamefont {Liang},
  \citenamefont {Meng},\ and\ \citenamefont {Zhou}}]{Liang2015_PR570-1}%
  \BibitemOpen
  \bibfield  {author} {\bibinfo {author} {\bibfnamefont {H.}~\bibnamefont
  {Liang}}, \bibinfo {author} {\bibfnamefont {J.}~\bibnamefont {Meng}},\ and\
  \bibinfo {author} {\bibfnamefont {S.-G.}\ \bibnamefont {Zhou}},\ }\href
  {https://doi.org/10.1016/j.physrep.2014.12.005} {\bibfield  {journal}
  {\bibinfo  {journal} {Phys. Rep.}\ }\textbf {\bibinfo {volume} {570}},\
  \bibinfo {pages} {1} (\bibinfo {year} {2015})}\BibitemShut {NoStop}%
\bibitem [{\citenamefont {Meng}\ \emph {et~al.}(2002)\citenamefont {Meng},
  \citenamefont {Toki}, \citenamefont {Zeng}, \citenamefont {Zhang},\ and\
  \citenamefont {Zhou}}]{Meng2002_PRC65-041302R}%
  \BibitemOpen
  \bibfield  {author} {\bibinfo {author} {\bibfnamefont {J.}~\bibnamefont
  {Meng}}, \bibinfo {author} {\bibfnamefont {H.}~\bibnamefont {Toki}}, \bibinfo
  {author} {\bibfnamefont {J.~Y.}\ \bibnamefont {Zeng}}, \bibinfo {author}
  {\bibfnamefont {S.~Q.}\ \bibnamefont {Zhang}},\ and\ \bibinfo {author}
  {\bibfnamefont {S.-G.}\ \bibnamefont {Zhou}},\ }\href
  {https://doi.org/10.1103/PhysRevC.65.041302} {\bibfield  {journal} {\bibinfo
  {journal} {Phys. Rev. C}\ }\textbf {\bibinfo {volume} {65}},\ \bibinfo
  {pages} {041302(R)} (\bibinfo {year} {2002})}\BibitemShut {NoStop}%
\bibitem [{\citenamefont {Lu}\ \emph {et~al.}(2013)\citenamefont {Lu},
  \citenamefont {Sun},\ and\ \citenamefont {Long}}]{Lu2013_PRC87-034311}%
  \BibitemOpen
  \bibfield  {author} {\bibinfo {author} {\bibfnamefont {X.~L.}\ \bibnamefont
  {Lu}}, \bibinfo {author} {\bibfnamefont {B.~Y.}\ \bibnamefont {Sun}},\ and\
  \bibinfo {author} {\bibfnamefont {W.~H.}\ \bibnamefont {Long}},\ }\href
  {https://doi.org/10.1103/PhysRevC.87.034311} {\bibfield  {journal} {\bibinfo
  {journal} {Phys. Rev. C}\ }\textbf {\bibinfo {volume} {87}},\ \bibinfo
  {pages} {034311} (\bibinfo {year} {2013})}\BibitemShut {NoStop}%
\bibitem [{\citenamefont {Yang}\ \emph {et~al.}(2021)\citenamefont {Yang},
  \citenamefont {Kubota}, \citenamefont {Corsi}, \citenamefont {Yoshida},
  \citenamefont {Sun}, \citenamefont {Li}, \citenamefont {Kimura},
  \citenamefont {Michel}, \citenamefont {Ogata}, \citenamefont {Yuan},
  \citenamefont {Yuan}, \citenamefont {Authelet}, \citenamefont {Baba},
  \citenamefont {Caesar}, \citenamefont {Calvet}, \citenamefont {Delbart},
  \citenamefont {Dozono}, \citenamefont {Feng}, \citenamefont {Flavigny},
  \citenamefont {Gheller}, \citenamefont {Gibelin}, \citenamefont {Giganon},
  \citenamefont {Gillibert}, \citenamefont {Hasegawa}, \citenamefont {Isobe},
  \citenamefont {Kanaya}, \citenamefont {Kawakami}, \citenamefont {Kim},
  \citenamefont {Kiyokawa}, \citenamefont {Kobayashi}, \citenamefont
  {Kobayashi}, \citenamefont {Kobayashi}, \citenamefont {Kondo}, \citenamefont
  {Korkulu}, \citenamefont {Koyama}, \citenamefont {Lapoux}, \citenamefont
  {Maeda}, \citenamefont {Marqu\'es}, \citenamefont {Motobayashi},
  \citenamefont {Miyazaki}, \citenamefont {Nakamura}, \citenamefont
  {Nakatsuka}, \citenamefont {Nishio}, \citenamefont {Obertelli}, \citenamefont
  {Ohkura}, \citenamefont {Orr}, \citenamefont {Ota}, \citenamefont {Otsu},
  \citenamefont {Ozaki}, \citenamefont {Panin}, \citenamefont {Paschalis},
  \citenamefont {Pollacco}, \citenamefont {Reichert}, \citenamefont {Rouss\'e},
  \citenamefont {Saito}, \citenamefont {Sakaguchi}, \citenamefont {Sako},
  \citenamefont {Santamaria}, \citenamefont {Sasano}, \citenamefont {Sato},
  \citenamefont {Shikata}, \citenamefont {Shimizu}, \citenamefont {Shindo},
  \citenamefont {Stuhl}, \citenamefont {Sumikama}, \citenamefont {Sun},
  \citenamefont {Tabata}, \citenamefont {Togano}, \citenamefont {Tsubota},
  \citenamefont {Xu}, \citenamefont {Yasuda}, \citenamefont {Yoneda},
  \citenamefont {Zenihiro}, \citenamefont {Zhou}, \citenamefont {Zuo},\ and\
  \citenamefont {Uesaka}}]{Yang2021_PRL126-082501}%
  \BibitemOpen
  \bibfield  {author} {\bibinfo {author} {\bibfnamefont {Z.~H.}\ \bibnamefont
  {Yang}}, \bibinfo {author} {\bibfnamefont {Y.}~\bibnamefont {Kubota}},
  \bibinfo {author} {\bibfnamefont {A.}~\bibnamefont {Corsi}}, \bibinfo
  {author} {\bibfnamefont {K.}~\bibnamefont {Yoshida}}, \bibinfo {author}
  {\bibfnamefont {X.-X.}\ \bibnamefont {Sun}}, \bibinfo {author} {\bibfnamefont
  {J.~G.}\ \bibnamefont {Li}}, \bibinfo {author} {\bibfnamefont
  {M.}~\bibnamefont {Kimura}}, \bibinfo {author} {\bibfnamefont
  {N.}~\bibnamefont {Michel}}, \bibinfo {author} {\bibfnamefont
  {K.}~\bibnamefont {Ogata}}, \bibinfo {author} {\bibfnamefont {C.~X.}\
  \bibnamefont {Yuan}}, \bibinfo {author} {\bibfnamefont {Q.}~\bibnamefont
  {Yuan}}, \bibinfo {author} {\bibfnamefont {G.}~\bibnamefont {Authelet}},
  \bibinfo {author} {\bibfnamefont {H.}~\bibnamefont {Baba}}, \bibinfo {author}
  {\bibfnamefont {C.}~\bibnamefont {Caesar}}, \bibinfo {author} {\bibfnamefont
  {D.}~\bibnamefont {Calvet}}, \bibinfo {author} {\bibfnamefont
  {A.}~\bibnamefont {Delbart}}, \bibinfo {author} {\bibfnamefont
  {M.}~\bibnamefont {Dozono}}, \bibinfo {author} {\bibfnamefont
  {J.}~\bibnamefont {Feng}}, \bibinfo {author} {\bibfnamefont {F.}~\bibnamefont
  {Flavigny}}, \bibinfo {author} {\bibfnamefont {J.-M.}\ \bibnamefont
  {Gheller}}, \bibinfo {author} {\bibfnamefont {J.}~\bibnamefont {Gibelin}},
  \bibinfo {author} {\bibfnamefont {A.}~\bibnamefont {Giganon}}, \bibinfo
  {author} {\bibfnamefont {A.}~\bibnamefont {Gillibert}}, \bibinfo {author}
  {\bibfnamefont {K.}~\bibnamefont {Hasegawa}}, \bibinfo {author}
  {\bibfnamefont {T.}~\bibnamefont {Isobe}}, \bibinfo {author} {\bibfnamefont
  {Y.}~\bibnamefont {Kanaya}}, \bibinfo {author} {\bibfnamefont
  {S.}~\bibnamefont {Kawakami}}, \bibinfo {author} {\bibfnamefont
  {D.}~\bibnamefont {Kim}}, \bibinfo {author} {\bibfnamefont {Y.}~\bibnamefont
  {Kiyokawa}}, \bibinfo {author} {\bibfnamefont {M.}~\bibnamefont {Kobayashi}},
  \bibinfo {author} {\bibfnamefont {N.}~\bibnamefont {Kobayashi}}, \bibinfo
  {author} {\bibfnamefont {T.}~\bibnamefont {Kobayashi}}, \bibinfo {author}
  {\bibfnamefont {Y.}~\bibnamefont {Kondo}}, \bibinfo {author} {\bibfnamefont
  {Z.}~\bibnamefont {Korkulu}}, \bibinfo {author} {\bibfnamefont
  {S.}~\bibnamefont {Koyama}}, \bibinfo {author} {\bibfnamefont
  {V.}~\bibnamefont {Lapoux}}, \bibinfo {author} {\bibfnamefont
  {Y.}~\bibnamefont {Maeda}}, \bibinfo {author} {\bibfnamefont {F.~M.}\
  \bibnamefont {Marqu\'es}}, \bibinfo {author} {\bibfnamefont {T.}~\bibnamefont
  {Motobayashi}}, \bibinfo {author} {\bibfnamefont {T.}~\bibnamefont
  {Miyazaki}}, \bibinfo {author} {\bibfnamefont {T.}~\bibnamefont {Nakamura}},
  \bibinfo {author} {\bibfnamefont {N.}~\bibnamefont {Nakatsuka}}, \bibinfo
  {author} {\bibfnamefont {Y.}~\bibnamefont {Nishio}}, \bibinfo {author}
  {\bibfnamefont {A.}~\bibnamefont {Obertelli}}, \bibinfo {author}
  {\bibfnamefont {A.}~\bibnamefont {Ohkura}}, \bibinfo {author} {\bibfnamefont
  {N.~A.}\ \bibnamefont {Orr}}, \bibinfo {author} {\bibfnamefont
  {S.}~\bibnamefont {Ota}}, \bibinfo {author} {\bibfnamefont {H.}~\bibnamefont
  {Otsu}}, \bibinfo {author} {\bibfnamefont {T.}~\bibnamefont {Ozaki}},
  \bibinfo {author} {\bibfnamefont {V.}~\bibnamefont {Panin}}, \bibinfo
  {author} {\bibfnamefont {S.}~\bibnamefont {Paschalis}}, \bibinfo {author}
  {\bibfnamefont {E.~C.}\ \bibnamefont {Pollacco}}, \bibinfo {author}
  {\bibfnamefont {S.}~\bibnamefont {Reichert}}, \bibinfo {author}
  {\bibfnamefont {J.-Y.}\ \bibnamefont {Rouss\'e}}, \bibinfo {author}
  {\bibfnamefont {A.~T.}\ \bibnamefont {Saito}}, \bibinfo {author}
  {\bibfnamefont {S.}~\bibnamefont {Sakaguchi}}, \bibinfo {author}
  {\bibfnamefont {M.}~\bibnamefont {Sako}}, \bibinfo {author} {\bibfnamefont
  {C.}~\bibnamefont {Santamaria}}, \bibinfo {author} {\bibfnamefont
  {M.}~\bibnamefont {Sasano}}, \bibinfo {author} {\bibfnamefont
  {H.}~\bibnamefont {Sato}}, \bibinfo {author} {\bibfnamefont {M.}~\bibnamefont
  {Shikata}}, \bibinfo {author} {\bibfnamefont {Y.}~\bibnamefont {Shimizu}},
  \bibinfo {author} {\bibfnamefont {Y.}~\bibnamefont {Shindo}}, \bibinfo
  {author} {\bibfnamefont {L.}~\bibnamefont {Stuhl}}, \bibinfo {author}
  {\bibfnamefont {T.}~\bibnamefont {Sumikama}}, \bibinfo {author}
  {\bibfnamefont {Y.~L.}\ \bibnamefont {Sun}}, \bibinfo {author} {\bibfnamefont
  {M.}~\bibnamefont {Tabata}}, \bibinfo {author} {\bibfnamefont
  {Y.}~\bibnamefont {Togano}}, \bibinfo {author} {\bibfnamefont
  {J.}~\bibnamefont {Tsubota}}, \bibinfo {author} {\bibfnamefont {F.~R.}\
  \bibnamefont {Xu}}, \bibinfo {author} {\bibfnamefont {J.}~\bibnamefont
  {Yasuda}}, \bibinfo {author} {\bibfnamefont {K.}~\bibnamefont {Yoneda}},
  \bibinfo {author} {\bibfnamefont {J.}~\bibnamefont {Zenihiro}}, \bibinfo
  {author} {\bibfnamefont {S.-G.}\ \bibnamefont {Zhou}}, \bibinfo {author}
  {\bibfnamefont {W.}~\bibnamefont {Zuo}},\ and\ \bibinfo {author}
  {\bibfnamefont {T.}~\bibnamefont {Uesaka}},\ }\href
  {https://doi.org/10.1103/physrevlett.126.082501} {\bibfield  {journal}
  {\bibinfo  {journal} {Phys. Rev. Lett.}\ }\textbf {\bibinfo {volume} {126}},\
  \bibinfo {pages} {082501} (\bibinfo {year} {2021})}\BibitemShut {NoStop}%
\bibitem [{\citenamefont {Sun}(2021)}]{Sun2021_PRC103-054315}%
  \BibitemOpen
  \bibfield  {author} {\bibinfo {author} {\bibfnamefont {X.-X.}\ \bibnamefont
  {Sun}},\ }\href {https://doi.org/10.1103/PhysRevC.103.054315} {\bibfield
  {journal} {\bibinfo  {journal} {Phys. Rev. C}\ }\textbf {\bibinfo {volume}
  {103}},\ \bibinfo {pages} {054315} (\bibinfo {year} {2021})}\BibitemShut
  {NoStop}%
\bibitem [{\citenamefont {Zhang}\ \emph {et~al.}(2020)\citenamefont {Zhang},
  \citenamefont {Cheoun}, \citenamefont {Choi}, \citenamefont {Chong},
  \citenamefont {Dong}, \citenamefont {Geng}, \citenamefont {Ha}, \citenamefont
  {He}, \citenamefont {Heo}, \citenamefont {Ho}, \citenamefont {In},
  \citenamefont {Kim}, \citenamefont {Kim}, \citenamefont {Lee}, \citenamefont
  {Lee}, \citenamefont {Li}, \citenamefont {Luo}, \citenamefont {Meng},
  \citenamefont {Mun}, \citenamefont {Niu}, \citenamefont {Pan}, \citenamefont
  {Papakonstantinou}, \citenamefont {Shang}, \citenamefont {Shen},
  \citenamefont {Shen}, \citenamefont {Sun}, \citenamefont {Sun}, \citenamefont
  {Tam}, \citenamefont {Thaivayongnou}, \citenamefont {Wang}, \citenamefont
  {Wong}, \citenamefont {Xia}, \citenamefont {Yan}, \citenamefont {Yeung},
  \citenamefont {Yiu}, \citenamefont {Zhang}, \citenamefont {Zhang},\ and\
  \citenamefont {Zhou}}]{Zhang2020_PRC102-024314}%
  \BibitemOpen
  \bibfield  {author} {\bibinfo {author} {\bibfnamefont {K.}~\bibnamefont
  {Zhang}}, \bibinfo {author} {\bibfnamefont {M.-K.}\ \bibnamefont {Cheoun}},
  \bibinfo {author} {\bibfnamefont {Y.-B.}\ \bibnamefont {Choi}}, \bibinfo
  {author} {\bibfnamefont {P.~S.}\ \bibnamefont {Chong}}, \bibinfo {author}
  {\bibfnamefont {J.}~\bibnamefont {Dong}}, \bibinfo {author} {\bibfnamefont
  {L.}~\bibnamefont {Geng}}, \bibinfo {author} {\bibfnamefont {E.}~\bibnamefont
  {Ha}}, \bibinfo {author} {\bibfnamefont {X.}~\bibnamefont {He}}, \bibinfo
  {author} {\bibfnamefont {C.}~\bibnamefont {Heo}}, \bibinfo {author}
  {\bibfnamefont {M.~C.}\ \bibnamefont {Ho}}, \bibinfo {author} {\bibfnamefont
  {E.~J.}\ \bibnamefont {In}}, \bibinfo {author} {\bibfnamefont
  {S.}~\bibnamefont {Kim}}, \bibinfo {author} {\bibfnamefont {Y.}~\bibnamefont
  {Kim}}, \bibinfo {author} {\bibfnamefont {C.-H.}\ \bibnamefont {Lee}},
  \bibinfo {author} {\bibfnamefont {J.}~\bibnamefont {Lee}}, \bibinfo {author}
  {\bibfnamefont {Z.}~\bibnamefont {Li}}, \bibinfo {author} {\bibfnamefont
  {T.}~\bibnamefont {Luo}}, \bibinfo {author} {\bibfnamefont {J.}~\bibnamefont
  {Meng}}, \bibinfo {author} {\bibfnamefont {M.-H.}\ \bibnamefont {Mun}},
  \bibinfo {author} {\bibfnamefont {Z.}~\bibnamefont {Niu}}, \bibinfo {author}
  {\bibfnamefont {C.}~\bibnamefont {Pan}}, \bibinfo {author} {\bibfnamefont
  {P.}~\bibnamefont {Papakonstantinou}}, \bibinfo {author} {\bibfnamefont
  {X.}~\bibnamefont {Shang}}, \bibinfo {author} {\bibfnamefont
  {C.}~\bibnamefont {Shen}}, \bibinfo {author} {\bibfnamefont {G.}~\bibnamefont
  {Shen}}, \bibinfo {author} {\bibfnamefont {W.}~\bibnamefont {Sun}}, \bibinfo
  {author} {\bibfnamefont {X.-X.}\ \bibnamefont {Sun}}, \bibinfo {author}
  {\bibfnamefont {C.~K.}\ \bibnamefont {Tam}}, \bibinfo {author} {\bibnamefont
  {Thaivayongnou}}, \bibinfo {author} {\bibfnamefont {C.}~\bibnamefont {Wang}},
  \bibinfo {author} {\bibfnamefont {S.~H.}\ \bibnamefont {Wong}}, \bibinfo
  {author} {\bibfnamefont {X.}~\bibnamefont {Xia}}, \bibinfo {author}
  {\bibfnamefont {Y.}~\bibnamefont {Yan}}, \bibinfo {author} {\bibfnamefont
  {R.~W.-Y.}\ \bibnamefont {Yeung}}, \bibinfo {author} {\bibfnamefont {T.~C.}\
  \bibnamefont {Yiu}}, \bibinfo {author} {\bibfnamefont {S.}~\bibnamefont
  {Zhang}}, \bibinfo {author} {\bibfnamefont {W.}~\bibnamefont {Zhang}},\ and\
  \bibinfo {author} {\bibfnamefont {S.-G.}\ \bibnamefont {Zhou}},\ }\href
  {https://doi.org/10.1103/physrevc.102.024314} {\bibfield  {journal} {\bibinfo
   {journal} {Phys. Rev. C}\ }\textbf {\bibinfo {volume} {102}},\ \bibinfo
  {pages} {024314} (\bibinfo {year} {2020})}\BibitemShut {NoStop}%
\bibitem [{\citenamefont {In}\ \emph {et~al.}(2021)\citenamefont {In},
  \citenamefont {Papakonstantinou}, \citenamefont {Kim},\ and\ \citenamefont
  {Hong}}]{In2021_IJMPE30-2150009}%
  \BibitemOpen
  \bibfield  {author} {\bibinfo {author} {\bibfnamefont {E.~J.}\ \bibnamefont
  {In}}, \bibinfo {author} {\bibfnamefont {P.}~\bibnamefont
  {Papakonstantinou}}, \bibinfo {author} {\bibfnamefont {Y.}~\bibnamefont
  {Kim}},\ and\ \bibinfo {author} {\bibfnamefont {S.-W.}\ \bibnamefont
  {Hong}},\ }\href {https://doi.org/10.1142/S0218301321500099} {\bibfield
  {journal} {\bibinfo  {journal} {Int. J. Mod. Phys. E}\ }\textbf {\bibinfo
  {volume} {30}},\ \bibinfo {pages} {2150009} (\bibinfo {year}
  {2021})}\BibitemShut {NoStop}%
\bibitem [{\citenamefont {Zhang}\ \emph {et~al.}()\citenamefont {Zhang},
  \citenamefont {He}, \citenamefont {Meng}, \citenamefont {Pan}, \citenamefont
  {Shen}, \citenamefont {Wang},\ and\ \citenamefont
  {Zhang}}]{Zhang2021_arXiv2103.08142}%
  \BibitemOpen
  \bibfield  {author} {\bibinfo {author} {\bibfnamefont {K.}~\bibnamefont
  {Zhang}}, \bibinfo {author} {\bibfnamefont {X.}~\bibnamefont {He}}, \bibinfo
  {author} {\bibfnamefont {J.}~\bibnamefont {Meng}}, \bibinfo {author}
  {\bibfnamefont {C.}~\bibnamefont {Pan}}, \bibinfo {author} {\bibfnamefont
  {C.}~\bibnamefont {Shen}}, \bibinfo {author} {\bibfnamefont {C.}~\bibnamefont
  {Wang}},\ and\ \bibinfo {author} {\bibfnamefont {S.}~\bibnamefont {Zhang}},\
  }\href@noop {} {\bibinfo {title} {Predictive power for superheavy nuclear
  mass and possible stability beyond the neutron drip line in deformed
  relativistic hartree-bogoliubov theory in continuum}},\ \bibinfo
  {howpublished} {arXiv:2103.08142 [nucl-th]},\ \Eprint
  {https://arxiv.org/abs/2103.08142} {2103.08142} \BibitemShut {NoStop}%
\bibitem [{\citenamefont {Pan}\ \emph {et~al.}(2021)\citenamefont {Pan},
  \citenamefont {Zhang}, \citenamefont {Chong}, \citenamefont {Heo},
  \citenamefont {Ho}, \citenamefont {Lee}, \citenamefont {Li}, \citenamefont
  {Sun}, \citenamefont {Tam}, \citenamefont {Wong}, \citenamefont {Yeung},
  \citenamefont {Yiu},\ and\ \citenamefont {Zhang}}]{Pan2021_arXiv2104.07337}%
  \BibitemOpen
  \bibfield  {author} {\bibinfo {author} {\bibfnamefont {C.}~\bibnamefont
  {Pan}}, \bibinfo {author} {\bibfnamefont {K.~Y.}\ \bibnamefont {Zhang}},
  \bibinfo {author} {\bibfnamefont {P.~S.}\ \bibnamefont {Chong}}, \bibinfo
  {author} {\bibfnamefont {C.}~\bibnamefont {Heo}}, \bibinfo {author}
  {\bibfnamefont {M.~C.}\ \bibnamefont {Ho}}, \bibinfo {author} {\bibfnamefont
  {J.}~\bibnamefont {Lee}}, \bibinfo {author} {\bibfnamefont {Z.~P.}\
  \bibnamefont {Li}}, \bibinfo {author} {\bibfnamefont {W.}~\bibnamefont
  {Sun}}, \bibinfo {author} {\bibfnamefont {C.~K.}\ \bibnamefont {Tam}},
  \bibinfo {author} {\bibfnamefont {S.~H.}\ \bibnamefont {Wong}}, \bibinfo
  {author} {\bibfnamefont {R.~W.-Y.}\ \bibnamefont {Yeung}}, \bibinfo {author}
  {\bibfnamefont {T.~C.}\ \bibnamefont {Yiu}},\ and\ \bibinfo {author}
  {\bibfnamefont {S.~Q.}\ \bibnamefont {Zhang}},\ }\href
  {https://arxiv.org/abs/2104.07337v1} {\bibinfo {title} {Possible bound nuclei
  beyond the two-neutron drip line in the $50\leqslant {Z} \leqslant 70$
  region}},\ \bibinfo {howpublished} {arXiv:2104,07337 [nucl-th]} (\bibinfo
  {year} {2021}),\ \Eprint {https://arxiv.org/abs/2104.07337} {2104.07337}
  \BibitemShut {NoStop}%
\bibitem [{\citenamefont {He}\ \emph {et~al.}(2021)\citenamefont {He},
  \citenamefont {Wang}, \citenamefont {Zhang},\ and\ \citenamefont
  {Shen}}]{He2021_arXiv2104.12987}%
  \BibitemOpen
  \bibfield  {author} {\bibinfo {author} {\bibfnamefont {X.-T.}\ \bibnamefont
  {He}}, \bibinfo {author} {\bibfnamefont {C.}~\bibnamefont {Wang}}, \bibinfo
  {author} {\bibfnamefont {K.-Y.}\ \bibnamefont {Zhang}},\ and\ \bibinfo
  {author} {\bibfnamefont {C.-W.}\ \bibnamefont {Shen}},\ }\href
  {https://arxiv.org/abs/2104.12987v1} {\bibinfo {title} {Possible existence of
  bound nuclei beyond neutron drip lines driven by deformation}},\ \bibinfo
  {howpublished} {arXiv:2104.12987 [nucl-th]} (\bibinfo {year} {2021}),\
  \Eprint {https://arxiv.org/abs/2104.12987} {2104.12987} \BibitemShut
  {NoStop}%
\bibitem [{\citenamefont {Nik{\v{s}}i{\'{c}}}\ \emph
  {et~al.}(2006{\natexlab{a}})\citenamefont {Nik{\v{s}}i{\'{c}}}, \citenamefont
  {Vretenar},\ and\ \citenamefont {Ring}}]{Niksic2006_PRC73-034308}%
  \BibitemOpen
  \bibfield  {author} {\bibinfo {author} {\bibfnamefont {T.}~\bibnamefont
  {Nik{\v{s}}i{\'{c}}}}, \bibinfo {author} {\bibfnamefont {D.}~\bibnamefont
  {Vretenar}},\ and\ \bibinfo {author} {\bibfnamefont {P.}~\bibnamefont
  {Ring}},\ }\href {https://doi.org/10.1103/PhysRevC.73.034308} {\bibfield
  {journal} {\bibinfo  {journal} {Phys. Rev. C}\ }\textbf {\bibinfo {volume}
  {73}},\ \bibinfo {pages} {034308} (\bibinfo {year}
  {2006}{\natexlab{a}})}\BibitemShut {NoStop}%
\bibitem [{\citenamefont {Nik{\v{s}}i{\'{c}}}\ \emph
  {et~al.}(2006{\natexlab{b}})\citenamefont {Nik{\v{s}}i{\'{c}}}, \citenamefont
  {Vretenar},\ and\ \citenamefont {Ring}}]{Niksic2006_PRC74-064309}%
  \BibitemOpen
  \bibfield  {author} {\bibinfo {author} {\bibfnamefont {T.}~\bibnamefont
  {Nik{\v{s}}i{\'{c}}}}, \bibinfo {author} {\bibfnamefont {D.}~\bibnamefont
  {Vretenar}},\ and\ \bibinfo {author} {\bibfnamefont {P.}~\bibnamefont
  {Ring}},\ }\href {https://doi.org/10.1103/PhysRevC.74.064309} {\bibfield
  {journal} {\bibinfo  {journal} {Phys. Rev. C}\ }\textbf {\bibinfo {volume}
  {74}},\ \bibinfo {pages} {064309} (\bibinfo {year}
  {2006}{\natexlab{b}})}\BibitemShut {NoStop}%
\bibitem [{\citenamefont {Yao}\ \emph {et~al.}(2015{\natexlab{a}})\citenamefont
  {Yao}, \citenamefont {Zhou},\ and\ \citenamefont
  {Li}}]{Yao2015_PRC92-041304R}%
  \BibitemOpen
  \bibfield  {author} {\bibinfo {author} {\bibfnamefont {J.~M.}\ \bibnamefont
  {Yao}}, \bibinfo {author} {\bibfnamefont {E.~F.}\ \bibnamefont {Zhou}},\ and\
  \bibinfo {author} {\bibfnamefont {Z.~P.}\ \bibnamefont {Li}},\ }\href
  {https://doi.org/10.1103/PhysRevC.92.041304} {\bibfield  {journal} {\bibinfo
  {journal} {Phys. Rev. C}\ }\textbf {\bibinfo {volume} {92}},\ \bibinfo
  {pages} {041304(R)} (\bibinfo {year} {2015}{\natexlab{a}})}\BibitemShut
  {NoStop}%
\bibitem [{\citenamefont {Lu}\ \emph {et~al.}(2012)\citenamefont {Lu},
  \citenamefont {Zhao},\ and\ \citenamefont {Zhou}}]{Lu2012_PRC85-011301R}%
  \BibitemOpen
  \bibfield  {author} {\bibinfo {author} {\bibfnamefont {B.-N.}\ \bibnamefont
  {Lu}}, \bibinfo {author} {\bibfnamefont {E.-G.}\ \bibnamefont {Zhao}},\ and\
  \bibinfo {author} {\bibfnamefont {S.-G.}\ \bibnamefont {Zhou}},\ }\href
  {https://doi.org/10.1103/PhysRevC.85.011301} {\bibfield  {journal} {\bibinfo
  {journal} {Phys. Rev. C}\ }\textbf {\bibinfo {volume} {85}},\ \bibinfo
  {pages} {011301(R)} (\bibinfo {year} {2012})}\BibitemShut {NoStop}%
\bibitem [{\citenamefont {Lu}\ \emph {et~al.}(2014)\citenamefont {Lu},
  \citenamefont {Zhao}, \citenamefont {Zhao},\ and\ \citenamefont
  {Zhou}}]{Lu2014_PRC89-014323}%
  \BibitemOpen
  \bibfield  {author} {\bibinfo {author} {\bibfnamefont {B.-N.}\ \bibnamefont
  {Lu}}, \bibinfo {author} {\bibfnamefont {J.}~\bibnamefont {Zhao}}, \bibinfo
  {author} {\bibfnamefont {E.-G.}\ \bibnamefont {Zhao}},\ and\ \bibinfo
  {author} {\bibfnamefont {S.-G.}\ \bibnamefont {Zhou}},\ }\href
  {https://doi.org/10.1103/PhysRevC.89.014323} {\bibfield  {journal} {\bibinfo
  {journal} {Phys. Rev. C}\ }\textbf {\bibinfo {volume} {89}},\ \bibinfo
  {pages} {014323} (\bibinfo {year} {2014})}\BibitemShut {NoStop}%
\bibitem [{\citenamefont {Zhao}\ \emph {et~al.}(2017)\citenamefont {Zhao},
  \citenamefont {Lu}, \citenamefont {Zhao},\ and\ \citenamefont
  {Zhou}}]{Zhao2017_PRC95-014320}%
  \BibitemOpen
  \bibfield  {author} {\bibinfo {author} {\bibfnamefont {J.}~\bibnamefont
  {Zhao}}, \bibinfo {author} {\bibfnamefont {B.-N.}\ \bibnamefont {Lu}},
  \bibinfo {author} {\bibfnamefont {E.-G.}\ \bibnamefont {Zhao}},\ and\
  \bibinfo {author} {\bibfnamefont {S.-G.}\ \bibnamefont {Zhou}},\ }\href
  {https://doi.org/10.1103/PhysRevC.95.014320} {\bibfield  {journal} {\bibinfo
  {journal} {Phys. Rev. C}\ }\textbf {\bibinfo {volume} {95}},\ \bibinfo
  {pages} {014320} (\bibinfo {year} {2017})}\BibitemShut {NoStop}%
\bibitem [{\citenamefont {Wang}\ and\ \citenamefont
  {Lu}()}]{Wang2021_MDC-RHB+AMP}%
  \BibitemOpen
  \bibinfo {author} {\bibfnamefont {K.}~\bibnamefont {Wang}}\ and\ \bibinfo
  {author} {\bibfnamefont {B.-N.}\ \bibnamefont {Lu}}\BibitemShut {NoStop}%
\bibitem [{\citenamefont {Sun}\ and\ \citenamefont
  {Zhou}(2021)}]{Sun2021_arXiv2103.10886}%
  \BibitemOpen
\bibfield  {author} {  }\bibfield  {author} {\bibinfo {author} {\bibfnamefont
  {X.-X.}\ \bibnamefont {Sun}}\ and\ \bibinfo {author} {\bibfnamefont {S.-G.}\
  \bibnamefont {Zhou}},\ }\href {https://arxiv.org/abs/2103.10886} {\bibinfo
  {title} {Rotating deformed halo nuclei and shape decoupling effects}},\
  \bibinfo {howpublished} {arXiv:2103.10886 [nucl-th]} (\bibinfo {year}
  {2021}),\ \Eprint {https://arxiv.org/abs/2103.10886v1} {2103.10886v1}
  \BibitemShut {NoStop}%
\bibitem [{\citenamefont {Chen}\ \emph {et~al.}(2012)\citenamefont {Chen},
  \citenamefont {Li}, \citenamefont {Liang},\ and\ \citenamefont
  {Meng}}]{Chen2012_PRC85-067301}%
  \BibitemOpen
  \bibfield  {author} {\bibinfo {author} {\bibfnamefont {Y.}~\bibnamefont
  {Chen}}, \bibinfo {author} {\bibfnamefont {L.}~\bibnamefont {Li}}, \bibinfo
  {author} {\bibfnamefont {H.}~\bibnamefont {Liang}},\ and\ \bibinfo {author}
  {\bibfnamefont {J.}~\bibnamefont {Meng}},\ }\href
  {https://doi.org/10.1103/PhysRevC.85.067301} {\bibfield  {journal} {\bibinfo
  {journal} {Phys. Rev. C}\ }\textbf {\bibinfo {volume} {85}},\ \bibinfo
  {pages} {067301} (\bibinfo {year} {2012})}\BibitemShut {NoStop}%
\bibitem [{\citenamefont {Kucharek}\ and\ \citenamefont
  {Ring}(1991)}]{Kucharek1991_ZPA339-23}%
  \BibitemOpen
  \bibfield  {author} {\bibinfo {author} {\bibfnamefont {H.}~\bibnamefont
  {Kucharek}}\ and\ \bibinfo {author} {\bibfnamefont {P.}~\bibnamefont
  {Ring}},\ }\href {https://doi.org/10.1007/BF01282930} {\bibfield  {journal}
  {\bibinfo  {journal} {Z. Phys. A}\ }\textbf {\bibinfo {volume} {339}},\
  \bibinfo {pages} {23} (\bibinfo {year} {1991})}\BibitemShut {NoStop}%
\bibitem [{\citenamefont {Koepf}\ and\ \citenamefont
  {Ring}(1991)}]{Koepf1991_ZPA339-81}%
  \BibitemOpen
  \bibfield  {author} {\bibinfo {author} {\bibfnamefont {W.}~\bibnamefont
  {Koepf}}\ and\ \bibinfo {author} {\bibfnamefont {P.}~\bibnamefont {Ring}},\
  }\href {https://doi.org/10.1007/BF01282936} {\bibfield  {journal} {\bibinfo
  {journal} {Z. Phys. A}\ }\textbf {\bibinfo {volume} {339}},\ \bibinfo {pages}
  {81} (\bibinfo {year} {1991})}\BibitemShut {NoStop}%
\bibitem [{\citenamefont {Blaizot}\ and\ \citenamefont
  {Ripka}(1985)}]{Blaizot1985_QTFS}%
  \BibitemOpen
  \bibfield  {author} {\bibinfo {author} {\bibfnamefont {J.~P.}\ \bibnamefont
  {Blaizot}}\ and\ \bibinfo {author} {\bibfnamefont {G.}~\bibnamefont
  {Ripka}},\ }\href@noop {} {\emph {\bibinfo {title} {Quantum Theory of Finite
  Systems}}}\ (\bibinfo  {publisher} {The MIT Press},\ \bibinfo {year}
  {1985})\BibitemShut {NoStop}%
\bibitem [{\citenamefont {Bender}\ \emph {et~al.}(2000)\citenamefont {Bender},
  \citenamefont {Rutz}, \citenamefont {Reinhard},\ and\ \citenamefont
  {Maruhn}}]{Bender2000_EPJA7-467}%
  \BibitemOpen
  \bibfield  {author} {\bibinfo {author} {\bibfnamefont {M.}~\bibnamefont
  {Bender}}, \bibinfo {author} {\bibfnamefont {K.}~\bibnamefont {Rutz}},
  \bibinfo {author} {\bibfnamefont {P.-G.}\ \bibnamefont {Reinhard}},\ and\
  \bibinfo {author} {\bibfnamefont {J.}~\bibnamefont {Maruhn}},\ }\href
  {https://doi.org/10.1007/PL00013645} {\bibfield  {journal} {\bibinfo
  {journal} {Eur. Phys. J. A}\ }\textbf {\bibinfo {volume} {7}},\ \bibinfo
  {pages} {467} (\bibinfo {year} {2000})}\BibitemShut {NoStop}%
\bibitem [{\citenamefont {Long}\ \emph {et~al.}(2004)\citenamefont {Long},
  \citenamefont {Meng}, \citenamefont {Giai},\ and\ \citenamefont
  {Zhou}}]{Long2004_PRC69-034319}%
  \BibitemOpen
  \bibfield  {author} {\bibinfo {author} {\bibfnamefont {W.}~\bibnamefont
  {Long}}, \bibinfo {author} {\bibfnamefont {J.}~\bibnamefont {Meng}}, \bibinfo
  {author} {\bibfnamefont {N.~V.}\ \bibnamefont {Giai}},\ and\ \bibinfo
  {author} {\bibfnamefont {S.-G.}\ \bibnamefont {Zhou}},\ }\href
  {https://doi.org/10.1103/PhysRevC.69.034319} {\bibfield  {journal} {\bibinfo
  {journal} {Phys. Rev. C}\ }\textbf {\bibinfo {volume} {69}},\ \bibinfo
  {pages} {034319} (\bibinfo {year} {2004})}\BibitemShut {NoStop}%
\bibitem [{\citenamefont {Peng-Wei}\ \emph {et~al.}(2009)\citenamefont
  {Peng-Wei}, \citenamefont {Bao-Yuan},\ and\ \citenamefont
  {Jie}}]{Zhao2009_CPL26-112102}%
  \BibitemOpen
  \bibfield  {author} {\bibinfo {author} {\bibfnamefont {Z.}~\bibnamefont
  {Peng-Wei}}, \bibinfo {author} {\bibfnamefont {S.}~\bibnamefont {Bao-Yuan}},\
  and\ \bibinfo {author} {\bibfnamefont {M.}~\bibnamefont {Jie}},\ }\href
  {https://doi.org/10.1088/0256-307x/26/11/112102} {\bibfield  {journal}
  {\bibinfo  {journal} {Chin. Phys. Lett.}\ }\textbf {\bibinfo {volume} {26}},\
  \bibinfo {pages} {112102} (\bibinfo {year} {2009})}\BibitemShut {NoStop}%
\bibitem [{\citenamefont {Serot}\ and\ \citenamefont
  {Walecka}(1986)}]{Serot1986_ANP16-1}%
  \BibitemOpen
  \bibfield  {author} {\bibinfo {author} {\bibfnamefont {B.}~\bibnamefont
  {Serot}}\ and\ \bibinfo {author} {\bibfnamefont {J.}~\bibnamefont
  {Walecka}},\ }\href@noop {} {\bibfield  {journal} {\bibinfo  {journal} {Adv.
  Nucl. Phys.}\ }\textbf {\bibinfo {volume} {16}},\ \bibinfo {pages} {1}
  (\bibinfo {year} {1986})}\BibitemShut {NoStop}%
\bibitem [{\citenamefont {Hara}\ and\ \citenamefont
  {Sun}(1995)}]{Hara1995_IJMPE4-637}%
  \BibitemOpen
  \bibfield  {author} {\bibinfo {author} {\bibfnamefont {K.}~\bibnamefont
  {Hara}}\ and\ \bibinfo {author} {\bibfnamefont {Y.}~\bibnamefont {Sun}},\
  }\href {https://doi.org/10.1142/S0218301395000250} {\bibfield  {journal}
  {\bibinfo  {journal} {Int. J. Mod. Phys. E}\ }\textbf {\bibinfo {volume}
  {4}},\ \bibinfo {pages} {637} (\bibinfo {year} {1995})}\BibitemShut {NoStop}%
\bibitem [{\citenamefont {Valor}\ \emph {et~al.}(2000)\citenamefont {Valor},
  \citenamefont {Heenen},\ and\ \citenamefont {Bonche}}]{Valor2000_NPA671-145}%
  \BibitemOpen
  \bibfield  {author} {\bibinfo {author} {\bibfnamefont {A.}~\bibnamefont
  {Valor}}, \bibinfo {author} {\bibfnamefont {P.-H.}\ \bibnamefont {Heenen}},\
  and\ \bibinfo {author} {\bibfnamefont {P.}~\bibnamefont {Bonche}},\ }\href
  {https://doi.org/10.1016/S0375-9474(99)00830-1} {\bibfield  {journal}
  {\bibinfo  {journal} {Nucl. Phys. A}\ }\textbf {\bibinfo {volume} {671}},\
  \bibinfo {pages} {145 } (\bibinfo {year} {2000})}\BibitemShut {NoStop}%
\bibitem [{\citenamefont {Balian}\ and\ \citenamefont
  {Brezin}(1969)}]{Balian1969_INCB164-37}%
  \BibitemOpen
  \bibfield  {author} {\bibinfo {author} {\bibfnamefont {R.}~\bibnamefont
  {Balian}}\ and\ \bibinfo {author} {\bibfnamefont {E.}~\bibnamefont
  {Brezin}},\ }\href {https://doi.org/10.1007/BF02710281} {\bibfield  {journal}
  {\bibinfo  {journal} {Il Nuovo Cimento B (1965-1970)}\ }\textbf {\bibinfo
  {volume} {64}},\ \bibinfo {pages} {37} (\bibinfo {year} {1969})}\BibitemShut
  {NoStop}%
\bibitem [{\citenamefont {Onishi}\ and\ \citenamefont
  {Yoshida}(1966)}]{Onishi1966_NP80-367}%
  \BibitemOpen
  \bibfield  {author} {\bibinfo {author} {\bibfnamefont {N.}~\bibnamefont
  {Onishi}}\ and\ \bibinfo {author} {\bibfnamefont {S.}~\bibnamefont
  {Yoshida}},\ }\href {https://doi.org/10.1016/0029-5582(66)90096-4} {\bibfield
   {journal} {\bibinfo  {journal} {Nucl. Phys.}\ }\textbf {\bibinfo {volume}
  {80}},\ \bibinfo {pages} {367 } (\bibinfo {year} {1966})}\BibitemShut
  {NoStop}%
\bibitem [{\citenamefont {Bonche}\ \emph {et~al.}(1990)\citenamefont {Bonche},
  \citenamefont {Dobaczewski}, \citenamefont {Flocard}, \citenamefont
  {Heenen},\ and\ \citenamefont {Meyer}}]{Bonche1990_NPA510-466}%
  \BibitemOpen
  \bibfield  {author} {\bibinfo {author} {\bibfnamefont {P.}~\bibnamefont
  {Bonche}}, \bibinfo {author} {\bibfnamefont {J.}~\bibnamefont {Dobaczewski}},
  \bibinfo {author} {\bibfnamefont {H.}~\bibnamefont {Flocard}}, \bibinfo
  {author} {\bibfnamefont {P.-H.}\ \bibnamefont {Heenen}},\ and\ \bibinfo
  {author} {\bibfnamefont {J.}~\bibnamefont {Meyer}},\ }\href
  {https://doi.org/10.1016/0375-9474(90)90062-Q} {\bibfield  {journal}
  {\bibinfo  {journal} {Nucl. Phys. A}\ }\textbf {\bibinfo {volume} {510}},\
  \bibinfo {pages} {466 } (\bibinfo {year} {1990})}\BibitemShut {NoStop}%
\bibitem [{\citenamefont {Rodr\'iguez-Guzm\'an}\ \emph
  {et~al.}(2002)\citenamefont {Rodr\'iguez-Guzm\'an}, \citenamefont {Egido},\
  and\ \citenamefont {Robledo}}]{Rodriguez-Guzman2002_NPA709-201}%
  \BibitemOpen
  \bibfield  {author} {\bibinfo {author} {\bibfnamefont {R.}~\bibnamefont
  {Rodr\'iguez-Guzm\'an}}, \bibinfo {author} {\bibfnamefont {J.~L.}\
  \bibnamefont {Egido}},\ and\ \bibinfo {author} {\bibfnamefont {L.~M.}\
  \bibnamefont {Robledo}},\ }\href
  {https://doi.org/10.1016/S0375-9474(02)01019-9} {\bibfield  {journal}
  {\bibinfo  {journal} {Nucl. Phys. A}\ }\textbf {\bibinfo {volume} {709}},\
  \bibinfo {pages} {201} (\bibinfo {year} {2002})}\BibitemShut {NoStop}%
\bibitem [{\citenamefont {Yao}\ \emph {et~al.}(2015{\natexlab{b}})\citenamefont
  {Yao}, \citenamefont {Bender},\ and\ \citenamefont
  {Heenen}}]{Yao2015_PRC91-024301}%
  \BibitemOpen
  \bibfield  {author} {\bibinfo {author} {\bibfnamefont {J.~M.}\ \bibnamefont
  {Yao}}, \bibinfo {author} {\bibfnamefont {M.}~\bibnamefont {Bender}},\ and\
  \bibinfo {author} {\bibfnamefont {P.-H.}\ \bibnamefont {Heenen}},\ }\href
  {https://doi.org/10.1103/PhysRevC.91.024301} {\bibfield  {journal} {\bibinfo
  {journal} {Phys. Rev. C}\ }\textbf {\bibinfo {volume} {91}},\ \bibinfo
  {pages} {024301} (\bibinfo {year} {2015}{\natexlab{b}})}\BibitemShut
  {NoStop}%
\bibitem [{\citenamefont {B{\"u}rvenich}\ \emph {et~al.}(2002)\citenamefont
  {B{\"u}rvenich}, \citenamefont {Madland}, \citenamefont {Maruhn},\ and\
  \citenamefont {Reinhard}}]{Burvenich2002_PRC65-044308}%
  \BibitemOpen
  \bibfield  {author} {\bibinfo {author} {\bibfnamefont {T.}~\bibnamefont
  {B{\"u}rvenich}}, \bibinfo {author} {\bibfnamefont {D.~G.}\ \bibnamefont
  {Madland}}, \bibinfo {author} {\bibfnamefont {J.~A.}\ \bibnamefont
  {Maruhn}},\ and\ \bibinfo {author} {\bibfnamefont {P.-G.}\ \bibnamefont
  {Reinhard}},\ }\href {https://doi.org/10.1103/PhysRevC.65.044308} {\bibfield
  {journal} {\bibinfo  {journal} {Phys. Rev. C}\ }\textbf {\bibinfo {volume}
  {65}},\ \bibinfo {pages} {044308} (\bibinfo {year} {2002})}\BibitemShut
  {NoStop}%
\bibitem [{\citenamefont {Yao}\ \emph {et~al.}(2011)\citenamefont {Yao},
  \citenamefont {Mei}, \citenamefont {Chen}, \citenamefont {Meng},
  \citenamefont {Ring},\ and\ \citenamefont {Vretenar}}]{Yao2011_PRC83-014308}%
  \BibitemOpen
  \bibfield  {author} {\bibinfo {author} {\bibfnamefont {J.~M.}\ \bibnamefont
  {Yao}}, \bibinfo {author} {\bibfnamefont {H.}~\bibnamefont {Mei}}, \bibinfo
  {author} {\bibfnamefont {H.}~\bibnamefont {Chen}}, \bibinfo {author}
  {\bibfnamefont {J.}~\bibnamefont {Meng}}, \bibinfo {author} {\bibfnamefont
  {P.}~\bibnamefont {Ring}},\ and\ \bibinfo {author} {\bibfnamefont
  {D.}~\bibnamefont {Vretenar}},\ }\href
  {https://doi.org/10.1103/PhysRevC.83.014308} {\bibfield  {journal} {\bibinfo
  {journal} {Phys. Rev. C}\ }\textbf {\bibinfo {volume} {83}},\ \bibinfo
  {pages} {014308} (\bibinfo {year} {2011})}\BibitemShut {NoStop}%
\bibitem [{\citenamefont {Rodr\'{i}guez-Guzm\'an}\ \emph
  {et~al.}(2000{\natexlab{b}})\citenamefont {Rodr\'{i}guez-Guzm\'an},
  \citenamefont {Egido},\ and\ \citenamefont
  {Robledo}}]{Rodriguez-Guzman2000_PLB474-15}%
  \BibitemOpen
  \bibfield  {author} {\bibinfo {author} {\bibfnamefont {R.~R.}\ \bibnamefont
  {Rodr\'{i}guez-Guzm\'an}}, \bibinfo {author} {\bibfnamefont {J.}~\bibnamefont
  {Egido}},\ and\ \bibinfo {author} {\bibfnamefont {L.}~\bibnamefont
  {Robledo}},\ }\href {https://doi.org/10.1016/S0370-2693(00)00015-0}
  {\bibfield  {journal} {\bibinfo  {journal} {Phys. Lett. B}\ }\textbf
  {\bibinfo {volume} {474}},\ \bibinfo {pages} {15 } (\bibinfo {year}
  {2000}{\natexlab{b}})}\BibitemShut {NoStop}%
\bibitem [{\citenamefont {Baumann}\ \emph {et~al.}(2007)\citenamefont
  {Baumann}, \citenamefont {Amthor}, \citenamefont {Bazin}, \citenamefont
  {Brown}, \citenamefont {III}, \citenamefont {Gade}, \citenamefont {Ginter},
  \citenamefont {Hausmann}, \citenamefont {Mato{\v{s}}}, \citenamefont
  {Morrissey}, \citenamefont {Portillo}, \citenamefont {Schiller},
  \citenamefont {Sherrill}, \citenamefont {Stolz}, \citenamefont {Tarasov},\
  and\ \citenamefont {Thoennessen}}]{Baumann2007_Nature449-1022}%
  \BibitemOpen
  \bibfield  {author} {\bibinfo {author} {\bibfnamefont {T.}~\bibnamefont
  {Baumann}}, \bibinfo {author} {\bibfnamefont {A.~M.}\ \bibnamefont {Amthor}},
  \bibinfo {author} {\bibfnamefont {D.}~\bibnamefont {Bazin}}, \bibinfo
  {author} {\bibfnamefont {B.~A.}\ \bibnamefont {Brown}}, \bibinfo {author}
  {\bibfnamefont {C.~M.~F.}\ \bibnamefont {III}}, \bibinfo {author}
  {\bibfnamefont {A.}~\bibnamefont {Gade}}, \bibinfo {author} {\bibfnamefont
  {T.~N.}\ \bibnamefont {Ginter}}, \bibinfo {author} {\bibfnamefont
  {M.}~\bibnamefont {Hausmann}}, \bibinfo {author} {\bibfnamefont
  {M.}~\bibnamefont {Mato{\v{s}}}}, \bibinfo {author} {\bibfnamefont {D.~J.}\
  \bibnamefont {Morrissey}}, \bibinfo {author} {\bibfnamefont {M.}~\bibnamefont
  {Portillo}}, \bibinfo {author} {\bibfnamefont {A.}~\bibnamefont {Schiller}},
  \bibinfo {author} {\bibfnamefont {B.~M.}\ \bibnamefont {Sherrill}}, \bibinfo
  {author} {\bibfnamefont {A.}~\bibnamefont {Stolz}}, \bibinfo {author}
  {\bibfnamefont {O.~B.}\ \bibnamefont {Tarasov}},\ and\ \bibinfo {author}
  {\bibfnamefont {M.}~\bibnamefont {Thoennessen}},\ }\href
  {https://doi.org/10.1038/nature06213} {\bibfield  {journal} {\bibinfo
  {journal} {Nature}\ }\textbf {\bibinfo {volume} {449}},\ \bibinfo {pages}
  {1022} (\bibinfo {year} {2007})}\BibitemShut {NoStop}%
\bibitem [{\citenamefont {Erler}\ \emph {et~al.}(2012)\citenamefont {Erler},
  \citenamefont {Birge}, \citenamefont {Kortelainen}, \citenamefont
  {Nazarewicz}, \citenamefont {Olsen}, \citenamefont {Perhac},\ and\
  \citenamefont {Stoitsov}}]{Erler2012_Nature486-509}%
  \BibitemOpen
  \bibfield  {author} {\bibinfo {author} {\bibfnamefont {J.}~\bibnamefont
  {Erler}}, \bibinfo {author} {\bibfnamefont {N.}~\bibnamefont {Birge}},
  \bibinfo {author} {\bibfnamefont {M.}~\bibnamefont {Kortelainen}}, \bibinfo
  {author} {\bibfnamefont {W.}~\bibnamefont {Nazarewicz}}, \bibinfo {author}
  {\bibfnamefont {E.}~\bibnamefont {Olsen}}, \bibinfo {author} {\bibfnamefont
  {A.~M.}\ \bibnamefont {Perhac}},\ and\ \bibinfo {author} {\bibfnamefont
  {M.}~\bibnamefont {Stoitsov}},\ }\href {https://doi.org/10.1038/nature11188}
  {\bibfield  {journal} {\bibinfo  {journal} {Nature}\ }\textbf {\bibinfo
  {volume} {486}},\ \bibinfo {pages} {509} (\bibinfo {year}
  {2012})}\BibitemShut {NoStop}%
\bibitem [{\citenamefont {Chai}\ \emph {et~al.}(2020)\citenamefont {Chai},
  \citenamefont {Pei}, \citenamefont {Fei},\ and\ \citenamefont
  {Guan}}]{Chai2020_PRC102-014312}%
  \BibitemOpen
  \bibfield  {author} {\bibinfo {author} {\bibfnamefont {Q.~Z.}\ \bibnamefont
  {Chai}}, \bibinfo {author} {\bibfnamefont {J.~C.}\ \bibnamefont {Pei}},
  \bibinfo {author} {\bibfnamefont {N.}~\bibnamefont {Fei}},\ and\ \bibinfo
  {author} {\bibfnamefont {D.~W.}\ \bibnamefont {Guan}},\ }\href
  {https://doi.org/10.1103/PhysRevC.102.014312} {\bibfield  {journal} {\bibinfo
   {journal} {Phys. Rev. C}\ }\textbf {\bibinfo {volume} {102}},\ \bibinfo
  {pages} {014312} (\bibinfo {year} {2020})}\BibitemShut {NoStop}%
\bibitem [{\citenamefont {Tsunoda}\ \emph {et~al.}(2020)\citenamefont
  {Tsunoda}, \citenamefont {Otsuka}, \citenamefont {Takayanagi}, \citenamefont
  {Shimizu}, \citenamefont {Suzuki}, \citenamefont {Utsuno}, \citenamefont
  {Yoshida},\ and\ \citenamefont {Ueno}}]{Tsunoda2020_Nature587-66}%
  \BibitemOpen
  \bibfield  {author} {\bibinfo {author} {\bibfnamefont {N.}~\bibnamefont
  {Tsunoda}}, \bibinfo {author} {\bibfnamefont {T.}~\bibnamefont {Otsuka}},
  \bibinfo {author} {\bibfnamefont {K.}~\bibnamefont {Takayanagi}}, \bibinfo
  {author} {\bibfnamefont {N.}~\bibnamefont {Shimizu}}, \bibinfo {author}
  {\bibfnamefont {T.}~\bibnamefont {Suzuki}}, \bibinfo {author} {\bibfnamefont
  {Y.}~\bibnamefont {Utsuno}}, \bibinfo {author} {\bibfnamefont
  {S.}~\bibnamefont {Yoshida}},\ and\ \bibinfo {author} {\bibfnamefont
  {H.}~\bibnamefont {Ueno}},\ }\href
  {https://doi.org/10.1038/s41586-020-2848-x} {\bibfield  {journal} {\bibinfo
  {journal} {Nature}\ }\textbf {\bibinfo {volume} {587}},\ \bibinfo {pages}
  {66} (\bibinfo {year} {2020})}\BibitemShut {NoStop}%
\bibitem [{\citenamefont {Stroberg}\ \emph {et~al.}(2021)\citenamefont
  {Stroberg}, \citenamefont {Holt}, \citenamefont {Schwenk},\ and\
  \citenamefont {Simonis}}]{Stroberg2021_PRL126-022501}%
  \BibitemOpen
  \bibfield  {author} {\bibinfo {author} {\bibfnamefont {S.~R.}\ \bibnamefont
  {Stroberg}}, \bibinfo {author} {\bibfnamefont {J.~D.}\ \bibnamefont {Holt}},
  \bibinfo {author} {\bibfnamefont {A.}~\bibnamefont {Schwenk}},\ and\ \bibinfo
  {author} {\bibfnamefont {J.}~\bibnamefont {Simonis}},\ }\href
  {https://doi.org/10.1103/PhysRevLett.126.022501} {\bibfield  {journal}
  {\bibinfo  {journal} {Phys. Rev. Lett.}\ }\textbf {\bibinfo {volume} {126}},\
  \bibinfo {pages} {022501} (\bibinfo {year} {2021})}\BibitemShut {NoStop}%
\bibitem [{\citenamefont {Randhawa}\ \emph {et~al.}(2019)\citenamefont
  {Randhawa}, \citenamefont {Kanungo}, \citenamefont {Holl}, \citenamefont
  {Holt}, \citenamefont {Navr\'atil}, \citenamefont {Stroberg}, \citenamefont
  {Hagen}, \citenamefont {Jansen}, \citenamefont {Alcorta}, \citenamefont
  {Andreoiu}, \citenamefont {Barnes}, \citenamefont {Burbadge}, \citenamefont
  {Burke}, \citenamefont {Chen}, \citenamefont {Chester}, \citenamefont
  {Christian}, \citenamefont {Cruz}, \citenamefont {Davids}, \citenamefont
  {Even}, \citenamefont {Hackman}, \citenamefont {Henderson}, \citenamefont
  {Ishimoto}, \citenamefont {Jassal}, \citenamefont {Kaur}, \citenamefont
  {Keefe}, \citenamefont {Kisliuk}, \citenamefont {Kr\"ucken}, \citenamefont
  {Liang}, \citenamefont {Lighthall}, \citenamefont {McGee}, \citenamefont
  {Measures}, \citenamefont {Moukaddam}, \citenamefont {Padilla-Rodal},
  \citenamefont {Shotter}, \citenamefont {Thompson}, \citenamefont {Turko},
  \citenamefont {Williams},\ and\ \citenamefont
  {Workman}}]{Randhawa2019_PRC99-021301R}%
  \BibitemOpen
  \bibfield  {author} {\bibinfo {author} {\bibfnamefont {J.~S.}\ \bibnamefont
  {Randhawa}}, \bibinfo {author} {\bibfnamefont {R.}~\bibnamefont {Kanungo}},
  \bibinfo {author} {\bibfnamefont {M.}~\bibnamefont {Holl}}, \bibinfo {author}
  {\bibfnamefont {J.~D.}\ \bibnamefont {Holt}}, \bibinfo {author}
  {\bibfnamefont {P.}~\bibnamefont {Navr\'atil}}, \bibinfo {author}
  {\bibfnamefont {S.~R.}\ \bibnamefont {Stroberg}}, \bibinfo {author}
  {\bibfnamefont {G.}~\bibnamefont {Hagen}}, \bibinfo {author} {\bibfnamefont
  {G.~R.}\ \bibnamefont {Jansen}}, \bibinfo {author} {\bibfnamefont
  {M.}~\bibnamefont {Alcorta}}, \bibinfo {author} {\bibfnamefont
  {C.}~\bibnamefont {Andreoiu}}, \bibinfo {author} {\bibfnamefont
  {C.}~\bibnamefont {Barnes}}, \bibinfo {author} {\bibfnamefont
  {C.}~\bibnamefont {Burbadge}}, \bibinfo {author} {\bibfnamefont
  {D.}~\bibnamefont {Burke}}, \bibinfo {author} {\bibfnamefont {A.~A.}\
  \bibnamefont {Chen}}, \bibinfo {author} {\bibfnamefont {A.}~\bibnamefont
  {Chester}}, \bibinfo {author} {\bibfnamefont {G.}~\bibnamefont {Christian}},
  \bibinfo {author} {\bibfnamefont {S.}~\bibnamefont {Cruz}}, \bibinfo {author}
  {\bibfnamefont {B.}~\bibnamefont {Davids}}, \bibinfo {author} {\bibfnamefont
  {J.}~\bibnamefont {Even}}, \bibinfo {author} {\bibfnamefont {G.}~\bibnamefont
  {Hackman}}, \bibinfo {author} {\bibfnamefont {J.}~\bibnamefont {Henderson}},
  \bibinfo {author} {\bibfnamefont {S.}~\bibnamefont {Ishimoto}}, \bibinfo
  {author} {\bibfnamefont {P.}~\bibnamefont {Jassal}}, \bibinfo {author}
  {\bibfnamefont {S.}~\bibnamefont {Kaur}}, \bibinfo {author} {\bibfnamefont
  {M.}~\bibnamefont {Keefe}}, \bibinfo {author} {\bibfnamefont
  {D.}~\bibnamefont {Kisliuk}}, \bibinfo {author} {\bibfnamefont
  {R.}~\bibnamefont {Kr\"ucken}}, \bibinfo {author} {\bibfnamefont
  {J.}~\bibnamefont {Liang}}, \bibinfo {author} {\bibfnamefont
  {J.}~\bibnamefont {Lighthall}}, \bibinfo {author} {\bibfnamefont
  {E.}~\bibnamefont {McGee}}, \bibinfo {author} {\bibfnamefont
  {J.}~\bibnamefont {Measures}}, \bibinfo {author} {\bibfnamefont
  {M.}~\bibnamefont {Moukaddam}}, \bibinfo {author} {\bibfnamefont
  {E.}~\bibnamefont {Padilla-Rodal}}, \bibinfo {author} {\bibfnamefont
  {A.}~\bibnamefont {Shotter}}, \bibinfo {author} {\bibfnamefont {I.~J.}\
  \bibnamefont {Thompson}}, \bibinfo {author} {\bibfnamefont {J.}~\bibnamefont
  {Turko}}, \bibinfo {author} {\bibfnamefont {M.}~\bibnamefont {Williams}},\
  and\ \bibinfo {author} {\bibfnamefont {O.}~\bibnamefont {Workman}},\ }\href
  {https://doi.org/10.1103/PhysRevC.99.021301} {\bibfield  {journal} {\bibinfo
  {journal} {Phys. Rev. C}\ }\textbf {\bibinfo {volume} {99}},\ \bibinfo
  {pages} {021301(R)} (\bibinfo {year} {2019})}\BibitemShut {NoStop}%
\bibitem [{\citenamefont {Ebran}\ \emph {et~al.}(2017)\citenamefont {Ebran},
  \citenamefont {Khan}, \citenamefont {Nik{\v{s}}i{\'{c}}},\ and\ \citenamefont
  {Vretenar}}]{Ebran2017_JPG44-103001}%
  \BibitemOpen
  \bibfield  {author} {\bibinfo {author} {\bibfnamefont {J.-P.}\ \bibnamefont
  {Ebran}}, \bibinfo {author} {\bibfnamefont {E.}~\bibnamefont {Khan}},
  \bibinfo {author} {\bibfnamefont {T.}~\bibnamefont {Nik{\v{s}}i{\'{c}}}},\
  and\ \bibinfo {author} {\bibfnamefont {D.}~\bibnamefont {Vretenar}},\ }\href
  {https://doi.org/10.1088/1361-6471/aa809b} {\bibfield  {journal} {\bibinfo
  {journal} {J. Phys. G: Nucl. Part. Phys.}\ }\textbf {\bibinfo {volume}
  {44}},\ \bibinfo {pages} {103001} (\bibinfo {year} {2017})}\BibitemShut
  {NoStop}%
\bibitem [{\citenamefont {Tohsaki}\ and\ \citenamefont
  {Itagaki}(2018)}]{Tohsaki2018_PRC97-011301R}%
  \BibitemOpen
  \bibfield  {author} {\bibinfo {author} {\bibfnamefont {A.}~\bibnamefont
  {Tohsaki}}\ and\ \bibinfo {author} {\bibfnamefont {N.}~\bibnamefont
  {Itagaki}},\ }\href {https://doi.org/10.1103/PhysRevC.97.011301} {\bibfield
  {journal} {\bibinfo  {journal} {Phys. Rev. C}\ }\textbf {\bibinfo {volume}
  {97}},\ \bibinfo {pages} {011301(R)} (\bibinfo {year} {2018})}\BibitemShut
  {NoStop}%
\bibitem [{\citenamefont {Motobayashi}\ \emph {et~al.}(1995)\citenamefont
  {Motobayashi}, \citenamefont {Ikeda}, \citenamefont {Ieki}, \citenamefont
  {Inoue}, \citenamefont {Iwasa}, \citenamefont {Kikuchi}, \citenamefont
  {Kurokawa}, \citenamefont {Moriya}, \citenamefont {Ogawa}, \citenamefont
  {Murakami}, \citenamefont {Shimoura}, \citenamefont {Yanagisawa},
  \citenamefont {Nakamura}, \citenamefont {Watanabe}, \citenamefont {Ishihara},
  \citenamefont {Teranishi}, \citenamefont {Okuno},\ and\ \citenamefont
  {Casten}}]{Motobayashi1995_PLB346-9}%
  \BibitemOpen
  \bibfield  {author} {\bibinfo {author} {\bibfnamefont {T.}~\bibnamefont
  {Motobayashi}}, \bibinfo {author} {\bibfnamefont {Y.}~\bibnamefont {Ikeda}},
  \bibinfo {author} {\bibfnamefont {K.}~\bibnamefont {Ieki}}, \bibinfo {author}
  {\bibfnamefont {M.}~\bibnamefont {Inoue}}, \bibinfo {author} {\bibfnamefont
  {N.}~\bibnamefont {Iwasa}}, \bibinfo {author} {\bibfnamefont
  {T.}~\bibnamefont {Kikuchi}}, \bibinfo {author} {\bibfnamefont
  {M.}~\bibnamefont {Kurokawa}}, \bibinfo {author} {\bibfnamefont
  {S.}~\bibnamefont {Moriya}}, \bibinfo {author} {\bibfnamefont
  {S.}~\bibnamefont {Ogawa}}, \bibinfo {author} {\bibfnamefont
  {H.}~\bibnamefont {Murakami}}, \bibinfo {author} {\bibfnamefont
  {S.}~\bibnamefont {Shimoura}}, \bibinfo {author} {\bibfnamefont
  {Y.}~\bibnamefont {Yanagisawa}}, \bibinfo {author} {\bibfnamefont
  {T.}~\bibnamefont {Nakamura}}, \bibinfo {author} {\bibfnamefont
  {Y.}~\bibnamefont {Watanabe}}, \bibinfo {author} {\bibfnamefont
  {M.}~\bibnamefont {Ishihara}}, \bibinfo {author} {\bibfnamefont
  {T.}~\bibnamefont {Teranishi}}, \bibinfo {author} {\bibfnamefont
  {H.}~\bibnamefont {Okuno}},\ and\ \bibinfo {author} {\bibfnamefont
  {R.}~\bibnamefont {Casten}},\ }\href
  {https://doi.org/10.1016/0370-2693(95)00012-A} {\bibfield  {journal}
  {\bibinfo  {journal} {Phys. Lett. B}\ }\textbf {\bibinfo {volume} {346}},\
  \bibinfo {pages} {9 } (\bibinfo {year} {1995})}\BibitemShut {NoStop}%
\bibitem [{\citenamefont {Takechi}\ \emph
  {et~al.}(2014{\natexlab{a}})\citenamefont {Takechi}, \citenamefont {Suzuki},
  \citenamefont {Nishimura}, \citenamefont {Fukuda}, \citenamefont {Ohtsubo},
  \citenamefont {Nagashima}, \citenamefont {Suzuki}, \citenamefont {Yamaguchi},
  \citenamefont {Ozawa}, \citenamefont {Moriguchi}, \citenamefont {Ohishi},
  \citenamefont {Sumikama}, \citenamefont {Geissel}, \citenamefont {Aoi},
  \citenamefont {Chen}, \citenamefont {Fang}, \citenamefont {Fukuda},
  \citenamefont {Fukuoka}, \citenamefont {Furuki}, \citenamefont {Inabe},
  \citenamefont {Ishibashi}, \citenamefont {Itoh}, \citenamefont {Izumikawa},
  \citenamefont {Kameda}, \citenamefont {Kubo}, \citenamefont {Lantz},
  \citenamefont {Lee}, \citenamefont {Ma}, \citenamefont {Matsuta},
  \citenamefont {Mihara}, \citenamefont {Momota}, \citenamefont {Nagae},
  \citenamefont {Nishikiori}, \citenamefont {Niwa}, \citenamefont {Ohnishi},
  \citenamefont {Okumura}, \citenamefont {Ohtake}, \citenamefont {Ogura},
  \citenamefont {Sakurai}, \citenamefont {Sato}, \citenamefont {Shimbara},
  \citenamefont {Suzuki}, \citenamefont {Takeda}, \citenamefont {Takeuchi},
  \citenamefont {Tanaka}, \citenamefont {Tanaka}, \citenamefont {Uenishi},
  \citenamefont {Winkler}, \citenamefont {Yanagisawa}, \citenamefont
  {Watanabe}, \citenamefont {Minomo}, \citenamefont {Tagami}, \citenamefont
  {Shimada}, \citenamefont {Kimura}, \citenamefont {Matsumoto}, \citenamefont
  {Shimizu},\ and\ \citenamefont {Yahiro}}]{Takechi2014_PRC90-061305R}%
  \BibitemOpen
  \bibfield  {author} {\bibinfo {author} {\bibfnamefont {M.}~\bibnamefont
  {Takechi}}, \bibinfo {author} {\bibfnamefont {S.}~\bibnamefont {Suzuki}},
  \bibinfo {author} {\bibfnamefont {D.}~\bibnamefont {Nishimura}}, \bibinfo
  {author} {\bibfnamefont {M.}~\bibnamefont {Fukuda}}, \bibinfo {author}
  {\bibfnamefont {T.}~\bibnamefont {Ohtsubo}}, \bibinfo {author} {\bibfnamefont
  {M.}~\bibnamefont {Nagashima}}, \bibinfo {author} {\bibfnamefont
  {T.}~\bibnamefont {Suzuki}}, \bibinfo {author} {\bibfnamefont
  {T.}~\bibnamefont {Yamaguchi}}, \bibinfo {author} {\bibfnamefont
  {A.}~\bibnamefont {Ozawa}}, \bibinfo {author} {\bibfnamefont
  {T.}~\bibnamefont {Moriguchi}}, \bibinfo {author} {\bibfnamefont
  {H.}~\bibnamefont {Ohishi}}, \bibinfo {author} {\bibfnamefont
  {T.}~\bibnamefont {Sumikama}}, \bibinfo {author} {\bibfnamefont
  {H.}~\bibnamefont {Geissel}}, \bibinfo {author} {\bibfnamefont
  {N.}~\bibnamefont {Aoi}}, \bibinfo {author} {\bibfnamefont {R.-J.}\
  \bibnamefont {Chen}}, \bibinfo {author} {\bibfnamefont {D.-Q.}\ \bibnamefont
  {Fang}}, \bibinfo {author} {\bibfnamefont {N.}~\bibnamefont {Fukuda}},
  \bibinfo {author} {\bibfnamefont {S.}~\bibnamefont {Fukuoka}}, \bibinfo
  {author} {\bibfnamefont {H.}~\bibnamefont {Furuki}}, \bibinfo {author}
  {\bibfnamefont {N.}~\bibnamefont {Inabe}}, \bibinfo {author} {\bibfnamefont
  {Y.}~\bibnamefont {Ishibashi}}, \bibinfo {author} {\bibfnamefont
  {T.}~\bibnamefont {Itoh}}, \bibinfo {author} {\bibfnamefont {T.}~\bibnamefont
  {Izumikawa}}, \bibinfo {author} {\bibfnamefont {D.}~\bibnamefont {Kameda}},
  \bibinfo {author} {\bibfnamefont {T.}~\bibnamefont {Kubo}}, \bibinfo {author}
  {\bibfnamefont {M.}~\bibnamefont {Lantz}}, \bibinfo {author} {\bibfnamefont
  {C.~S.}\ \bibnamefont {Lee}}, \bibinfo {author} {\bibfnamefont {Y.-G.}\
  \bibnamefont {Ma}}, \bibinfo {author} {\bibfnamefont {K.}~\bibnamefont
  {Matsuta}}, \bibinfo {author} {\bibfnamefont {M.}~\bibnamefont {Mihara}},
  \bibinfo {author} {\bibfnamefont {S.}~\bibnamefont {Momota}}, \bibinfo
  {author} {\bibfnamefont {D.}~\bibnamefont {Nagae}}, \bibinfo {author}
  {\bibfnamefont {R.}~\bibnamefont {Nishikiori}}, \bibinfo {author}
  {\bibfnamefont {T.}~\bibnamefont {Niwa}}, \bibinfo {author} {\bibfnamefont
  {T.}~\bibnamefont {Ohnishi}}, \bibinfo {author} {\bibfnamefont
  {K.}~\bibnamefont {Okumura}}, \bibinfo {author} {\bibfnamefont
  {M.}~\bibnamefont {Ohtake}}, \bibinfo {author} {\bibfnamefont
  {T.}~\bibnamefont {Ogura}}, \bibinfo {author} {\bibfnamefont
  {H.}~\bibnamefont {Sakurai}}, \bibinfo {author} {\bibfnamefont
  {K.}~\bibnamefont {Sato}}, \bibinfo {author} {\bibfnamefont {Y.}~\bibnamefont
  {Shimbara}}, \bibinfo {author} {\bibfnamefont {H.}~\bibnamefont {Suzuki}},
  \bibinfo {author} {\bibfnamefont {H.}~\bibnamefont {Takeda}}, \bibinfo
  {author} {\bibfnamefont {S.}~\bibnamefont {Takeuchi}}, \bibinfo {author}
  {\bibfnamefont {K.}~\bibnamefont {Tanaka}}, \bibinfo {author} {\bibfnamefont
  {M.}~\bibnamefont {Tanaka}}, \bibinfo {author} {\bibfnamefont
  {H.}~\bibnamefont {Uenishi}}, \bibinfo {author} {\bibfnamefont
  {M.}~\bibnamefont {Winkler}}, \bibinfo {author} {\bibfnamefont
  {Y.}~\bibnamefont {Yanagisawa}}, \bibinfo {author} {\bibfnamefont
  {S.}~\bibnamefont {Watanabe}}, \bibinfo {author} {\bibfnamefont
  {K.}~\bibnamefont {Minomo}}, \bibinfo {author} {\bibfnamefont
  {S.}~\bibnamefont {Tagami}}, \bibinfo {author} {\bibfnamefont
  {M.}~\bibnamefont {Shimada}}, \bibinfo {author} {\bibfnamefont
  {M.}~\bibnamefont {Kimura}}, \bibinfo {author} {\bibfnamefont
  {T.}~\bibnamefont {Matsumoto}}, \bibinfo {author} {\bibfnamefont {Y.~R.}\
  \bibnamefont {Shimizu}},\ and\ \bibinfo {author} {\bibfnamefont
  {M.}~\bibnamefont {Yahiro}},\ }\href
  {https://doi.org/10.1103/PhysRevC.90.061305} {\bibfield  {journal} {\bibinfo
  {journal} {Phys. Rev. C}\ }\textbf {\bibinfo {volume} {90}},\ \bibinfo
  {pages} {061305(R)} (\bibinfo {year} {2014}{\natexlab{a}})}\BibitemShut
  {NoStop}%
\bibitem [{\citenamefont {Watanabe}\ \emph {et~al.}(2014)\citenamefont
  {Watanabe}, \citenamefont {Minomo}, \citenamefont {Shimada}, \citenamefont
  {Tagami}, \citenamefont {Kimura}, \citenamefont {Takechi}, \citenamefont
  {Fukuda}, \citenamefont {Nishimura}, \citenamefont {Suzuki}, \citenamefont
  {Matsumoto}, \citenamefont {Shimizu},\ and\ \citenamefont
  {Yahiro}}]{Watanabe2014_PRC89-044610}%
  \BibitemOpen
  \bibfield  {author} {\bibinfo {author} {\bibfnamefont {S.}~\bibnamefont
  {Watanabe}}, \bibinfo {author} {\bibfnamefont {K.}~\bibnamefont {Minomo}},
  \bibinfo {author} {\bibfnamefont {M.}~\bibnamefont {Shimada}}, \bibinfo
  {author} {\bibfnamefont {S.}~\bibnamefont {Tagami}}, \bibinfo {author}
  {\bibfnamefont {M.}~\bibnamefont {Kimura}}, \bibinfo {author} {\bibfnamefont
  {M.}~\bibnamefont {Takechi}}, \bibinfo {author} {\bibfnamefont
  {M.}~\bibnamefont {Fukuda}}, \bibinfo {author} {\bibfnamefont
  {D.}~\bibnamefont {Nishimura}}, \bibinfo {author} {\bibfnamefont
  {T.}~\bibnamefont {Suzuki}}, \bibinfo {author} {\bibfnamefont
  {T.}~\bibnamefont {Matsumoto}}, \bibinfo {author} {\bibfnamefont {Y.~R.}\
  \bibnamefont {Shimizu}},\ and\ \bibinfo {author} {\bibfnamefont
  {M.}~\bibnamefont {Yahiro}},\ }\href
  {https://doi.org/10.1103/PhysRevC.89.044610} {\bibfield  {journal} {\bibinfo
  {journal} {Phys. Rev. C}\ }\textbf {\bibinfo {volume} {89}},\ \bibinfo
  {pages} {044610} (\bibinfo {year} {2014})}\BibitemShut {NoStop}%
\bibitem [{\citenamefont {Kondev}\ \emph {et~al.}(2021)\citenamefont {Kondev},
  \citenamefont {Wang}, \citenamefont {Huang}, \citenamefont {Naimi},\ and\
  \citenamefont {Audi}}]{Kondev2021_ChinPC45-030001}%
  \BibitemOpen
  \bibfield  {author} {\bibinfo {author} {\bibfnamefont {F.}~\bibnamefont
  {Kondev}}, \bibinfo {author} {\bibfnamefont {M.}~\bibnamefont {Wang}},
  \bibinfo {author} {\bibfnamefont {W.}~\bibnamefont {Huang}}, \bibinfo
  {author} {\bibfnamefont {S.}~\bibnamefont {Naimi}},\ and\ \bibinfo {author}
  {\bibfnamefont {G.}~\bibnamefont {Audi}},\ }\href
  {https://doi.org/10.1088/1674-1137/abddae} {\bibfield  {journal} {\bibinfo
  {journal} {Chin. Phys. C}\ }\textbf {\bibinfo {volume} {45}},\ \bibinfo
  {pages} {030001} (\bibinfo {year} {2021})}\BibitemShut {NoStop}%
\bibitem [{\citenamefont {Huang}\ \emph {et~al.}(2021)\citenamefont {Huang},
  \citenamefont {Wang}, \citenamefont {Kondev}, \citenamefont {Audi},\ and\
  \citenamefont {Naimi}}]{Huang2021_ChinPC45-030002}%
  \BibitemOpen
  \bibfield  {author} {\bibinfo {author} {\bibfnamefont {W.}~\bibnamefont
  {Huang}}, \bibinfo {author} {\bibfnamefont {M.}~\bibnamefont {Wang}},
  \bibinfo {author} {\bibfnamefont {F.}~\bibnamefont {Kondev}}, \bibinfo
  {author} {\bibfnamefont {G.}~\bibnamefont {Audi}},\ and\ \bibinfo {author}
  {\bibfnamefont {S.}~\bibnamefont {Naimi}},\ }\href
  {https://doi.org/10.1088/1674-1137/abddb0} {\bibfield  {journal} {\bibinfo
  {journal} {Chin. Phys. C}\ }\textbf {\bibinfo {volume} {45}},\ \bibinfo
  {pages} {030002} (\bibinfo {year} {2021})}\BibitemShut {NoStop}%
\bibitem [{\citenamefont {Wang}\ \emph {et~al.}(2021)\citenamefont {Wang},
  \citenamefont {Huang}, \citenamefont {Kondev}, \citenamefont {Audi},\ and\
  \citenamefont {Naimi}}]{Wang2021_ChinPC45-030003}%
  \BibitemOpen
  \bibfield  {author} {\bibinfo {author} {\bibfnamefont {M.}~\bibnamefont
  {Wang}}, \bibinfo {author} {\bibfnamefont {W.}~\bibnamefont {Huang}},
  \bibinfo {author} {\bibfnamefont {F.}~\bibnamefont {Kondev}}, \bibinfo
  {author} {\bibfnamefont {G.}~\bibnamefont {Audi}},\ and\ \bibinfo {author}
  {\bibfnamefont {S.}~\bibnamefont {Naimi}},\ }\href
  {https://doi.org/10.1088/1674-1137/abddaf} {\bibfield  {journal} {\bibinfo
  {journal} {Chin. Phys. C}\ }\textbf {\bibinfo {volume} {45}},\ \bibinfo
  {pages} {030003} (\bibinfo {year} {2021})}\BibitemShut {NoStop}%
\bibitem [{\citenamefont {Crawford}\ \emph {et~al.}(2019)\citenamefont
  {Crawford}, \citenamefont {Fallon}, \citenamefont {Macchiavelli},
  \citenamefont {Doornenbal}, \citenamefont {Aoi}, \citenamefont {Browne},
  \citenamefont {Campbell}, \citenamefont {Chen}, \citenamefont {Clark},
  \citenamefont {Cort\'es}, \citenamefont {Cromaz}, \citenamefont {Ideguchi},
  \citenamefont {Jones}, \citenamefont {Kanungo}, \citenamefont {MacCormick},
  \citenamefont {Momiyama}, \citenamefont {Murray}, \citenamefont {Niikura},
  \citenamefont {Paschalis}, \citenamefont {Petri}, \citenamefont {Sakurai},
  \citenamefont {Salathe}, \citenamefont {Schrock}, \citenamefont
  {Steppenbeck}, \citenamefont {Takeuchi}, \citenamefont {Tanaka},
  \citenamefont {Taniuchi}, \citenamefont {Wang},\ and\ \citenamefont
  {Wimmer}}]{Crawford2019_PRL122-052501}%
  \BibitemOpen
  \bibfield  {author} {\bibinfo {author} {\bibfnamefont {H.~L.}\ \bibnamefont
  {Crawford}}, \bibinfo {author} {\bibfnamefont {P.}~\bibnamefont {Fallon}},
  \bibinfo {author} {\bibfnamefont {A.~O.}\ \bibnamefont {Macchiavelli}},
  \bibinfo {author} {\bibfnamefont {P.}~\bibnamefont {Doornenbal}}, \bibinfo
  {author} {\bibfnamefont {N.}~\bibnamefont {Aoi}}, \bibinfo {author}
  {\bibfnamefont {F.}~\bibnamefont {Browne}}, \bibinfo {author} {\bibfnamefont
  {C.~M.}\ \bibnamefont {Campbell}}, \bibinfo {author} {\bibfnamefont
  {S.}~\bibnamefont {Chen}}, \bibinfo {author} {\bibfnamefont {R.~M.}\
  \bibnamefont {Clark}}, \bibinfo {author} {\bibfnamefont {M.~L.}\ \bibnamefont
  {Cort\'es}}, \bibinfo {author} {\bibfnamefont {M.}~\bibnamefont {Cromaz}},
  \bibinfo {author} {\bibfnamefont {E.}~\bibnamefont {Ideguchi}}, \bibinfo
  {author} {\bibfnamefont {M.~D.}\ \bibnamefont {Jones}}, \bibinfo {author}
  {\bibfnamefont {R.}~\bibnamefont {Kanungo}}, \bibinfo {author} {\bibfnamefont
  {M.}~\bibnamefont {MacCormick}}, \bibinfo {author} {\bibfnamefont
  {S.}~\bibnamefont {Momiyama}}, \bibinfo {author} {\bibfnamefont
  {I.}~\bibnamefont {Murray}}, \bibinfo {author} {\bibfnamefont
  {M.}~\bibnamefont {Niikura}}, \bibinfo {author} {\bibfnamefont
  {S.}~\bibnamefont {Paschalis}}, \bibinfo {author} {\bibfnamefont
  {M.}~\bibnamefont {Petri}}, \bibinfo {author} {\bibfnamefont
  {H.}~\bibnamefont {Sakurai}}, \bibinfo {author} {\bibfnamefont
  {M.}~\bibnamefont {Salathe}}, \bibinfo {author} {\bibfnamefont
  {P.}~\bibnamefont {Schrock}}, \bibinfo {author} {\bibfnamefont
  {D.}~\bibnamefont {Steppenbeck}}, \bibinfo {author} {\bibfnamefont
  {S.}~\bibnamefont {Takeuchi}}, \bibinfo {author} {\bibfnamefont {Y.~K.}\
  \bibnamefont {Tanaka}}, \bibinfo {author} {\bibfnamefont {R.}~\bibnamefont
  {Taniuchi}}, \bibinfo {author} {\bibfnamefont {H.}~\bibnamefont {Wang}},\
  and\ \bibinfo {author} {\bibfnamefont {K.}~\bibnamefont {Wimmer}},\ }\href
  {https://doi.org/10.1103/PhysRevLett.122.052501} {\bibfield  {journal}
  {\bibinfo  {journal} {Phys. Rev. Lett.}\ }\textbf {\bibinfo {volume} {122}},\
  \bibinfo {pages} {052501} (\bibinfo {year} {2019})}\BibitemShut {NoStop}%
\bibitem [{\citenamefont {Nakada}(2013)}]{Nakada2013_PRC87-014336}%
  \BibitemOpen
  \bibfield  {author} {\bibinfo {author} {\bibfnamefont {H.}~\bibnamefont
  {Nakada}},\ }\href {https://doi.org/10.1103/PhysRevC.87.014336} {\bibfield
  {journal} {\bibinfo  {journal} {Phys. Rev. C}\ }\textbf {\bibinfo {volume}
  {87}},\ \bibinfo {pages} {014336} (\bibinfo {year} {2013})}\BibitemShut
  {NoStop}%
\bibitem [{\citenamefont {Dong}\ \emph {et~al.}(2013)\citenamefont {Dong},
  \citenamefont {Wang}, \citenamefont {Liu},\ and\ \citenamefont
  {Xu}}]{Dong2013_PRC88-024328}%
  \BibitemOpen
  \bibfield  {author} {\bibinfo {author} {\bibfnamefont {G.~X.}\ \bibnamefont
  {Dong}}, \bibinfo {author} {\bibfnamefont {X.~B.}\ \bibnamefont {Wang}},
  \bibinfo {author} {\bibfnamefont {H.~L.}\ \bibnamefont {Liu}},\ and\ \bibinfo
  {author} {\bibfnamefont {F.~R.}\ \bibnamefont {Xu}},\ }\href
  {https://doi.org/10.1103/PhysRevC.88.024328} {\bibfield  {journal} {\bibinfo
  {journal} {Phys. Rev. C}\ }\textbf {\bibinfo {volume} {88}},\ \bibinfo
  {pages} {024328} (\bibinfo {year} {2013})}\BibitemShut {NoStop}%
\bibitem [{\citenamefont {Rodr{\'{i}}guez}(2016)}]{Rodriguez2016_EPJA52-190}%
  \BibitemOpen
  \bibfield  {author} {\bibinfo {author} {\bibfnamefont {T.~R.}\ \bibnamefont
  {Rodr{\'{i}}guez}},\ }\href {https://doi.org/10.1140/epja/i2016-16190-2}
  {\bibfield  {journal} {\bibinfo  {journal} {Eur. Phys. J. A}\ }\textbf
  {\bibinfo {volume} {52}},\ \bibinfo {pages} {190} (\bibinfo {year}
  {2016})}\BibitemShut {NoStop}%
\bibitem [{\citenamefont {Shimada}\ \emph {et~al.}(2016)\citenamefont
  {Shimada}, \citenamefont {Watanabe}, \citenamefont {Tagami}, \citenamefont
  {Matsumoto}, \citenamefont {Shimizu},\ and\ \citenamefont
  {Yahiro}}]{Shimada2016_PRC93-064314}%
  \BibitemOpen
  \bibfield  {author} {\bibinfo {author} {\bibfnamefont {M.}~\bibnamefont
  {Shimada}}, \bibinfo {author} {\bibfnamefont {S.}~\bibnamefont {Watanabe}},
  \bibinfo {author} {\bibfnamefont {S.}~\bibnamefont {Tagami}}, \bibinfo
  {author} {\bibfnamefont {T.}~\bibnamefont {Matsumoto}}, \bibinfo {author}
  {\bibfnamefont {Y.~R.}\ \bibnamefont {Shimizu}},\ and\ \bibinfo {author}
  {\bibfnamefont {M.}~\bibnamefont {Yahiro}},\ }\href
  {https://doi.org/10.1103/PhysRevC.93.064314} {\bibfield  {journal} {\bibinfo
  {journal} {Phys. Rev. C}\ }\textbf {\bibinfo {volume} {93}},\ \bibinfo
  {pages} {064314} (\bibinfo {year} {2016})}\BibitemShut {NoStop}%
\bibitem [{\citenamefont {Wu}\ and\ \citenamefont
  {Zhou}(2015)}]{Wu2015_PRC92-054321}%
  \BibitemOpen
  \bibfield  {author} {\bibinfo {author} {\bibfnamefont {X.-Y.}\ \bibnamefont
  {Wu}}\ and\ \bibinfo {author} {\bibfnamefont {X.-R.}\ \bibnamefont {Zhou}},\
  }\href {https://doi.org/10.1103/PhysRevC.92.054321} {\bibfield  {journal}
  {\bibinfo  {journal} {Phys. Rev. C}\ }\textbf {\bibinfo {volume} {92}},\
  \bibinfo {pages} {054321} (\bibinfo {year} {2015})}\BibitemShut {NoStop}%
\bibitem [{\citenamefont {Takechi}\ \emph
  {et~al.}(2014{\natexlab{b}})\citenamefont {Takechi}, \citenamefont {Suzuki},
  \citenamefont {Nishimura}, \citenamefont {Fukuda}, \citenamefont {Ohtsubo},
  \citenamefont {Nagashima}, \citenamefont {Suzuki}, \citenamefont {Yamaguchi},
  \citenamefont {Ozawa}, \citenamefont {Moriguchi}, \citenamefont {Ohishi},
  \citenamefont {Sumikama}, \citenamefont {Geissel}, \citenamefont {Ishihara},
  \citenamefont {Aoi}, \citenamefont {Chen}, \citenamefont {Fang},
  \citenamefont {Fukuda}, \citenamefont {Fukuoka}, \citenamefont {Furuki},
  \citenamefont {Inabe}, \citenamefont {Ishibashi}, \citenamefont {Itoh},
  \citenamefont {Izumikawa}, \citenamefont {Kameda}, \citenamefont {Kubo},
  \citenamefont {Lee}, \citenamefont {Lantz}, \citenamefont {Ma}, \citenamefont
  {Matsuta}, \citenamefont {Mihara}, \citenamefont {Momota}, \citenamefont
  {Nagae}, \citenamefont {Nishikiori}, \citenamefont {Niwa}, \citenamefont
  {Ohnishi}, \citenamefont {Okumura}, \citenamefont {Ogura}, \citenamefont
  {Sakurai}, \citenamefont {Sato}, \citenamefont {Shimbara}, \citenamefont
  {Suzuki}, \citenamefont {Takeda}, \citenamefont {Takeuchi}, \citenamefont
  {Tanaka}, \citenamefont {Uenishi}, \citenamefont {Winkler}, \citenamefont
  {Yanagisawa}, \citenamefont {Watanabe}, \citenamefont {Minomo}, \citenamefont
  {Tagami}, \citenamefont {Shimada}, \citenamefont {Kimura}, \citenamefont
  {Matsumoto}, \citenamefont {Shimizu},\ and\ \citenamefont
  {Yahiro}}]{Takechi2014_EWC66-02101}%
  \BibitemOpen
  \bibfield  {author} {\bibinfo {author} {\bibfnamefont {M.}~\bibnamefont
  {Takechi}}, \bibinfo {author} {\bibfnamefont {S.}~\bibnamefont {Suzuki}},
  \bibinfo {author} {\bibfnamefont {D.}~\bibnamefont {Nishimura}}, \bibinfo
  {author} {\bibfnamefont {M.}~\bibnamefont {Fukuda}}, \bibinfo {author}
  {\bibfnamefont {T.}~\bibnamefont {Ohtsubo}}, \bibinfo {author} {\bibfnamefont
  {M.}~\bibnamefont {Nagashima}}, \bibinfo {author} {\bibfnamefont
  {T.}~\bibnamefont {Suzuki}}, \bibinfo {author} {\bibfnamefont
  {T.}~\bibnamefont {Yamaguchi}}, \bibinfo {author} {\bibfnamefont
  {A.}~\bibnamefont {Ozawa}}, \bibinfo {author} {\bibfnamefont
  {T.}~\bibnamefont {Moriguchi}}, \bibinfo {author} {\bibfnamefont
  {H.}~\bibnamefont {Ohishi}}, \bibinfo {author} {\bibfnamefont
  {T.}~\bibnamefont {Sumikama}}, \bibinfo {author} {\bibfnamefont
  {H.}~\bibnamefont {Geissel}}, \bibinfo {author} {\bibfnamefont
  {M.}~\bibnamefont {Ishihara}}, \bibinfo {author} {\bibfnamefont
  {N.}~\bibnamefont {Aoi}}, \bibinfo {author} {\bibfnamefont {R.-J.}\
  \bibnamefont {Chen}}, \bibinfo {author} {\bibfnamefont {D.-Q.}\ \bibnamefont
  {Fang}}, \bibinfo {author} {\bibfnamefont {N.}~\bibnamefont {Fukuda}},
  \bibinfo {author} {\bibfnamefont {S.}~\bibnamefont {Fukuoka}}, \bibinfo
  {author} {\bibfnamefont {H.}~\bibnamefont {Furuki}}, \bibinfo {author}
  {\bibfnamefont {N.}~\bibnamefont {Inabe}}, \bibinfo {author} {\bibfnamefont
  {Y.}~\bibnamefont {Ishibashi}}, \bibinfo {author} {\bibfnamefont
  {T.}~\bibnamefont {Itoh}}, \bibinfo {author} {\bibfnamefont {T.}~\bibnamefont
  {Izumikawa}}, \bibinfo {author} {\bibfnamefont {D.}~\bibnamefont {Kameda}},
  \bibinfo {author} {\bibfnamefont {T.}~\bibnamefont {Kubo}}, \bibinfo {author}
  {\bibfnamefont {C.~S.}\ \bibnamefont {Lee}}, \bibinfo {author} {\bibfnamefont
  {M.}~\bibnamefont {Lantz}}, \bibinfo {author} {\bibfnamefont {Y.-G.}\
  \bibnamefont {Ma}}, \bibinfo {author} {\bibfnamefont {K.}~\bibnamefont
  {Matsuta}}, \bibinfo {author} {\bibfnamefont {M.}~\bibnamefont {Mihara}},
  \bibinfo {author} {\bibfnamefont {S.}~\bibnamefont {Momota}}, \bibinfo
  {author} {\bibfnamefont {D.}~\bibnamefont {Nagae}}, \bibinfo {author}
  {\bibfnamefont {R.}~\bibnamefont {Nishikiori}}, \bibinfo {author}
  {\bibfnamefont {T.}~\bibnamefont {Niwa}}, \bibinfo {author} {\bibfnamefont
  {T.}~\bibnamefont {Ohnishi}}, \bibinfo {author} {\bibfnamefont
  {K.}~\bibnamefont {Okumura}}, \bibinfo {author} {\bibfnamefont
  {T.}~\bibnamefont {Ogura}}, \bibinfo {author} {\bibfnamefont
  {H.}~\bibnamefont {Sakurai}}, \bibinfo {author} {\bibfnamefont
  {K.}~\bibnamefont {Sato}}, \bibinfo {author} {\bibfnamefont {Y.}~\bibnamefont
  {Shimbara}}, \bibinfo {author} {\bibfnamefont {H.}~\bibnamefont {Suzuki}},
  \bibinfo {author} {\bibfnamefont {H.}~\bibnamefont {Takeda}}, \bibinfo
  {author} {\bibfnamefont {S.}~\bibnamefont {Takeuchi}}, \bibinfo {author}
  {\bibfnamefont {K.}~\bibnamefont {Tanaka}}, \bibinfo {author} {\bibfnamefont
  {H.}~\bibnamefont {Uenishi}}, \bibinfo {author} {\bibfnamefont
  {M.}~\bibnamefont {Winkler}}, \bibinfo {author} {\bibfnamefont
  {Y.}~\bibnamefont {Yanagisawa}}, \bibinfo {author} {\bibfnamefont
  {S.}~\bibnamefont {Watanabe}}, \bibinfo {author} {\bibfnamefont
  {K.}~\bibnamefont {Minomo}}, \bibinfo {author} {\bibfnamefont
  {S.}~\bibnamefont {Tagami}}, \bibinfo {author} {\bibfnamefont
  {M.}~\bibnamefont {Shimada}}, \bibinfo {author} {\bibfnamefont
  {M.}~\bibnamefont {Kimura}}, \bibinfo {author} {\bibfnamefont
  {T.}~\bibnamefont {Matsumoto}}, \bibinfo {author} {\bibfnamefont {Y.~R.}\
  \bibnamefont {Shimizu}},\ and\ \bibinfo {author} {\bibfnamefont
  {M.}~\bibnamefont {Yahiro}},\ }\href
  {https://doi.org/10.1051/epjconf/20146602101} {\bibfield  {journal} {\bibinfo
   {journal} {EPJ Web of Conferences}\ }\textbf {\bibinfo {volume} {66}},\
  \bibinfo {pages} {02101} (\bibinfo {year} {2014}{\natexlab{b}})}\BibitemShut
  {NoStop}%
\bibitem [{\citenamefont {Wang}\ \emph {et~al.}(2017)\citenamefont {Wang},
  \citenamefont {Audi}, \citenamefont {Kondev}, \citenamefont {Huang},
  \citenamefont {Naimi},\ and\ \citenamefont
  {Xu}}]{Wang2017_ChinPhysC41-030003}%
  \BibitemOpen
  \bibfield  {author} {\bibinfo {author} {\bibfnamefont {M.}~\bibnamefont
  {Wang}}, \bibinfo {author} {\bibfnamefont {G.}~\bibnamefont {Audi}}, \bibinfo
  {author} {\bibfnamefont {F.}~\bibnamefont {Kondev}}, \bibinfo {author}
  {\bibfnamefont {W.}~\bibnamefont {Huang}}, \bibinfo {author} {\bibfnamefont
  {S.}~\bibnamefont {Naimi}},\ and\ \bibinfo {author} {\bibfnamefont
  {X.}~\bibnamefont {Xu}},\ }\href
  {https://doi.org/10.1088/1674-1137/41/3/030003} {\bibfield  {journal}
  {\bibinfo  {journal} {Chin. Phys. C}\ }\textbf {\bibinfo {volume} {41}},\
  \bibinfo {eid} {30003} (\bibinfo {year} {2017})}\BibitemShut {NoStop}%
\bibitem [{\citenamefont {Kasuya}\ and\ \citenamefont
  {Yoshida}(2020)}]{Kasuya2020_PTEP2021-013D01}%
  \BibitemOpen
  \bibfield  {author} {\bibinfo {author} {\bibfnamefont {H.}~\bibnamefont
  {Kasuya}}\ and\ \bibinfo {author} {\bibfnamefont {K.}~\bibnamefont
  {Yoshida}},\ }\href {https://doi.org/10.1093/ptep/ptaa163} {\bibfield
  {journal} {\bibinfo  {journal} {Prog. Theo. Exp. Phys.}\ }\textbf {\bibinfo
  {volume} {2021}} (\bibinfo {year} {2020})}\BibitemShut {NoStop}%
\bibitem [{\citenamefont {Sun}\ \emph {et~al.}(2020{\natexlab{b}})\citenamefont
  {Sun}, \citenamefont {Qian}, \citenamefont {Chen}, \citenamefont {Ring},\
  and\ \citenamefont {Li}}]{Sun2020_PRC101-014321}%
  \BibitemOpen
  \bibfield  {author} {\bibinfo {author} {\bibfnamefont {T.-T.}\ \bibnamefont
  {Sun}}, \bibinfo {author} {\bibfnamefont {L.}~\bibnamefont {Qian}}, \bibinfo
  {author} {\bibfnamefont {C.}~\bibnamefont {Chen}}, \bibinfo {author}
  {\bibfnamefont {P.}~\bibnamefont {Ring}},\ and\ \bibinfo {author}
  {\bibfnamefont {Z.~P.}\ \bibnamefont {Li}},\ }\href
  {https://doi.org/10.1103/PhysRevC.101.014321} {\bibfield  {journal} {\bibinfo
   {journal} {Phys. Rev. C}\ }\textbf {\bibinfo {volume} {101}},\ \bibinfo
  {pages} {014321} (\bibinfo {year} {2020}{\natexlab{b}})}\BibitemShut
  {NoStop}%
\bibitem [{\citenamefont {Zhao}\ \emph {et~al.}(2010)\citenamefont {Zhao},
  \citenamefont {Li}, \citenamefont {Yao},\ and\ \citenamefont
  {Meng}}]{Zhao2010_PRC82-054319}%
  \BibitemOpen
  \bibfield  {author} {\bibinfo {author} {\bibfnamefont {P.~W.}\ \bibnamefont
  {Zhao}}, \bibinfo {author} {\bibfnamefont {Z.~P.}\ \bibnamefont {Li}},
  \bibinfo {author} {\bibfnamefont {J.~M.}\ \bibnamefont {Yao}},\ and\ \bibinfo
  {author} {\bibfnamefont {J.}~\bibnamefont {Meng}},\ }\href
  {https://doi.org/10.1103/PhysRevC.82.054319} {\bibfield  {journal} {\bibinfo
  {journal} {Phys. Rev. C}\ }\textbf {\bibinfo {volume} {82}},\ \bibinfo
  {pages} {054319} (\bibinfo {year} {2010})}\BibitemShut {NoStop}%
\bibitem [{\citenamefont {Taninah}\ \emph {et~al.}(2020)\citenamefont
  {Taninah}, \citenamefont {Agbemava}, \citenamefont {Afanasjev},\ and\
  \citenamefont {Ring}}]{Taninah2020_PLB800-135065}%
  \BibitemOpen
  \bibfield  {author} {\bibinfo {author} {\bibfnamefont {A.}~\bibnamefont
  {Taninah}}, \bibinfo {author} {\bibfnamefont {S.~E.}\ \bibnamefont
  {Agbemava}}, \bibinfo {author} {\bibfnamefont {A.~V.}\ \bibnamefont
  {Afanasjev}},\ and\ \bibinfo {author} {\bibfnamefont {P.}~\bibnamefont
  {Ring}},\ }\href {https://doi.org/10.1016/j.physletb.2019.135065} {\bibfield
  {journal} {\bibinfo  {journal} {Phys. Lett. B}\ }\textbf {\bibinfo {volume}
  {800}},\ \bibinfo {pages} {135065} (\bibinfo {year} {2020})}\BibitemShut
  {NoStop}%
\bibitem [{\citenamefont {Nik\v{s}i\'{c}}\ \emph {et~al.}(2008)\citenamefont
  {Nik\v{s}i\'{c}}, \citenamefont {Vretenar},\ and\ \citenamefont
  {Ring}}]{Niksic2008_PRC78-034318}%
  \BibitemOpen
  \bibfield  {author} {\bibinfo {author} {\bibfnamefont {T.}~\bibnamefont
  {Nik\v{s}i\'{c}}}, \bibinfo {author} {\bibfnamefont {D.}~\bibnamefont
  {Vretenar}},\ and\ \bibinfo {author} {\bibfnamefont {P.}~\bibnamefont
  {Ring}},\ }\href {https://doi.org/10.1103/PhysRevC.78.034318} {\bibfield
  {journal} {\bibinfo  {journal} {Phys. Rev. C}\ }\textbf {\bibinfo {volume}
  {78}},\ \bibinfo {pages} {034318} (\bibinfo {year} {2008})}\BibitemShut
  {NoStop}%
\end{thebibliography}
%apsrev4-2.bst 2019-01-14 (MD) hand-edited version of apsrev4-1.bst
%Control: key (0)
%Control: author (72) initials jnrlst
%Control: editor formatted (1) identically to author
%Control: production of article title (-1) disabled
%Control: page (0) single
%Control: year (1) truncated
%Control: production of eprint (0) enabled
%

\end{document}